\documentclass[aps,prb,twocolumn,showpacs,floats]{revtex4}
\usepackage{latexsym}
\usepackage{amsmath}
\usepackage{amssymb}
\usepackage{bm}
\usepackage{wasysym}
\usepackage{graphicx}
\usepackage{graphicx,color}
\usepackage{subfigure}
\usepackage{epsfig}

\newcommand{\veps}{\varepsilon}

\def\mbfE{\mathbf{E}}
\def\mbfe{\mathbf{e}}

\def\g{\mathrm{g}}
\def\e{\mathrm{e}}

\def\mbfk{\mathbf{k}}

\def\mbfr{\mathbf{r}}
\def\mbfR{\mathbf{R}}

\def\mbfS{\mathbf{S}}

\def\1S0{$^{1}$S$_{0}$}
\def\3P2{$^{3}$P$_{2}$}

\usepackage{amsmath,stmaryrd,graphicx}
\makeatletter
\newcommand{\fixed@sra}{$\vrule height 2\fontdimen22\textfont2 width 0pt\shortrightarrow$}
\newcommand{\shortarrow}[1]{%
  \mathrel{\text{\rotatebox[origin=c]{\numexpr#1*45}{\fixed@sra}}}
}
\makeatother

\begin{document}

\title{Multipolar Kondo Effect in $^1$S$_0$-$^3$P$_2$ Mixture of
       $^{173}$Yb Atoms}

\author{Igor Kuzmenko$^1$, Tetyana Kuzmenko$^1$,
    Yshai Avishai$^{1,2,4}$ and Gyu-Boong Jo$^3$}

\affiliation{$^1$Department of Physics,
  Ben-Gurion University of the Negev, Beer-Sheva, Israel \\
  $^2$NYU-Shanghai, Pudong, Shanghai, China,\\
  $^3$Department of Physics,
  The Hong Kong University of Science and Technology,
  Clear Water Bay, Kowloon, Hong Kong, China\\
  $^4$Yukawa Institute for Theoretical Physics, Kyoto University,
  Kyoto 606-8502, Japan}

\begin{abstract}
Whereas in the familiar Kondo effect the exchange interaction
is {\it dipolar}, there are systems in which the exchange
interaction  is {\it multipolar}, as has been
realized in a recent experiment.
Here  we study multipolar Kondo effect in a Fermi gas of
cold $^{173}$Yb atoms. Making use of different AC
polarizabilities of the electronic ground state
Yb($^{1}$S$_{0}$) and the long-lived metastable
state Yb$^{*}$($^{3}$P$_{2}$), it is suggested
that the latter atoms can be localized and serve as a dilute
concentration of magnetic impurities while the former ones
remain itinerant. The exchange mechanism between the itinerant
Yb and the localized Yb$^{*}$ atoms
is analyzed and shown to be antiferromagnetic.
The quadrupole and octupole interactions act to enhance the Kondo
temperature $T_K$ that is found to be experimentally accessible.
The bare exchange Hamiltonian needs to be decomposed into
dipole (${\mathrm{d}}$), quadrupole (${\mathrm{q}}$) and
octupole (${\mathrm{o}}$) interactions in order to
retain its form under renormalization group (RG)
analysis, in which the corresponding exchange constants
($\lambda_{\mathrm{d}}$, $\lambda_{\mathrm{q}}$ and
$\lambda_{\mathrm{o}}$) flow independently. Numerical solution of
the RG scaling equations reveals a few finite fixed points.
Arguments are presented that the Fermi liquid fixed point at low
temperature is unstable, indicating that the impurity is
over-screened, which suggests a non-Fermi liquid phase.
The impurity contributions to the specific heat, entropy and
the magnetic susceptibility are calculated in the weak coupling
regime (${T}\gg{T}_{K}$), and are compared with the analogous
results obtained for the standard case of dipolar exchange
interaction (the $s-d$ Hamiltonian).
\end{abstract}

\maketitle

\section{Introduction}
\label{sec-intro}
\noindent
\underline{\it Background} In its most elementary form,
the (single channel) Kondo model describes the physics of
a magnetic impurity (of spin operator ${\bf S}$), immersed
in a host metal with a single continuous band of noninteracting
electrons (of spin operator ${\bf s}$, with $s=\tfrac{1}{2} \hbar$)
\cite{Kondo-64,Coleman-PhysWorld-95,Anderson-PhysWorld-95,%
Hewson-book}.
The impurity and the band electrons are coupled via
an antiferromagnetic exchange interaction $J{\bf s} \cdot {\bf S}$
of strength $J>0$. The corresponding Hamiltonian $H_{\mathrm{s-d}}$
has an SU(2) symmetry.  The Kondo model can naturally be
generalized into the Coqblin-Schrieffer model whose Hamiltonian
$H_{\mathrm{C-S}}$ has an SU(N) symmetry
\cite{Hewson-book,Rajan-prl-83,Schlottmann-83,KE-prb-98}.
Renormalization group (RG) analysis shows that the respective low
energy fixed point in either model is stable and that
the corresponding fixed point Hamiltonian describes regular or
singular Fermi liquid (FL), in which the impurity is fully or
under screened.\\
\ \ \ \ A seminal paper by Nozi\`eres and Blandin (NB) back in 1980,
discusses the fixed points of $H_{\mathrm{s-d}}$
under the assumption that a few electron channels  participate in
the impurity screening\cite{NB-JPh-80}. More precisely, suppose that
by some mechanism, there are $N$ independent electron channels
contributing to screening.  Then, under favourable conditions on
the corresponding exchange constants, together with the (hereafter
referred as NB inequality) $N>2S$, the impurity is
{\it over-screened}. The Hamiltonian of such over-screened
Kondo system flows at low temperature to a new fixed point in
which it displays a non-Fermi liquid (NFL) behaviour. Searching for
an experimental manifestation of over-screened Kondo effect in
solid state systems was notoriously frustrating, but eventually
it was demonstrated in a specifically designed quantum-dot system
\cite{Oreg-prl-03}.

Another route to over-screening in the Kondo effect is {\it single
channel over-screening by large spin Fermi-sea} \cite{Kim-prb-96}.
It is expected to occur when a magnetic impurity  is immersed in
a host Fermi-sea with a continuous band of noninteracting fermions
of spin operator ${\bf s}$, with $s>\tfrac{1}{2}\hbar$.
This system is shown to be equivalent to that with $N(s)$
independent electron channels where
\begin{equation}
  \label{Ns}
  N(s) = \frac{2}{3}~s(s+1)(2s+1).
\end{equation}
Since $N(s)$ is a cubic function of $s$, (for example
$N(\tfrac{5}{2})=35$), the NB inequality $N>2S$ (that is
a necessary but not sufficient condition for over-screening) can
easily be satisfied.   While it is hard (albeit possible) to perceive its
realization in solid state systems, the revelation and
the possible control of a gas composed of cold fermionic atoms
within a periodic optical lattice potential turned this scenario
to be realistic also outside the realm of
solid-state systems\cite{KE-opt-latt-prl-05,KE-quadrupole-prb-16}.
Indeed, we  have recently suggested a general framework for
a pertinent experiment to test this scenario of over-screening
and analyzed the conditions and parameter range for its
realization \cite{KKAK-prb-15}. It can be performed in a few
laboratories that specialize in controlling cold fermionic atoms.

In both cases (multi-channel and/or large spin over screening),
the corresponding Hamiltonian has an SU(2) symmetry, and
the exchange interaction is {\it dipolar}. An extension of
the SU(2) multi-channel over-screening scenario into over-screened
multi-channel SU($N$) Kondo model is discussed in Ref.
\cite{SUN-overscr-KE-prb-98}.
\\

\noindent
\underline {\it In the present work} we focus on an experimentally
accessible cold atom system and examine the concept of large spin
Kondo over-screening beyond  SU(2), in case where the exchange
interaction is {\it multipolar}.   This is motivated by a recent
experiment where a multipolar Kondo effect
is realized in solid state system \cite{KE-quadrupole-prb-16}. For
the cold atom arena, a concrete experimental candidate system is
that of fermionic alkali-earth-like isotopes such as $^{173}$Yb
atoms \cite{Fukuhara07,Zhang-Yb-prl15,Riegger-Yb-arxiv17,%
Zhang-Yb-KE-pra16,exchange-Yb-PRL-15}.
The underlying idea is to localize
an $^{173}$Yb atom in its long lived excited $^{3}$P$_{2}$ state
with atomic spin $F=\frac{3}{2}$, in a Fermi sea of non (or
weakly)-interacting itinerant $^{173}$Yb atoms in their ground
state $^{1}$S$_{0}$  with atomic spin $I=\frac{5}{2}$. For that
purpose, it should be demonstrated that an antiferromagnetic
exchange interaction exists between the impurity
Yb$^{*}$($^{3}$P$_{2}$) and the itinerant
Yb($^{1}$S$_{0}$) atoms. Intuitively, in that case
we might expect an over-screening by large spin scenario, since
the angular momentum of the itinerant atoms is larger than that
of the impurity atom ($I=\tfrac{5}{2} > F=\tfrac{3}{2}$).
However, quantitative analysis turns out to be extremely
complicated, due to several factors. First, elucidation and
calculation of the exchange interaction is rather involved, and
requires sophisticated multipole expansions to handle the pertinent
angular momentum algebra. Second, identifying and constructing
the explicit form of the exchange term is rather tedious, and,
unfortunately, the pertinent Kondo Hamiltonian does not have
a definite symmetry. Third, in order to identify the NFL fixed
points, perturbative RG calculations within the poor-man scaling
procedure must go at least up to third order and the relevant
expressions are long and involve multipole summations on angular
momentum quantum numbers. Finally, elucidating the NFL physics
requires the use of non-perturbative techniques (Bethe ansatz or
conformal field theory) which are still not developed for
this class of Hamiltonians.

It is worthwhile stressing at this early stage a central point
distinguishing multipolar from an SU(2) over-screening resulting
from our analysis: If the spin $F$ of the impurity and the spin
$I$ of itinerant fermions satisfy the inequalities
$$
  F \geq 3/2,
  \ \ \
  I \geq 3/2,
  \ \ \
  N(I) > 2S,
$$
the NB fixed point ${j}^{*}=1/N(I)$ is unstable.
The stable fixed points exposed here are
distinct from the NB fixed point, and
correspond to  different NFL  phases.
The reason is that in the process of carrying out
Schrieffer-Wolf transformations, one usually restricts oneself to
second order perturbation theory. However, when
quadrupole, octupole and higher exchange interactions are
present, new interactions are generated within the Schrieffer-Wolf
 procedure. At high temperature
these interactions are weaker than the lowest order (dipole)
interaction. But at low temperature, these interactions turn
the NB fixed point to be unstable.

\noindent
\underline{\it Organization:} The paper is organized as follows:
In section \ref{Exp} the question of experimental realization is
addressed. Specifically, we substantiate the feasibility of
fabricating a system consisting of a Fermi gas of $^{173}$Yb atoms
in their ground state (electronic configuration $^1$S$_0$) and
a small concentration of $^{173}$Yb atoms in their long lived
excited state (electronic configuration $^3$P$_2$) trapped in
a suitably designed optical potential. The rest of the paper is
devoted to theoretical analysis.
The Kondo Hamiltonian $H_K$ is derived
in Sec. \ref{sec-HK}. The main
technical endeavours are related to the decomposition of $H_K$
into $2^n$ poles components ($n=1,2,3$), and the numerical
estimates of the pertinent coupling constants.
The single atomic energies are estimated
in Sec. \ref{sec-model}. Exchange interaction between
$^{173}$Yb(\1S0) and  $^{173}$Yb$^{*}$(\3P2)
atoms is analyzed in section \ref{sec-exchange}.
In Section
\ref{sec-poor-man-scaling-2nd} the perturbative RG calculations
pertaining to $H_K$ are detailed up to second order. Although
the derivation of these corrections is rather technical, we find
it useful to present it within the main text because it starts
from the standard diagrams of poor-man's scaling analysis, and
the analysis that enables us to overcome the complexities
stemming from the relevant spin algebra is quite instructive.
At the end of this section we write down and solve the relevant
scaling equations (up to second order). The solutions enable us
to elucidate the Kondo temperature $T_K$, as explained in section
\ref{sec-TK}. However, to find stable fixed points that are
candidates for NFL behaviour, one must expand the perturbative RG
calculations up to third order. These calculations are carried
out in section \ref{sec-third-order}. Despite its highly technical
nature, we  include it in the main text, for the same reasons as
for section \ref{sec-poor-man-scaling-2nd}. The most significant
result that emerges is a list of seven possible fixed points. Yet,
further stability analysis is required in order to sort out the stable ones,
by linearizing the RG equations and identifying relevant and
irrelevant exponents. At the end of this procedure, only three
stable points are left. Analysis of the relation between
the fixed points and the interaction parameters is carried out in
section \ref{sec-analysis}, and the claim that the infinite fixed point in
the strong coupling limit is unstable is detailed in section \ref{sec-ground-state}.

\begin{figure}[htb]
\centering
  \includegraphics[width=70 mm,angle=0] {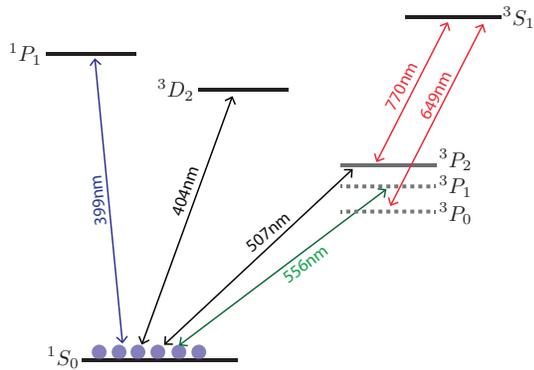}
\caption{(color online) 
    The electronic level structure of  $^{173}$Yb atoms. 
    A small portion of the atoms are optically pumped
    into the $^3$P$_2$ excited state via a resonant  $^1$S$_0$-$^3$P$_2$
    transition or an intermediate  $^1$S$_0$-$^3$D$_2$ transition.}
 \label{Fig-expSchematic}
\end{figure}

In section \ref{sec-suscept} we derive
(in the weak coupling regime $T>T_K$)
expressions for the impurity contributions to a few
thermodynamic observables related to this system,
and compare them with the corresponding quantities
for the standard Kondo effect based on the $s-d$ exchange
Hamiltonian. These include the impurity contribution to
the specific heat, entropy and magnetic susceptibility.
A short summary listing our main achievements is presented
in section \ref{sec-conclusion}. Numerous technical issues are
discussed in the Appendices.

\section{Experimental Feasibility}
\label{Exp}
Recent development of producing degenerate Bose and Fermi
gases of alkaline-earth-like atoms has attracted a great deal of
interests in utilizing such atoms for the study of many-body
physics (in the context of quantum simulation)
\cite{Pagano-Nat-Phys-14,Cappellini-PRL-14,Scazza-Nat-Phys-14} and
the realization of quantum computation
\cite{Ladd-Nat-10,Daley-QIP-11}. The enlarged SU($N$) spin symmetry for
the fermionic isotopes of alkaline-earth-like atoms expands our
capability in exploring large spin physics in low dimensions
\cite{Pagano-Nat-Phys-14} and a two-orbital Fermi gas with SU($N$)
interactions \cite{Cappellini-PRL-14,Scazza-Nat-Phys-14,Zhang-Sci-14}
in which the interactions can be tuned by the orbital Feshbach
resonance~\cite{Zhang-Yb-prl15}.
In addition, a narrow optical transition between the singlet and
the triplet state enables to realize a spin-orbit coupled Fermi
gas with minimal heating~\cite{Song-arxiv-16,Kolkowitz-arXiv-16,livi-arxiv-16,Song-arxiv-17,Bromley-arxiv-17},
and a long-lived triplet state holds the promise in studying
the Kondo effect~\cite{Riegger-Yb-arxiv17}. It has been demonstrated that in such a systems,
the Kondo temperature can be increased due to
the spin-exchange interaction and the confinement-induced resonance
effect\cite{Zhang-Yb-KE-pra16}.

More concretely,  making use of different AC polarizabilities of 
the ground state and the metastable state, the latter can be
localized and serve as local moments. Kondo effect is then
expected to occur due to an exchange interaction between
the atoms in the ground state $^1$S$_0$ and the atoms in
the metastable $^3$P$_2$ state. Such scenario, pertaining to
spin-exchange interactions between $^1$S$_0$ and $^3$P$_0$ states
has been explored in our previous work where it is shown
to realize an SU(6) Coqblin-Schrieffer model
\cite{IK-TK-YA-GBJ-PRB-16}.

Our starting point is a degenerate Fermi gas of
the ground-state $^{173}$Yb atoms. 
Subsequently, a small portion,
(a few $\%$) of the ground
state ${}^1$S${}_0$ atoms will be directly excited to the
${}^3$P${}_2$ level using a narrow line-width ${}^1$S${}_0-$${}^3$P${}_2$
transition at 507~nm~\cite{Yamaguchi-New-JPhys-10}.
Alternatively, the ground state atoms may be pumped into the state
${}^3$D${}_2$ with a 404~nm light and then spontaneously decay to
the ${}^3$P${}_2$ state~\cite{Khramov-PRL-14}. During the pumping
process, a three-dimensional optical lattice potential may be
applied to suppress the recoil kick in the Lamb Dicke regime.

To realize the Kondo model in a mixture of ${}^1$S${}_0$-${}^3$P${}_2$
atoms, it is critical to minimize the anisotropy of the trapping potential for
the localized ${}^3$P${}_2$ atoms whose atomic polarizability
$\alpha_j$ depends on the magnetic quantum number $j$
(${j}\equiv{J_z}$ with $J=2$).
Such anisotropic polarizability would lift
the degeneracy of the Kondo state of localized ${}^3$P${}_2$
atoms. For this reason, we propose to use the trapping light at
the double-magic wave-length $\lambda_0\simeq$~546~nm
which results\cite{Khramov-PRL-14} $\alpha_{|j|}=\alpha_e(\lambda_0)$ for
$|j| \leqslant$2. After preparing
a mixture of ${}^1$S${}_0$-${}^3$P${}_2$ atoms,
a three-dimensional optical lattice generated by double-magic
wavelength lights is adiabatically switched on in such a way
that the Fermi energy of ${}^1$S${}_0$ atoms is larger than
the lattice depth while the low-density ${}^3$P${}_2$ atoms
are localized by the lattice potential.

In the proposed experiment, ytterbium atoms in different orbitals,
Yb and Yb$^*$ atoms, can be selectively detected to extract
the thermodynamic quantities that will be discussed later. The Yb
atom in the ${}^1$S${}_0$ state are imaged using the 399~nm
${}^1$S${}_0$-${}^1$P${}_1$ transition. To image Yb$^*$ atoms,
we first blast Yb atoms with 399~nm light followed by optical
pumping Yb$^*$ atoms into the $^3$P$_1$ state with 770~nm and
649~nm lights (see Fig.~\ref{Fig-expSchematic}). The pumped atoms
in the $^3$P$_1$ state then decay to the $^1$S$_0$ state that can
be imaged by 399~nm light~\cite{Yamaguchi-New-JPhys-10}.
\vspace{-0.1in}
\section{Multipolar Kondo Hamiltonian}
  \label{sec-HK}
\noindent
The total Hamiltonian of the system includes the kinetic energy part $H_{\mathrm c}$
of the itinerant $^{173}$Yb($^1$S$_0$) fermion atoms, the internal excitation $H_{\mathrm {imp}}$ of the 
impurity (that is the trapped $^{173}$Yb($^3$P$_2$) atom),  and their mutual (exchange) interaction
 $H_{\mathrm K}$,  hereafter referred to as the exchange (or Kondo) Hamiltonian. 
 The structures of the various parts  are encoded by their
 (1) operator content (creation, annihilation or Hubbard), 
(2) spin coupling geometric tensors, (3) energies of the 
itinerant and the impurity atoms, and (4), 
strength of the exchange constant.  For pedagogical reasons,
this section focuses on the operator content. 
Definition and explicit expressions for the geometrical tensors 
is detailed in Appendix \ref{append-multipole}, 
while the atomic energies are defined and calculated in section \ref{sec-model}. 
Finally, the estimate of the exchange constant 
is  discussed in section \ref{sec-exchange}. 
 
The expressions for $H_{\mathrm c}$ and $H_{\mathrm imp}$ are 
simple and self-evident as they are diagonal 
in the appropriate bases [see Eq.(\ref{H-gas-def}) below].
On the other hand,  a proper treatment of $H_{\mathrm K}$ requires  some care.
Its expression  in terms of creation, annihilation and 
Hubbard operators and Clebsch-Gordan coefficients 
[see Eq.~(\ref {HK-def}) bellow] is relatively simple, and 
involves {\it a single bare coupling constant} $\lambda$.
However, 
in this form, the Hamiltonian does not keep its structure
under the poor-man scaling RG transformation (introduced in Sections
\ref{sec-poor-man-scaling-2nd} and
\ref{sec-third-order}). To alleviate this riddle, it is 
necessary to decompose the bare Hamiltonian (\ref {HK-def}) into a sum of terms
with {\it different multi-polarities}, $H_{\mathrm{d}}$ (dipole), $H_{\mathrm{q}}$ (quadrupole),
and $H_{\mathrm{o}}$ (octupole), whose strengths are determined by corresponding 
pre-factors $\lambda_{\mathrm{d}}$, $\lambda_{\mathrm{q}}$, and $\lambda_{\mathrm{o}}$.  
Initially, the ratios $\lambda_{\mathrm{d,q,o}}/\lambda$ are geometrical factors  
explicitly calculated below.
However, as will be shown in Sections
\ref{sec-poor-man-scaling-2nd} and
\ref{sec-third-order},
the three coefficients $\lambda_{\mathrm{d,q,o}}$
are renormalized {\it differently}, but the multipolar terms
 $H_{\mathrm {d,q,o}}$ keep their initial form, as is required by the RG procedure. 
This is the reason for the multipolar decomposition. The price is that 
 the spin algebra  required for manipulating the multipolar terms is complicated.  

\subsection{The total Hamiltonian}
\label{subsec-Htot}

The system's Hamiltonian reads
\begin{equation}
  \label{H-tot-def}
  H =
  \underbrace{H_{\mathrm{c}}+
  H_{\mathrm{imp}}}_{H_0}+H_K.
\end{equation}
Here
\begin{equation}
  H_{\mathrm{c}} =
  \sum_{n,i}
  \veps_n
  c_{n,i}^{\dag}
  c_{n,i},
  \ \
  H_{\mathrm{imp}} =
  \veps_{\mathrm{imp}}
  \sum_{f}
  X^{f,f}.
 \label{H-gas-def}
\end{equation}
The structure of $H_c$ is evident:
The itinerant atoms are trapped in a shallow harmonic potential and 
form a Fermi gas. Their wave functions 
(determined by the harmonic quantum number $n$
and the (nuclear) magnetic quantum number $i$),
are derived in Appendix \ref{append-trap}.
Correspondingly,  $c_{n,i}$ and $c_{n,i}^{\dag}$ are creation and annihilation
operators of the Yb(\1S0) atoms with these prescribed quantum numbers. 
Expressions for the energies  $\{ \veps_n \}$  are given in
eqs. (\ref{veps-itinerant})
 [see Appendix \ref{append-trap}
for further details].

The  localized impurity Hamiltonian $H_{\mathrm{imp}}$ is expressed in terms of
the Hubbard operators $X^{f=F_z,f'=F'_z}$,
\begin{equation}
  X^{f,f'} =
  \big|
      F,f
  \big\rangle
  \big\langle
      F,f'
  \big|,
\end{equation}
where $|F,f\rangle$ is the ket state of the localized impurity
with total spin $F$ and magnetic quantum number $f$. 
Expression for the energy $\veps_{\mathrm{imp}}$   is given in  Eq.~(\ref{veps-imp}).

\subsection{The Kondo Hamiltonian}
The Kondo Hamiltonian $H_K$ describes 
the scattering of itinerant $^{173}$Yb($^1$S$_0$) and localized
$^{173}$Yb($^3$P$_2$) atoms. Recall that the itinerant atom is in
its ground state with  total atomic spin $I=\frac{5}{2}$ (which is
the nuclear spin since the electronic spin is zero), and the trapped atom is in the long lived
excited state with its total atomic spin being $F=\frac{3}{2}$.
In its bare form, the interaction between the atoms, consisting of potential
scattering and  exchange terms, reads \cite{IK-TK-YA-GBJ-PRB-16},
\begin{eqnarray}
  H_K &=&
 \lambda
  \sum_{j}
  \sum_{f,f'}
  \sum_{i,i'}
  \sum_{n,n'}
  C_{J,j;I,i}^{F,f}
  C_{J,j;I,i'}^{F,f'}
  \times \nonumber \\ && \times
  X^{f,f'}~
  c_{n',i'}^{\dag}
  c_{n,i}.
  \label{HK-def}
\end{eqnarray}
Here the total electronic spin configuration of the  excited 
Yb($^{3}$P$_{2}$) atom is $|J j \rangle$ with $J=2$ and magnetic 
quantum number $|j|<J$, and
$C_{J,j;I,i}^{F,f}=\langle JjIi|Ff \rangle$ are Clebsch-Gordan coefficients. 
The exchange coupling constant 
$\lambda$, a central quantity in this context, is estimated below, see
eq. (\ref{exchange-strength-def}).
The Kondo Hamiltonian (\ref{HK-def}) conserves the $z$
component of the total angular momentum that is,
$$
  i+f'
  ~=~
  i'+f.
$$
The selection rules for
the Hamiltonian (\ref{HK-def}) are,
\begin{eqnarray}
  \Delta{f} &=& f-f'=
  0,~
  \pm1,~
  \pm2,~
  \pm3.
  \label{selection-rules-HK}
\end{eqnarray}
We shall now rewrite this {\it same Kondo Hamiltonian} as a sum of
terms representing potential, dipole, quadrupole and octupole
interactions,
\begin{eqnarray}
  H_{K} &=&
  H_{\mathrm{p}}+
  H_{\mathrm{d}}+
  H_{\mathrm{q}}+
  H_{\mathrm{o}}.
  \label{HK-Hd-Hq-Ho}
\end{eqnarray}
The precise expressions for these multipolar components
are given in the next subsection.
As will be evident from the discussion below, this form of
the Hamiltonian is more complicated than its initial form
(\ref{HK-def}). The reason for using the equivalent form
(\ref{HK-Hd-Hq-Ho}) is that if one applies the 
RG analysis on its bare form, 
Eq.~(\ref{HK-def}), {\it the structure of
the Hamiltonian  changes under the poor-man's scaling procedure}.
In other words, in addition to the fact that  
the single coupling constant $\lambda$ is renormalized, 
the Hamiltonian acquires a different structure,  
and that violates the spirit of the RG formalism.
Consequently, we need to
express the Hamiltonian in such a way that its structure is
unchanged under the poor-man's scaling procedure, whereas only
the coupling constants renormalize. As we shall show below, this
is indeed the case once we use the form (\ref{HK-Hd-Hq-Ho}).

\subsection{Multipolar terms and bare coefficients $\lambda_{\mathrm{d,q,o}}$}
  \label{subsec-HK-fit}

The four terms in the decomposition (\ref{HK-Hd-Hq-Ho}) involve
four exchange coefficients $\lambda_{\mathrm{p,d,q,o}}$, each one being
proportional to  $\lambda$, where the respective proportionality constants are
just simple geometric factors to be written down below in Eq.~(\ref{Jd-Jq-Jo-res}). 
In addition, we  
need to introduce three pairs of dipole, quadrupole  and octupole  tensors 
$\{ F^{\alpha}_{f,f'},
  I^{\alpha}_{i',i} \}, \{  F^{\alpha,\alpha'}_{f,f'},
  I^{\alpha',\alpha}_{i',i} \}$ and 
  $\{  F^{\alpha,\alpha',\alpha''}_{f,f'}, 
  I^{\alpha'',\alpha',\alpha}_{i',i} \}$ 
  ($F^{\alpha}_{f,f'}$ for the itinerant atoms and $I^{\alpha}_{i',i}$ for the impurity atoms), 
in order to manipulate the complex spin algebra. 
These tensors are defined in Appendix \ref{append-multipole}.

We are now in a position to write down the structure of 
the four parts of $H_K$. 
The first term on the right hand side of eq. (\ref{HK-Hd-Hq-Ho}),
$H_{\mathrm{p}}$, is due to potential scattering,
\begin{eqnarray}
  H_{\mathrm{p}} =
\lambda_{\mathrm{p}}
  \sum_{f}
  \sum_{i}
  \sum_{n,n'}
  X^{f,f}~
  c_{n',i}^{\dag}
  c_{n,i}.
  \label{H-pot}
\end{eqnarray}
The second term on the right hand side of eq. (\ref{HK-Hd-Hq-Ho}),
$H_{\mathrm{d}}$, is the dipole exchange interaction,
\begin{eqnarray}
  H_{\mathrm{d}} =
  \lambda_{\mathrm{d}}
  \sum_{\alpha}
  \sum_{f,f'}
  \sum_{i,i'}
  \sum_{n,n'}
  F^{\alpha}_{f,f'}
  I^{\alpha}_{i',i}~
  X^{f,f'}~
  c_{n',i'}^{\dag}
  c_{n,i}.
  \label{H-d-d}
\end{eqnarray}
The dipole tensors
$F^{\alpha}_{f,f'}$ and $I^{\alpha}_{f,f'}$ are
explicitly defined in Eq.~(\ref{Fdipole-def}) 
of Appendix \ref{append-multipole}, and
$\lambda_{\mathrm{d}}$ is the coupling strength of the dipole 
interaction.
The third term on the right hand side of eq. (\ref{HK-Hd-Hq-Ho}),
$H_{\mathrm{q}}$, is the quadrupole exchange interaction,
\begin{eqnarray}
  H_{\mathrm{q}} &=&
  \lambda_{\mathrm{q}}
  \sum_{\alpha,\alpha'}
  \sum_{f,f'}
  \sum_{i,i'}
  \sum_{n,n'}
  F^{\alpha,\alpha'}_{f,f'}
  I^{\alpha',\alpha}_{i',i}
  \times \nonumber \\ && \times
  X^{f,f'}~
  c_{n',i'}^{\dag}
  c_{n,i},
  \label{H-q-q}
\end{eqnarray}
where
$\hat{F}^{\alpha,\alpha'}_{f,f'}$ or $\hat{I}^{\alpha,\alpha'}_{f,f'}$ are
explicitly defined  in
Eq. (\ref{quadrupole-def}) of Appendix \ref{append-multipole},
and $\lambda_{\mathrm{q}}$ is the coupling strength of the quadrupole
interaction.  
Finally, the fourth term on the right hand side of eq.
(\ref{HK-Hd-Hq-Ho}), $H_{\mathrm{o}}$, is the octupole interaction,
\begin{eqnarray}
  H_{\mathrm{o}} &=&
  \lambda_{\mathrm{o}}
  \sum_{\alpha,\alpha',\alpha''}
  \sum_{f,f'}
  \sum_{i,i'}
  \sum_{n,n'}
  F^{\alpha,\alpha',\alpha''}_{f,f'}
  I^{\alpha'',\alpha',\alpha}_{i',i}
  \times \nonumber \\ && \times
  X^{f,f'}~
  c_{n',i'}^{\dag}
  c_{n,i},
  \label{H-o-o}
\end{eqnarray}
where $\hat{F}^{\alpha,\alpha',\alpha''}_{f,f'}$ or
$\hat{I}^{\alpha,\alpha',\alpha''}_{f,f'}$ are 
explicitly defined in
Eq. (\ref{octupole-def}) of Appendix \ref{append-multipole},
and
$\lambda_{\mathrm{o}}$ is the coupling strength of the octupole interaction.

On the bare level (before starting the poor-man's scaling
procedure), $H_K$ is determined by a {\it single coupling constant} $\lambda$.
Therefore, all the four coefficients $\lambda_{\mathrm{p}}$,
$\lambda_{\mathrm{d}}$,
$\lambda_{\mathrm{q}}$ and $\lambda_{\mathrm{o}}$ are simply
related to $\lambda$ through geometric factors.
Straightforward  analysis shows that the Hamiltonians
(\ref{HK-def}) and (\ref{HK-Hd-Hq-Ho}) are identical provided,
\begin{eqnarray}
  &&
  \lambda_{\mathrm{p}} = \frac{\lambda}{6},
  \ \ \ \ \
  \ \ \ \
  \lambda_{\mathrm{d}} =\frac{26\lambda}{525},
  \nonumber
  \\
  &&
  \lambda_{\mathrm{q}} =-\frac{\lambda}{840},
  \ \ \ \ \
  \lambda_{\mathrm{o}} =
  -\frac{\lambda}{1890}. 
  \label{Jd-Jq-Jo-res}
\end{eqnarray}

Note that on deriving the exchange Hamiltonian (\ref{HK-def}), we
neglect the spin-orbit and hyperfine interactions. It can be shown
that taking into account the spin-orbit and hyperfine interactions
modify the Hamiltonian (\ref{HK-def}), but leave the Hamiltonian
(\ref{HK-Hd-Hq-Ho}), [as well  as the dipole, quadrupole and octupole
interactions (\ref{H-d-d}), (\ref{H-q-q}) and (\ref{H-o-o})]
unchanged, except for slight modifications of the couplings
$\lambda_{\mathrm{d}}$, $\lambda_{\mathrm{q}}$ and
$\lambda_{\mathrm{o}}$. Indeed, as it is shown in Appendix
\ref{append-eigenstates-dd-qq-oo}, the Hamiltonian
(\ref{HK-Hd-Hq-Ho}) obeys spin-rotation SU(2) symmetry.
The spin-orbit and hyperfine interactions satisfy the same SU(2)
symmetry \cite{Landau-Lifshitz-III}. As a result, the spin-orbit
and hyperfine interactions cannot change the Hamiltonian
(\ref{HK-Hd-Hq-Ho}).


\section{Atomic energies $\veps_n$ and $\veps_{\mathrm {imp}}$}
  \label{sec-model}
In this section the atomic energies $\{ \veps_n \}$ 
of the itinerant atoms and
$\veps_{\mathrm{imp}}$ of the impurity atom 
(see Eq.~(\ref{H-gas-def})),  are computed. The main steps are: 
1) Derivation of the optical potentials in which the atoms are trapped. 
This should be carried out separately for the ground state atoms Yb($^1$S$_0$) 
and the excited state atoms Yb($^3$P$_2$).  
2) Solving the corresponding Schr\"odinger equations for the 
atoms in the pertinent optical potentials. 
 
Ultra-cold Yb atoms are trapped by an optical
dipole trap that is formed due to the interaction between
an induced dipole moment in an atom and
an external (laser) electric field ${\mathbf{E}}(\mbfr,t)$.
The oscillating electric  field induces
an oscillating dipole moment in the atom.
Since the Yb(\3P2) atom has electronic 
angular momentum ${\bf J}$ with $J=2$, 
the atomic polarizability depends
on the projection of ${\bf J}$ 
on the direction of the electric field.
Specifically, we consider an optical potential generated by
an electromagnetic wave with double magic
wavelength $\lambda_0=546$~{nm}, such that
the polarizabilities $\alpha_\g(\lambda_0)$ and 
$\alpha_\e(\lambda_0)$ for the atoms in the ground (g) and excited (e) 
states do not depend on
the magnetic quantum numbers (see
Appendix \ref{append-trap} for details).
The optical potential depends on the wave
number $k_0=\frac{2\pi}{\lambda_0}$ and
the waist radius $L$ of the laser beams.
Formally, it is written as,
\begin{eqnarray}
  V_{\nu}(\mbfr) &=&
  -\alpha_{\nu}(\lambda_0)
  \lim_{T\to\infty}
  \int\limits_{0}^{T}
  \big|
      {\mathbf{E}}(\mbfr,t)
  \big|^{2}
  dt
  = \nonumber \\ &=&
  V_{\nu}^{\mathrm{(slow)}}(\mbfr)+
  V_{\nu}^{\mathrm{(fast)}}(\mbfr).
  \label{opt-potential-def}
\end{eqnarray}
 Here $\nu=\e$ (for the excited state \3P2 state) 
 and $\nu=\g$ (for the ground state 
\1S0). The explicit expressions for the slow and fast components read,
\begin{eqnarray*}
  V_{\nu}^{\mathrm{(slow)}}(\mbfr)
  &=&
  -V_{0,\nu}
  \bigg\{
       e^{-\frac{x^2+y^2}{L^2}}+
       e^{-\frac{y^2+z^2}{L^2}}+
       e^{-\frac{z^2+x^2}{L^2}}
  \bigg\},
  \\
  V_{\nu}^{\mathrm{(fast)}}(\mbfr)
  &=&
  -V_{1,\nu}
  \bigg\{
       \cos(2k_0 x)
       e^{-\frac{y^2+z^2}{L^2}}+
  \nonumber \\ && +
       \cos(2k_0 y)
       e^{-\frac{z^2+x^2}{L^2}}+
       \cos(2k_0 z)
       e^{-\frac{x^2+y^2}{L^2}}
  \bigg\},
\end{eqnarray*}
where $V_{1,\nu} \ll V_{0,\nu}$ are
amplitudes of the fast-oscillating and
slow-changing potentials, see
eq. (\ref{V0-V1-def-append}) in
Appendix \ref{append-trap}.
\begin{figure}[htb]
\centering
  \includegraphics[width=60 mm,angle=0]
   {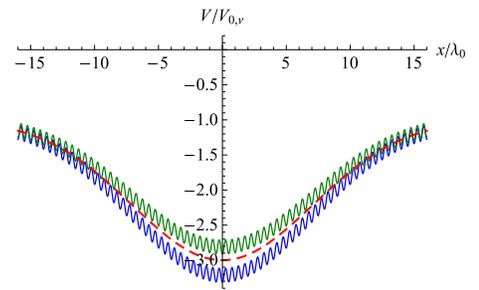}
 \caption{\footnotesize
   {\color{blue}(Color online)}
   $V_{\nu}(x,0,0)$ [solid blue curve],
   $V_{\nu}(x,\lambda_0/4,\lambda_0/4)$
   [solid green curve]
   and $V_{\nu}^{\mathrm{(slow)}}(x,0,0)$
   [dashed red curve] as functions of $x$
   for $L=10\lambda_0$ and $E_v=0.05E_u$.}
 \label{Fig-V-Gauss-cos}
\end{figure}
The optical potential (\ref{opt-potential-def})
is illustrated in Fig. \ref{Fig-V-Gauss-cos}
for $L=10\lambda_0$ and $V_{1,\nu}=0.1V_{0,\nu}$
[blue and green solid curves]. The dashed red curve
show $V_{\nu}^{\mathrm{(slow)}}(x,0,0)$.  

In order to find the wave functions and energies
of the Yb(\1S0) and Yb(\3P2) atoms, we solve
the Schr\"odinger equation for $\Psi^{\nu}(\mbfr)$, ($\nu$=e,g), 
\begin{eqnarray}
  -\frac{\hbar^2}{2 M}~
  \nabla^2
  \Psi^{\nu}(\mbfr)+
  V_{\nu}(\mbfr)~
  \Psi^{\nu}(\mbfr)
  =
  \veps_{\nu}~
  \Psi^{\nu}(\mbfr).
  \label{eq-Schrodinger}
\end{eqnarray}
Consider first $\Psi^{\text{e}}(\mbfr)$.
We assume that the Yb(\3P2) atom is trapped
by the potential $V_{\text{e}}^{\mathrm{(fast)}}(\mbfr)$.
When the corresponding energy level
$\veps_{\text{imp}}$ is deep enough, the wave
function of the radial wave function of the bound state near the potential
minimum at $\mbfr=0$ can be approximated within
the harmonic potential picture as
\begin{eqnarray}
  \Psi^{\e}(r) &=&
  \frac{1}{\big(\pi a_{\e}^{2}\big)^{3/4}}~
  \exp
  \bigg(
       -\frac{r^2}{2 a_{\e}^{2}}
  \bigg).
  \label{WF-trapped-3P2}
\end{eqnarray}
The harmonic length and frequency are,
\begin{eqnarray}
  k_0 a_{\e} ~=~
  \bigg(
       \frac{{\mathcal{E}}_{0}}{2 V_{1,\e}}
  \bigg)^{\frac{1}{4}},
  \ \ \ \ \
  \hbar \Omega_{\e} ~=~
  2
  \sqrt{2 {\mathcal{E}}_{0} V_{1,\e}},
  \label{a-omega-e-def-text}
\end{eqnarray}
while the recoil energy ${\mathcal{E}}_0$ is
\begin{eqnarray}
   {\mathcal{E}}_0 &=&
   \frac{\hbar^2 k_{0}^{2}}{2 M},
   \label{recoiling-energy}
\end{eqnarray}
where $M$ is the atomic mass.
The energy $\veps_{\mathrm{imp}}$ measured
from the bottom of the well is then,
\begin{eqnarray}
  \veps_{\mathrm{imp}} &=&
  \frac{3}{2}~
  \hbar \Omega_{\e}.
  \label{veps-imp}
\end{eqnarray}

Next, we consider wave functions and energy
levels of Yb(\1S0) atoms. Assume that
the Fermi energy $\epsilon_F$ satisfies
the inequalities,
$$
  V_{1,\g} ~\ll~
  \epsilon_F ~\ll~
  V_{0,\g}.
$$
Then for atoms with energies close to
$\epsilon_F$, we can neglect
$V_{\g}^{\mathrm{(fast)}}(\mbfr)$ and
write
$$
  V_{\g}(\mbfr)
  ~\approx~
  V_{\g}^{\mathrm{(slow)}}(\mbfr).
$$
Moreover, we can approximate
$V_{\g}^{\mathrm{(slow)}}(\mbfr)$ by
an isotropic harmonic oscillator,
see appendix \ref{append-trap}
for details. Atoms trapped by
the isotropic harmonic oscillator
potential are described by
the radial quantum number
$n=0,1,2,\ldots$, angular moment
$l=0,1,2,\ldots$ and projection
$m$ of the angular moment on
the axis $z$. 
Because of centrifugal barrier,
only the atoms with $l=0$ can
approach the impurity and be
involved in the exchange interaction
with it. The energy levels of the states
with $l=0$ are
\begin{eqnarray}
  \veps_{n} &=&
  \hbar \Omega_{\g}~
  \bigg(
       2 n+\frac{3}{2}
  \bigg).
  \label{veps-itinerant}
\end{eqnarray}
Here the harmonic length $a_{\g}$ and frequency
$\Omega_{\g}$ are defined as
\begin{eqnarray}
  \frac{a_{\g}}{L} ~=~
  \bigg(
       \frac{{\mathcal{E}}_{L}}{2 V_{0,\g}}
  \bigg)^{1/4},
  \ \ \ \ \
  \hbar \Omega_{\g} ~=~
  2
  \sqrt{2 {\mathcal{E}}_{L} V_{0,\g}},
  \label{a-omega-g-def}
\end{eqnarray}
where ${\mathcal{E}}_{L}$ is defined as,
\begin{eqnarray}
  {\mathcal{E}}_{L} &=&
  \frac{\hbar^2}{2 M L^2}.
  \label{EL-def}
\end{eqnarray}
In the following, we assume that
\begin{eqnarray}
  \Omega_{\e}
  ~\gg~
  \Omega_{\g}.
  \label{inequality}
\end{eqnarray}
Within this framework, the spectrum is nearly continuous
and the ytterbium atoms in the ground-state form a
Fermi gas. The Fermi energy $\epsilon_F$ is such that
$\epsilon_F\gg\hbar\Omega_{\g}$, hence the Fermi gas
is 3D.
The density of states (DOS) pertaining to the energy dispersion
(\ref{veps-itinerant}) is
\begin{eqnarray}
  \rho(\epsilon) &=&
  \frac{\Theta(\epsilon)}{2 \hbar \Omega_{\g}},
  \label{DOS-def}
\end{eqnarray}
where $\Theta(\xi)$ is the Heaviside theta function
equal to 0 for $\xi<0$, 1 for $\xi>0$ and $\frac{1}{2}$
for $\xi=0$.

\section{Exchange Interaction}
\label{sec-exchange}
\noindent
In this section, the origin of the
exchange interaction between the two atoms
$^{173}$Yb($^3$P$_2$)-$^{173}$Yb($^1$S$_0$) is explained, 
and the strength of the exchange constant $\lambda$ is computed. 
This exchange mechanism is similar {\it but not identical} to the one
derived within our analysis of the exchange interaction between the 
$^{173}$Yb($^3$P$_0$)-$^{173}$Yb($^1$S$_0$) atoms\cite{IK-TK-YA-GBJ-PRB-16}. 
The difference is that here, the total electronic
spin of one of the atoms, $^{173}$Yb($^3$P$_2$),  is not zero.

\begin{figure}[htb]
\centering
\subfigure[]
  {\includegraphics[width=55 mm,angle=0]
   {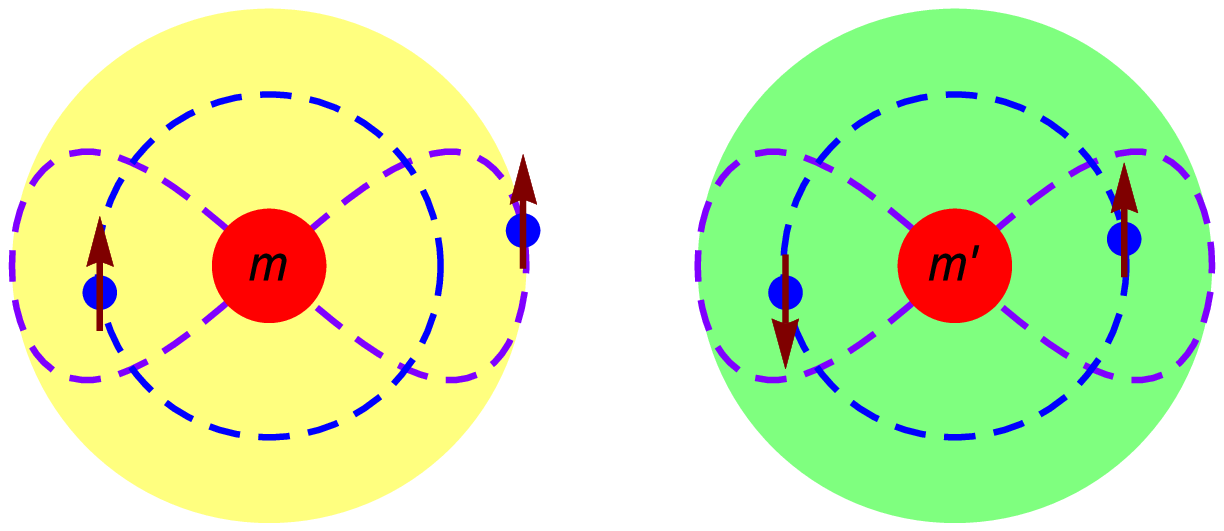}}
\subfigure[]
  {\includegraphics[width=55 mm,angle=0]
   {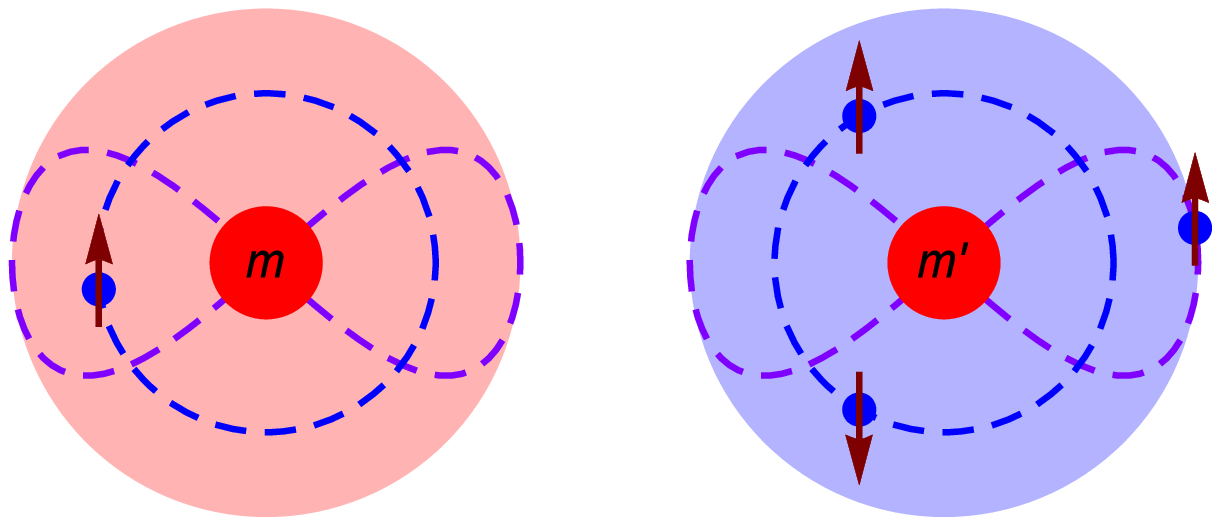}}
\subfigure[]
  {\includegraphics[width=55 mm,angle=0]
   {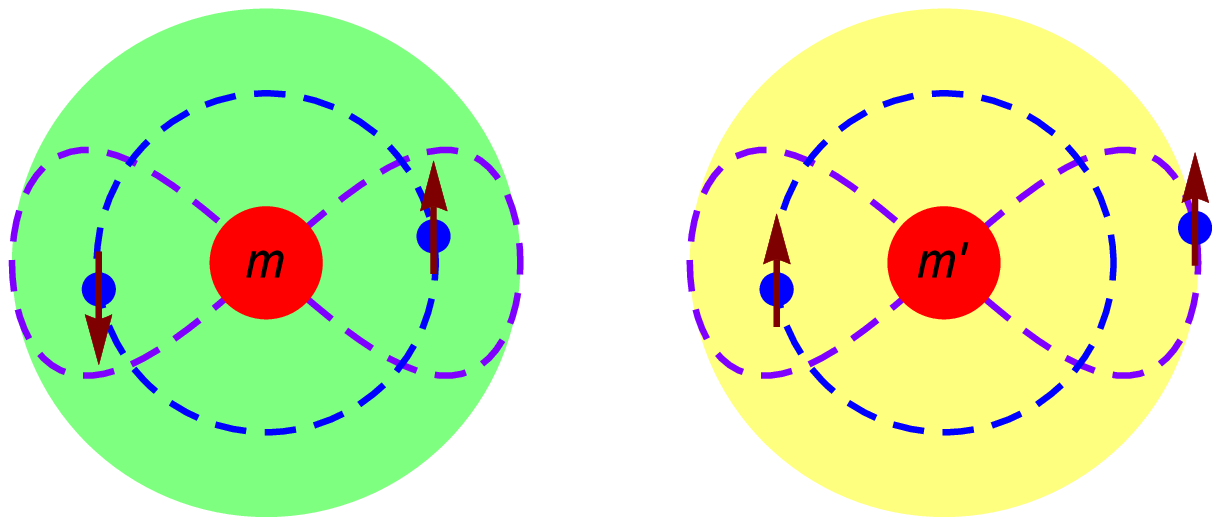}}
 \caption{\footnotesize
   ({\color{blue}Color online}) Illustration of exchange interaction
   between ytterbium atoms.
   Panel (a): Initial quantum state - the first atom is
   in the meta-stable state (light yellow disk) and the second
   one is in the ground state (light green disk);
   panel (b): virtual state - the first atom is positively ionized
   (light red disk), and the second one is negatively charged
   (light blue disk);
   panel (c): final state - the first atom is in the ground
   state and the other one is in the meta-stable state.
   For all the panels, arrows denote the electronic spin,
   $m$ or $m'$ is nuclear spin of the first or second atom.}
 \label{Fig-GX-II-XG}
\end{figure}

When the distance $R$ between an itinerant atom Yb($^1$S$_0$)
 (whose electronic configuration is $6s^2$) and the impurity
atom Yb($^3$P$_2$) (whose electronic configuration is $6s6p$)
  is of order $R_0$ (the atomic size), there is an {\it indirect exchange interaction}
between them \cite{IK-TK-YA-GBJ-PRB-16}. Heuristically it is
described in two steps [see Fig. \ref{Fig-GX-II-XG} for
illustration]:
1) The $6p$ electron tunnels from the Yb($^3$P$_2$) atom
to the Yb($^1$S$_0$) atom,  
$$ \mbox{Yb($^3$P$_2$)+Yb($^1$S$_0$) $\to$ Yb$^+(6s)$+Yb$^-(6s^26p)$}.$$
As a result, we have an intermediate state with 
two oppositely charged ions with parallel electronic orbital
angular momenta. 2) Then, one electron in a $6s$ orbital 
tunnels from the negatively charged ion to the $6s$ orbital of 
the positively charged ion. 
$$ \mbox{Yb$^+(6s)$+Yb$^-(6s^26p)$ $\to$ Yb($^1$S$_0$)+Yb($^3$P$_2$)}.$$
The net outcome is that
the atoms ``exchange their identities'' specified
by their electronic quantum states: one atom transforms
from the ground state to the excited state, whereas
the other atom transforms from the excited state to the
ground state.
The detailed calculations of the pertinent exchange interaction is
relegated to Appendix \ref{append-exchange}
(see also Ref. \cite{IK-TK-YA-GBJ-PRB-16}).
They employ two-particle wave functions describing the motion of
two atoms in the optical potential, taking into account
the atom-atom interaction. For short distance between the atoms,
(where the exchange interaction is essential), the two-atom
wave function is determined mainly by the inter-atomic
van der Waals potential \cite{IK-TK-YA-GBJ-PRB-16,Flambaum-WKB-93,vdW-pra-08}.
The  exchange interaction strength parameter $\lambda$ is given by,
\begin{eqnarray}
  \lambda &=&
 \frac{ \sqrt{3} k_F}{a_\g^2}~
  \Gamma^{2}\bigg(\frac{3}{2}\bigg)~
  \int\limits_{r_0}^{\infty}
  {\mathfrak{g}}(R)
  R^2 dR.
  \label{exchange-strength-def}
\end{eqnarray}
Here 
$$
 k_F ~=~
\frac{\sqrt{2 M \epsilon_F}}{\hbar}.
$$
is the Fermi wave number expressed in terms of the Fermi energy $\epsilon_F$ 
and the atomic mass $M$, while $a_\g$ is 
the harmonic length explicitly defined in Eq.(\ref{a-omega-g-def}).
The function ${\mathfrak{g}}(R)$ is decomposed as,
\begin{eqnarray}
  {\mathfrak{g}}(R) &=&
  \frac{t_{\mathrm{s}}(R)~
        t_{\mathrm{p}}(R)}
       {3~\Delta\epsilon}~
  {\mathfrak{R}}(R).
  \label{g(R)}
\end{eqnarray}
[see Appendix \ref{append-exchange} for details]. 
Here $t_{\mathrm{s}}$ and $t_{\mathrm{p}}$
are tunneling rates for the $6s$ and $6p$ electrons
\cite{IK-TK-YA-GBJ-PRB-16},
\begin{eqnarray}
  t_{\mu}(R) &=&
  t_{\mu}^{(0)}~
  \bigg(
       \frac{R}{r_0}
  \bigg)^{\frac{2}{\beta_{\mu}}+\frac{1}{2}}
  e^{-\kappa_{\mu}(R-r_0)},
  \label{tunneling-rate}
\end{eqnarray}
where $\mu={\mathrm{s,p}}$,
\begin{eqnarray*}
  t_{\mathrm{s}}^{(0)} ~=~
  1.09~{\text{eV}},
  \ \ \ \ \
  t_{\mathrm{p}}^{(0)} ~=~
  1.82~{\text{eV}}.
\end{eqnarray*}
The parameters $\kappa_{\nu}$ and $\beta_{\nu}$ are
\begin{eqnarray*}
  &&
  \kappa_{\mathrm{s}} ~=~
  1.28122~{\text{\AA}}^{-1},
  \ \ \ \ \
  \beta_{\mathrm{s}} ~=~
  0.677994,
  \\
  &&
  \kappa_{\mathrm{p}} ~=~
  1.00005~{\text{\AA}}^{-1},
  \ \ \ \ \
  \beta_{\mathrm{p}} ~=~
  0.529206.
\end{eqnarray*}
$\Delta\epsilon=\veps_{\mathrm{ion}}+\veps_{\mathrm{ea}}-%
\epsilon_{\mathrm{x}}=4.1104$~{eV} is the energy which should be
paid to get positively and negatively charged ions from two
neutral atoms, where $\veps_{\mathrm{ion}}=6.2542$~{eV} is
the ionization energy \cite{e-ion-Yb-78},
$\veps_{\mathrm{ea}}=0.3$~{eV} is the electron affinity
\cite{e-affin-Yb-04} and $\epsilon_{\mathrm{x}}=2.4438$~{eV} is
the excitation energy of the $^{3}$P$_{2}$ state
\cite{e-ion-Yb-78}.

The function ${\mathfrak{R}}(R)$ in eq. (\ref{g(R)})
encodes the deformation of the wave function of the itinerant
fermions at short distance from the impurity where the van der
Waals interaction is significant \cite{IK-TK-YA-GBJ-PRB-16},
\begin{eqnarray*}
  {\mathfrak{R}}(R) &=&
  \frac{8c}{R^2 K(R)}~
  \Bigg\{
       1+
       \bigg(
            \frac{a_w-\bar{a}}{\bar{a}}
       \bigg)^{2}
  \Bigg\}.
\end{eqnarray*}
Here
$$
  K(R) ~=~
  \frac{1}{\hbar}~
  \sqrt{-MW(R)},
$$
where $a_w=20.9973$~{\AA}, $\bar{a}=42.9984$~{\AA}
and $c=89.9569$~{\AA}.
We approximate the van der Waals interaction
by the Lennard-Jones potential as \cite{Yb-vdW-PRA-14},
\begin{eqnarray}
  W(R) &=&
  -\frac{C_6}{R^6}-
  \frac{C_8}{R^8}+
  \frac{C_{12}}{R^{12}}.
  \label{vdW-def-append}
\end{eqnarray}
Here $C_6=2.649\cdot10^{3}~E_ha_B^6$,
$C_8=3.21097\cdot10^{5}~E_ha_B^8$
and $C_{12}=1.41808\cdot10^{9}~E_h a_{B}^{12}$,
where $E_h=27.2114$~{eV} and
$a_B=0.529177$~{\AA}.
The parameter $r_0=3.6673$~{\AA} on the right hand
side of eq. (\ref{exchange-strength-def}) is found from
the condition $W(r_0)=0$.

Anticipating the use of scaling analysis,  
it is useful to define 
 the dimensionless exchange coupling constant
\begin{equation}
  \Lambda_0
  ~=~
\lambda \rho_0,
\label{Lambda}
\end{equation}
wherein
$\rho_0=\rho(\epsilon_F)$ is the density of
states of the itinerant atoms at the Fermi energy,
as defined in Eq. (\ref{DOS-def}).
Note that $\Lambda_0$ has a finite limit as $\Omega_{\g}\to0$
and $a_{\g}\to\infty$.
To show it, recall Eq.~(\ref{exchange-strength-def}) for $\lambda$, 
in which $a_\g$ and $\Omega_\g$ are 
the harmonic length and frequency 
explicitly defined in Eq.(\ref{a-omega-g-def}), 
Letting
$\Omega_{\g}\to0$ and $a_{\g}\to\infty$,
we can write
\begin{eqnarray}
  \frac{1}{2 \hbar \Omega_{\g} a_{\g}^{2}}
  &=&
  \frac{M}{4 \hbar^2},
  \label{continuous-limit}
\end{eqnarray}
where $M$ is the atomic mass.
 Notationally, we shall write the differential scaling equations in terms 
of the dimensionless parameters $\Lambda_\beta$ but 
keep the use of $\lambda_\beta$ as coupling constants with dimension of energy when 
computing the corrections to the Hamiltonian.

\section{Second Order Poor-Man's Scaling Analysis}
  \label{sec-poor-man-scaling-2nd}
\noindent
In this section we derive the poor-man scaling equations to
second-order in the exchange constant.  This procedure is quite
standard, and yet, there are important  differences between
the procedure applied here and that employed in the standard Kondo
effect. First, we have {\it three coupling constants} $\Lambda_{\mathrm{d}}, 
\Lambda_{\mathrm{q}}$, and $\Lambda_{\mathrm{o}}$ that are to be renormalized. 
As we see from Eqs.~(\ref{subeqs-scaling-2nd}) below,
these three coupling constants satisfy a set of three
coupled non-linear scaling equations. Second, within
the underlying representation of $SU(2)$ the number of spin
projections is $2 F+1 \geq 4$, compared with $2 s+1=2$ in the electronic version. 

Applying the poor man's scaling RG procedure to second order
enables one to determine  the Kondo
temperature. (In order to derive
scaling equations for the exchange coefficients and identifying
the fixed points one has to advance to third order). For the dipolar interaction,
the second order calculation procedure is straightforward and already well documented 
(at least for particles with spin $\tfrac{1}{2}$). 
For the multipolar exchange interactions,
 some technical modifications are required, as worked out below.
 
The "conduction" band of the (neutral)
itinerant atoms is defined by their energies $\{\epsilon\}$. 
Before starting the RG procedure the bandwidth is $D_0$. 
Within the RG framework, the bandwidth is narrowed 
in steps to be $D<D_0$ 
such that the energies of the atoms are constrained to be
$ |\epsilon-\epsilon_F|<D$ ( where $\epsilon_F$ is the Fermi
energy). As usual, at an intermediate stage, 
this conduction band is divided
 into three parts. The first part contains energies of
particle and hole states within a reduced  bandwidth
$|\epsilon-\epsilon_F|<D'$,
where $D'=D-\delta{D}$ with $\frac{\delta D}{D} \ll 1$,  which are retained. 
The second and third parts contains energies $\{ \epsilon \}$ of particle and
hole states at the band edges, within narrow intervals $D'<|\epsilon-\epsilon_F|<D$.
Within the RG procedure these states are to be integrated out
\cite{Hewson-book}.

\begin{figure}[htb]
\centering \subfigure[]
  {\includegraphics[width=50 mm,angle=0]
   {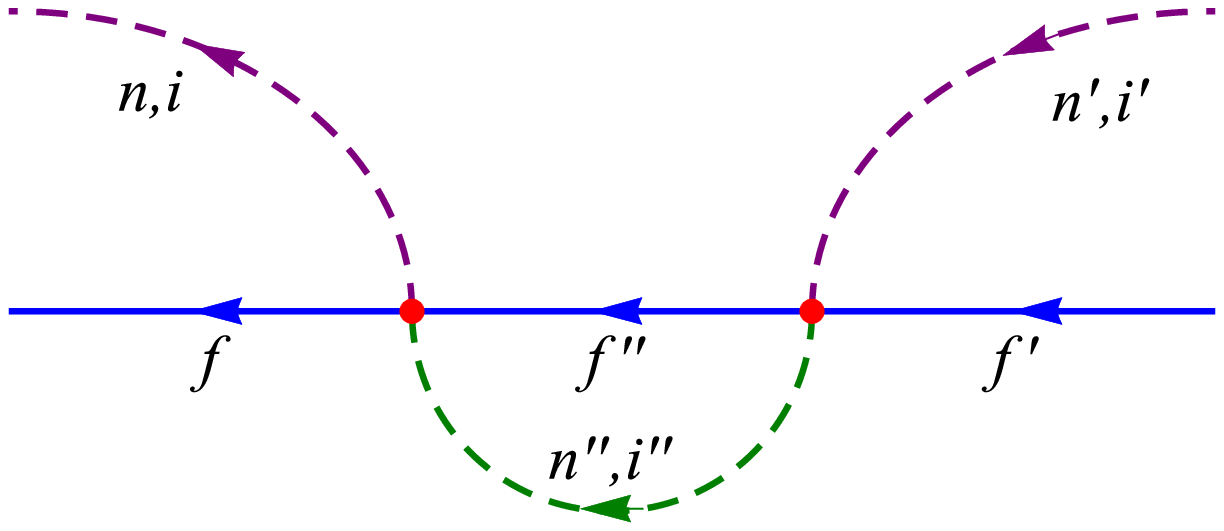}
   \label{Fig-2nd-particle}}
\centering \subfigure[]
  {\includegraphics[width=50 mm,angle=0]
   {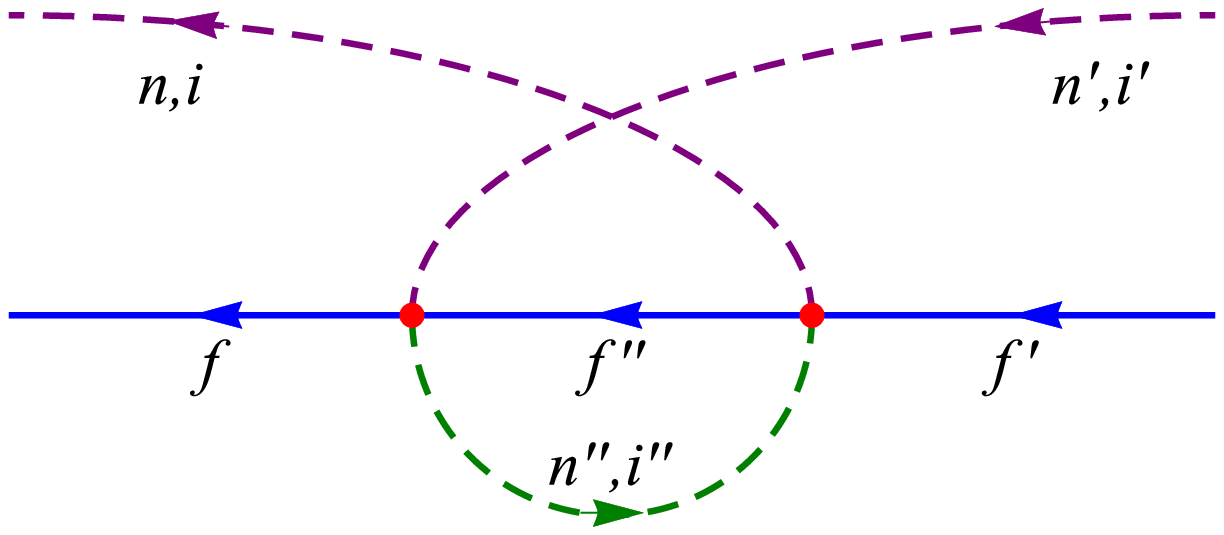}
   \label{Fig-2nd-hole}}
 \caption{\footnotesize
   ({\color{blue}Color online})
   ``Particle'' [panel (a)] and ``hole'' [panel (b)]
   second order diagrams for the Kondo Hamiltonian (\ref{HK-def}).
   The solid lines correspond to the localized impurity atom,
   the purple dashed curves describes itinerant atoms before or
   after the scattering, and the green dashed curves describe
   itinerant atom in the virtual state near the top edge
   [panel (a)] or bottom edge [panel (b)] of the conduction band.}
 \label{Fig-2nd-diagrams}
\end{figure}

The second order corrections to the Kondo Hamiltonian are
schematically illustrated in Fig. \ref{Fig-2nd-diagrams}. Here
the solid blue line describes the quantum state of the localized
impurity. The dashed purple curve restricted from one side by
the red dot describes itinerant atom (before or after scattering)
whose energy is close to the Fermi energy. The dashed green curves
restricted by red dots from both sides describe itinerant atom in
the virtual state with the energy within the interval
$D'<\epsilon-\epsilon_F<D$ (as in Fig. \ref{Fig-2nd-particle}) or
$-D<\epsilon-\epsilon_F<-D'$ (as in Fig.
\ref{Fig-2nd-hole}). The red dots denote the Kondo Hamiltonian
(\ref{HK-Hd-Hq-Ho}). Since $H_K$ has three terms, the second order
corrections to $H_K$ can formally be written as,
\begin{eqnarray}
  \delta{H}_{2} &=&
  \sum_{\beta,\beta'}
  \delta{H}_{\beta,\beta'}^{(2)}.
  \label{dH2-def}
\end{eqnarray}
Here $\beta,\beta'={\mathrm{d}},{\mathrm{q}},{\mathrm{o}}$
for dipole, quadrupole and octupole interaction,
\begin{eqnarray}
  \delta{H}_{\beta,\beta'}^{(2)} &=&
  \frac{1}{2}~
  \sum_{{\mathrm{e}},{\mathrm{e}}'}
  H_{\beta}
  \Big|
      {\mathrm{e}}
  \Big\rangle
  ~
  \Big\langle
      {\mathrm{e}}
  \Big|
      \frac{1}{\epsilon_0-H_{\mathrm{c}}}
  \Big|
      {\mathrm{e}}'
  \Big\rangle
  ~
  \Big\langle
      {\mathrm{e}}'
  \Big|
  H_{\beta'}+
  \nonumber \\ &+&
  \frac{1}{2}~
  \sum_{{\mathrm{e}},{\mathrm{e}}'}
  H_{\beta'}
  \Big|
      {\mathrm{e}}
  \Big\rangle
  ~
  \Big\langle
      {\mathrm{e}}
  \Big|
      \frac{1}{\epsilon_0-H_{\mathrm{c}}}
  \Big|
      {\mathrm{e}}'
  \Big\rangle
  ~
  \Big\langle
      {\mathrm{e}}'
  \Big|
  H_{\beta},
  \label{dH2-alpha-alpha}
\end{eqnarray}
where $|{\mathrm{e}}\rangle$ and $|{\mathrm{e}}'\rangle$
are quantum states with a hole near the Fermi energy and
an atom with energy in the interval $D'<\epsilon<D$,
or a hole on an energy level in the interval $-D<\epsilon<-D'$
and an additional atom near the Fermi level. Recall that $H_{\mathrm{c}}$ is
the Hamiltonian of itinerant atoms, see Eq.~(\ref{H-gas-def}). Since
$\delta{D}\ll{D}$, we can use the approximation,
$$
  \Big\langle
      {\mathrm{e}}
  \Big|
  \frac{1}{\epsilon_0-H_{\mathrm{c}}}
  \Big|
      {\mathrm{e}}'
  \Big\rangle
  ~\approx~
  -\frac{1}{D}~
  \delta_{{\mathrm{e}},{\mathrm{e}}'}.
$$
Explicit expressions for the operators
$\delta{H}_{\beta,\beta'}^{(2)}$ are derived in Appendix
\ref{append-derivation-2nd-PMS}. Combining all the differentials
$\delta{H}_{\beta,\beta'}^{(2)}$, we see that integrating out
the virtual states near the band edges to lowest order results in
a new Hamiltonian of the {\it same form} as eq.
(\ref{HK-Hd-Hq-Ho}) but with renormalized coupling constants
$\lambda_{\beta}(D)\to\lambda_{\beta}(D')=\lambda_{\beta}(D)+%
\delta\lambda_{\beta}$, where $\beta={\mathrm{d,q,o}}$.
Consequently, we arrive at the following second order poor man's
scaling equations for the dimensionless couplings
$\Lambda_{\beta}=\lambda_{\beta}\rho_0$,
\begin{subequations}
\begin{eqnarray}
  \frac{\partial \Lambda_{\mathrm{d}}}
       {\partial \ln D}
  &=&
  -\Lambda_{\mathrm{d}}^{2}-
  \frac{9216}{25}~
  \Lambda_{\mathrm{q}}^{2}-
  \frac{1469664}{25}~
  \Lambda_{\mathrm{o}}^{2},
  \label{eq1-scaling-2nd}
  \\
  \frac{\partial \Lambda_{\mathrm{q}}}
       {\partial \ln D}
  &=&
  -12
  \Lambda_{\mathrm{d}}
  \Lambda_{\mathrm{q}}-
  \frac{1458}{5}~
  \Lambda_{\mathrm{q}}
  \Lambda_{\mathrm{o}},
  \label{eq2-scaling-2nd}
  \\
  \frac{\partial \Lambda_{\mathrm{o}}}
       {\partial \ln D}
  &=&
  -18
  \Lambda_{\mathrm{d}}
  \Lambda_{\mathrm{o}}-
  \frac{64}{9}~
  \Lambda_{\mathrm{q}}^{2}+
  \frac{306}{5}~
  \Lambda_{\mathrm{o}}^{2}.
  \label{eq3-scaling-2nd}
\end{eqnarray}
  \label{subeqs-scaling-2nd}
\end{subequations}

\noindent
Recall that $\rho_0=\rho(0)$ is the density of states (\ref{DOS-def})
of itinerant atoms at the Fermi energy $\epsilon_F$ (we set $\epsilon_F=0$).
The initial values $\Lambda_{\beta}^{(0)}=\Lambda_{\beta}(D_0)$ of
the couplings $\Lambda_{\beta}$ (where $\beta={\mathrm{d,q,o}}$)
are obtained from Eq.~(\ref{Jd-Jq-Jo-res}) after multiplying both sides by $\rho_0$, that is, 
\begin{equation}
  \Lambda_{\mathrm{d}}^{(0)} =
    \frac{26}{525} \lambda \rho_0,
  \Lambda_{\mathrm{q}}^{(0)} =
  -\frac{1}{840} \lambda \rho_0,
  \Lambda_{\mathrm{o}}^{(0)} =
   -\frac{1}{1890} \lambda \rho_0. 
   \label{subeqs-Lambda-initial}
\end{equation}
Note that when the initial values of $\Lambda_{\mathrm{q}}$
and $\Lambda_{\mathrm{o}}$ are zero, the right hand sides of
eqs. (\ref{eq2-scaling-2nd}) and ({\ref{eq3-scaling-2nd})
vanish and the set of equations (\ref{subeqs-scaling-2nd})
reduces  to the standard scaling equation for the s-d Kondo model
\cite{Hewson-book},
\begin{eqnarray}
  \frac{\partial \Lambda_{\mathrm{d}}}
       {\partial \ln D}
  &=&
  -\Lambda_{\mathrm{d}}^{2}.
  \label{scaling-2nd-sd}
\end{eqnarray}
Finally, it should be noted that the scaling procedure is carried out until
the effective bandwidth $D$ essentially exceeds
$\hbar\Omega_{\mathrm{g}}$ and $T_K$ [where $T_K$ is the Kondo
temperature defined below]. In the following, we assume that
$\hbar\Omega_{\mathrm{g}}<T_K$, and therefore the Kondo
temperature is the infrared cutoff parameter of our theory.

In the next section we elucidate the effect of the quadrupole and
octupole interactions on the scaling invariant of the RG equations,
that is, the Kondo temperature \cite{Hewson-book} and show that it
is rather significant.

\section{Kondo Temperature}
  \label{sec-TK}
  \noindent
 In this section, the Kondo temperature
  is calculated and numerically estimated, based on the results of the previous section.
  The important conclusion from this analysis is that this central energy scale
   is within an experimental reach.

The Kondo temperature is defined as the value of $D$ for which
the running coupling constants $\Lambda_{\beta}(D)$ diverge
($\beta={\mathrm{d,q,o}}$). To elucidate it, we solve the set of equations
(\ref{subeqs-scaling-2nd}) numerically for different initial
values $\Lambda_{\beta}^{(0)}$. Using eq.
(\ref{subeqs-Lambda-initial}), it is possible to express
$\Lambda_{\mathrm{q}}^{(0)}$ and $\Lambda_{\mathrm{o}}^{(0)}$
in terms of $\Lambda_{\mathrm{d}}^{(0)}$. Then the Kondo
temperature becomes a function of a single parameter, that is
$\Lambda_{\mathrm{d}}^{(0)}$. The results of these numerical
calculations for the Kondo temperature are displayed in Fig.
\ref{Fig-TK-num}, (solid curve). In fact, $T_K$ can be approximated by
the following expression,
\begin{eqnarray}
  T_K ~=~
  D_0~
  \exp
  \bigg(
       -\frac{1}{A\Lambda_{\mathrm{d}}^{(0)}}
  \bigg),
  \ \ \
  A ~=~
  13.9594.
  \label{TK-num-approx}
\end{eqnarray}
This approximation (\ref{TK-num-approx}) for $T_K$ is compared with 
the numerical results as displayed in
Fig. \ref{Fig-TK-num}, (dashed curve). It is clear 
that the approximation (\ref{TK-num-approx}) is
excellent.
\begin{figure}[htb]
\centering
  \includegraphics[width=60 mm,angle=0]
   {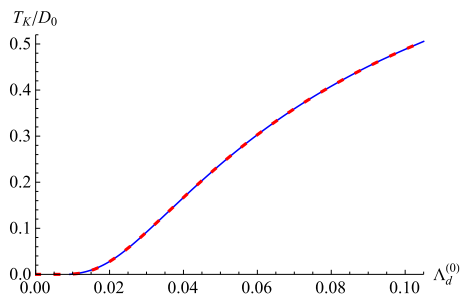}
 \caption{\footnotesize
   ({\color{blue}Color online})
   Kondo temperature calculated numerically from the set
   of equations (\ref{subeqs-scaling-2nd}) [solid blue curve]
   and the approximation (\ref{TK-num-approx})
   [dashed red curve].}
 \label{Fig-TK-num}
\end{figure}
Note that the scaling equation (\ref{scaling-2nd-sd}) yields
the following expressions for the Kondo temperature,
\begin{eqnarray}
  T_{K}^{\mathrm{(d)}} &=&
  D_0
  \exp
  \bigg(
       -\frac{1}{\Lambda_{\mathrm{d}}^{(0)}}
  \bigg).
  \label{TK-dipole}
\end{eqnarray}
The factor $A\gg1$ in eq. (\ref{TK-num-approx}) indicates that
the quadrupole and octupole interactions are important and act to
enhance $T_K$.
Numerical calculations yield $0.03D_0<T_K<0.1D_0$ for
$400$~{nK}~$<T_F<1000$~{nK}. When $D_0=300$~{nK},
the Kondo temperature is $9$~{nK}~$<T_K<30$~{nK}. 
With today's cooling techniques, it is concluded 
that $T_K$ is experimentally accessible 
so that the multipolar Kondo effect can be measured.

\section{Third Order Poor-Man's Scaling Analysis}
  \label{sec-third-order}
\noindent
A necessary (but not sufficient) condition
for arriving at a novel  fixed point (at which over-screening occurs),  is to
check that such point is a {\it finite solution of third order
scaling equations}. Derivation and solutions of these equations
is carried out in this section. As expected, the calculations are
rather involved due to the occurrence of higher multipoles, third
order diagrams and spin $s>\tfrac{1}{2}$. Nevertheless, these
cumbersome calculations should not mask the important physical
consequence exposed here: There are three candidates for stable finite
fixed points $P_4,P_5$ and $P_7$ (see below), that correspond to non-Fermi
liquid ground-states.

In order to
derive the third order correction to the poor-man's scaling
equations (\ref{subeqs-scaling-2nd}), we need to consider
the second order correction to the energy of the system, as
encoded in the self energy diagrams shown in Fig.
\ref{Fig-2nd-energy-diagrams}, as well as the third order vertex
diagrams shown in Fig. \ref{Fig-3rd-diagrams} (see Ref.
\cite{Hewson-book}). These
diagrams are considered below each one in its turn.

\begin{figure}[htb]
\centering \subfigure[]
  {\includegraphics[width=50 mm,angle=0]
   {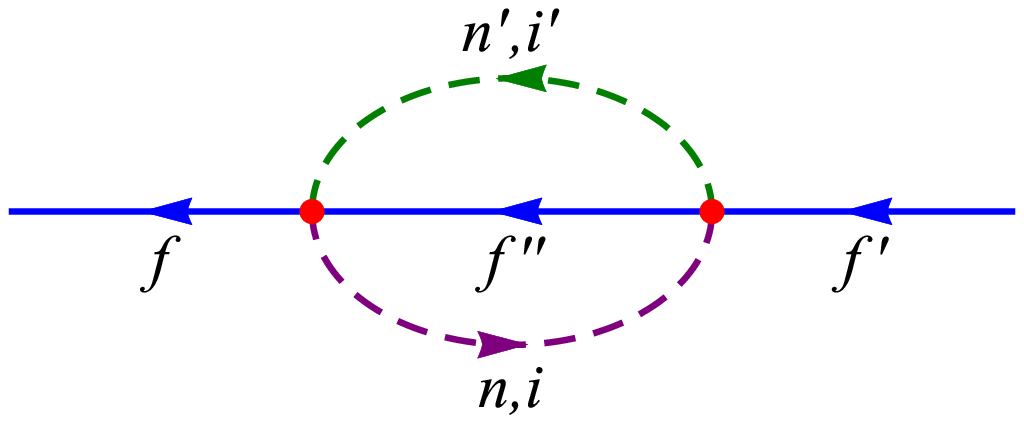}
   \label{Fig-2nd-E-particle}}
\centering \subfigure[]
  {\includegraphics[width=50 mm,angle=0]
   {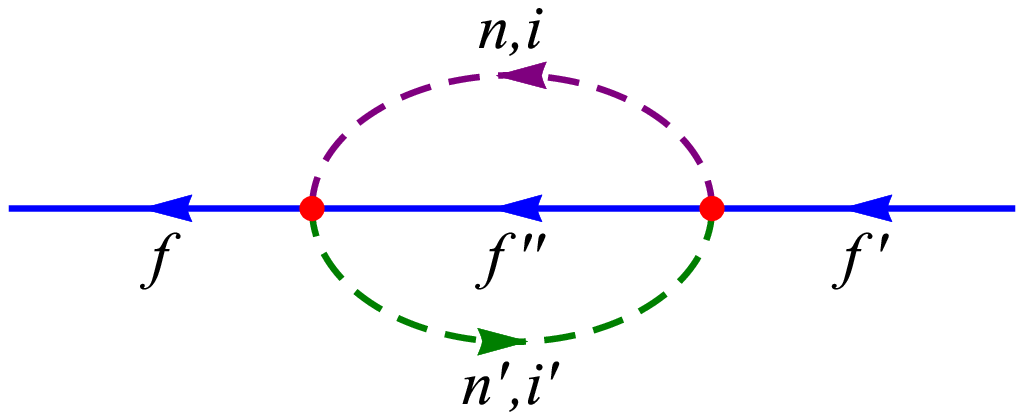}
   \label{Fig-2nd-E-hole}}
 \caption{\footnotesize
   ({\color{blue}Color online})
   ``Particle'' [panel (a)] and ``hole'' [panel (b)] second order
   self energy diagrams.
   The solid lines correspond to the localized impurity atom,
   the purple dashed curves restricted by the vertex from one
   side describe itinerant atoms before or after the scattering,
   the purple dashed curves restricted by vertex from both sides
   describe itinerant atoms in the virtual state in the reduced
   energy band and the green dashed curves describe itinerant
   atom in the virtual state near the top edge [panel (a)] or
   bottom edge [panel (b)] of the conduction band.}
 \label{Fig-2nd-energy-diagrams}
\end{figure}

\subsection{Second Order Self Energy Diagrams}
  \label{subsec-self-energy-sec-third-order}

Second order corrections to the self energy are  illustrated by the
diagrams displayed in Fig. \ref{Fig-2nd-energy-diagrams} and calculated
in  Appendix \ref{apppend-derivation-dE-2nd}. Taking into account
eqs. (\ref{dE-d-res}), (\ref{dE-q-res}) and (\ref{dE-o-res}), we
get
\begin{eqnarray}
  \delta{E} &=&
  -\frac{\delta{D}}{D}~
  E~
  \bigg\{
       \frac{525}{4}~
       \Lambda_{\mathrm{d}}^{2}+
       13440~
       \Lambda_{\mathrm{q}}^{2}+
  \nonumber \\ && +~
       3306744~
       \Lambda_{\mathrm{o}}^{2}
  \bigg\}.
  \label{dE-res}
\end{eqnarray}

\subsection{Third Order Vertex Diagrams}
  \label{subsec-third-order}

The third order contributions to
the scaling equations are given by diagrams in Fig.
\ref{Fig-3rd-diagrams}. The corresponding
correction to the Kondo Hamiltonian is decomposed as,
\begin{eqnarray}
  \delta{H}_{3} &=&
  \sum_{\beta,\beta'}
  \delta{H}_{\beta',\beta,\beta'}^{(3)},
  \label{dH3-def}
\end{eqnarray}
where $\beta,\beta'={\mathrm{d,q,o}}$ for the dipole, quadrupole
and octupole interactions. Explicitly, we get, 
\begin{eqnarray}
  \delta{H}_{\beta',\beta,\beta'}^{(3)} &=&
  \frac{\lambda_{\beta}\lambda_{\beta'}^{2}}{D^2}
  \sum_{\vec{\alpha}_{\beta},\vec{\alpha}_{\beta'},\vec{\alpha}'_{\beta'}}
  \sum_{f,f'}
  \Big(
      \hat{F}^{\vec{\alpha}_{\beta'}}
      \hat{F}^{\vec{\alpha}_{\beta}}
      \hat{F}^{\vec{\alpha}'_{\beta'}}
  \Big)_{f,f'}
  \times \nonumber \\ && \times
  X^{f,f'}~
  \sum_{i,i'}
  \sum_{n,n'}
  I^{\vec{\alpha}_{\beta}}_{i,i'}~
  c_{n,i}^{\dag}
  c_{n',i'}
  \times \nonumber \\ && \times
  {\mathrm{Tr}}
  \Big(
      \hat{I}^{\vec{\alpha}_{\beta'}}
      \hat{I}^{\vec{\alpha}'_{\beta'}}
  \Big)~
  2 \rho_0^2 D \delta D.
\label{dH3-beta-beta-beta}
\end{eqnarray}
Recall that $\hat{F}^{\vec{\alpha}_{\beta}}$ or
$\hat{I}^{\vec{\alpha}_{\beta}}$ are dipole ($\beta={\mathrm{d}}$),
quadrupole ($\beta={\mathrm{q}}$) and octupole ($\beta={\mathrm{o}}$)
tensors for a localized impurity or itinerant atoms,
$\vec{\alpha}_{\mathrm{d}}\equiv\alpha$,
$\vec{\alpha}_{\mathrm{q}}\equiv(\alpha,\alpha')$ and
$\vec{\alpha}_{\mathrm{o}}\equiv(\alpha,\alpha',\alpha'')$,
where $\alpha$'s are Cartesian indices [see
eqs. (\ref{subeqs-dipole}), (\ref{quadrupole-def}) and
(\ref{octupole-def})].

\begin{figure}[htb]
\centering \subfigure[]
  {\includegraphics[width=50 mm,angle=0]
   {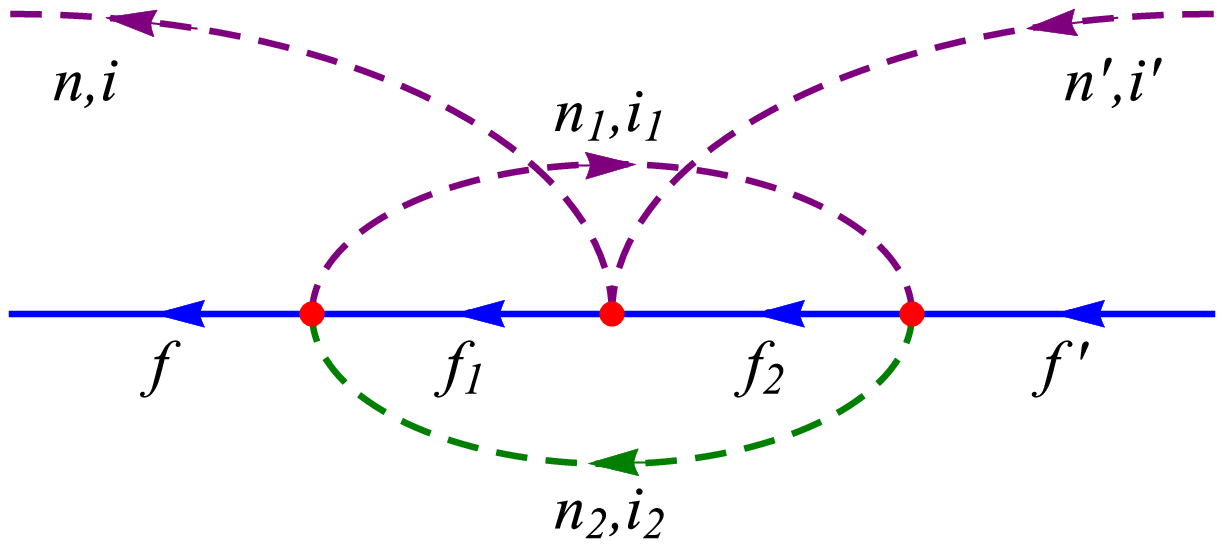}
   \label{Fig-3rd-particle}}
\centering \subfigure[]
  {\includegraphics[width=50 mm,angle=0]
   {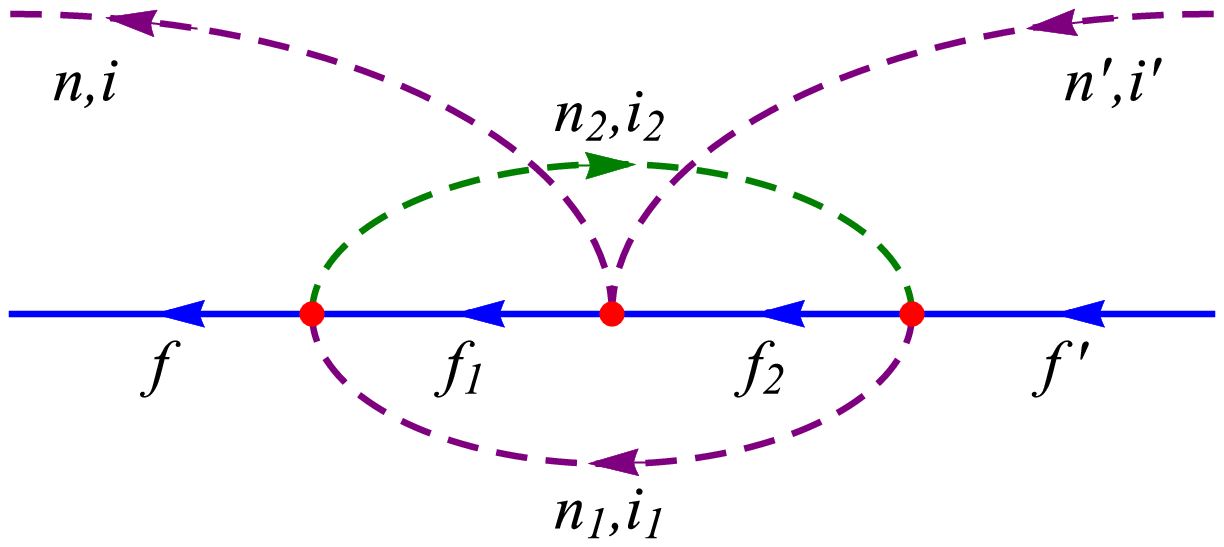}
   \label{Fig-3rd-hole}}
 \caption{\footnotesize
   ({\color{blue}Color online})
   ``Particle'' [panel (a)] and ``hole'' [panel (b)]
   third order diagrams for the Kondo Hamiltonian (\ref{HK-def}).
   The solid lines correspond to the localized impurity atom,
   the purple dashed curves restricted by the vertex from one side
   describe itinerant atoms before or after the scattering,
   the purple dashed curves restricted by vertex from both sides
   describe itinerant atoms in the virtual state in the reduced
   energy band and the green dashed curves describe itinerant atom
   in the virtual state near the top edge [panel (a)] or bottom
   edge [panel (b)] of the conduction band.}
 \label{Fig-3rd-diagrams}
\end{figure}
Explicit expressions for the operators
$\delta{H}_{\beta',\beta,\beta'}^{(3)}$ are derived
in Appendix \ref{apppend-derivation-3rd-PMS}. It is shown there
that $H_K+\delta{H}_{2}+\delta{H}_{3}$ has the {\it same form}
as $H_K$, albeit with proper corrections to the coupling constants
$\lambda_{\beta}+\delta \lambda_{\beta}$. Therefore we conclude
that inclusion of $\delta{H}_{3}$ does not change the structure
of the initial Kondo Hamiltonian {\it in its decomposed form} 
Eq.(\ref{HK-Hd-Hq-Ho}), but it causes
a renormalization of the coupling constants $\lambda_{\beta=\mathrm{d,q,o}}$.

\subsection{Third Order Poor Man's Scaling Equations}
  \label{subsec-third-order-eqs}

The effective Hamiltonian (which includes energy and vertex
renormalization) depends on the energy\cite{Hewson-book}
which is determined  through the Schr\"odinger equation
$$
  H \Psi ~=~ E \Psi,
$$
and this dependence is given by \cite{Hewson-book}
$$
  \tilde{H}_{\mathrm{eff}}(E) ~=~
  \tilde{H}_{\mathrm{eff}}(0)-
  E S,
$$
where the parameter $S=\delta{E}/E$ does not depend on $E$, (see
eq. (\ref{dE-res})), and
\begin{eqnarray}
  \tilde{H}_{\mathrm{eff}}(0) ~=~
  H_K+
  \delta{H}_{2}+
  \delta{H}_{3},
  \label{Heff=HK+dH2+dH3-tilde}
\end{eqnarray}
in which $\delta{H}_{2}$ and $\delta{H}_{3}$ are given by eqs.
(\ref{dH2-def}) and (\ref{dH3-def}). In order to get an effective
Hamiltonian, we solve the (implicit) secular equation,
$$
  \big|
      \tilde{H}_{\mathrm{eff}}(E)-E
  \big|
  ~=~ 0,
$$
(where $|A|$ denotes the determinant of the square matrix $A$)
which leads to
\begin{eqnarray*}
  \big|
      \tilde{H}_{\mathrm{eff}}(0)-
      (1+S)E
  \big|
  &=& 0.
\end{eqnarray*}
This equation yields an energy-independent effective Hamiltonian,
$$
  H_{\mathrm{eff}} ~=~
  \big(1+S\big)^{-1/2}~
  \tilde{H}_{\mathrm{eff}}(0)~
  \big(1+S\big)^{-1/2}.
$$
Taking into account that $S\sim\lambda^2$ [see eq. (\ref{dE-res})]
and keeping the terms up to $\lambda^3$,
we can write $H_{\mathrm{eff}}$ as,
\begin{eqnarray}
  H_{\mathrm{eff}} &=&
  H_K+
  \delta{H}_{2}+
  \delta{\tilde{H}}_{3},
  \label{H-eff-3rd-res}
\end{eqnarray}
where
\begin{eqnarray}
  \delta{\tilde{H}}_{3}
  &=&
  \delta{H}_{3}+
  S
  H_K.
  \label{tilde-dH3-def}
\end{eqnarray}
Employing the results of subsections
\ref{subsec-self-energy-sec-third-order} and
\ref{subsec-third-order}, we can see that the operator
$\delta{\tilde{H}}_{3}$ has the same form as the Hamiltonian
$H_K$, (see eq. (\ref{HK-Hd-Hq-Ho}) and the equations below it).
Therefore, it gives rise to renormalization of the couplings
constants $\lambda_{\beta}$.
The third order poor-man scaling equations for the dimensionless
couplings $\Lambda_{\beta}$ are then,
\begin{eqnarray}
  \frac{\partial \Lambda_{\beta}}
       {\partial \ln D}
  &=&
  {\mathfrak{F}}_{\beta}
  \big(
      \Lambda_{\mathrm{d}},~
      \Lambda_{\mathrm{q}},~
      \Lambda_{\mathrm{o}}
  \big), \ \ (\beta={\mathrm{d,q,o}}).
  \label{scaling-3rd}
\end{eqnarray}
The functions
${\mathfrak{F}}_{\mathrm{d,q,o}}$ on the RHS are,
\begin{subequations}
\begin{eqnarray}
  {\mathfrak{F}}_{\mathrm{d}}
  &=&
  -\Lambda_{\mathrm{d}}^{2}-
  \frac{9216}{25}~
  \Lambda_{\mathrm{q}}^{2}-
  \frac{1469664}{25}~
  \Lambda_{\mathrm{o}}^{2}+
  35~
  \Lambda_{\mathrm{d}}^{3}+
  \nonumber \\ && +
  12096~
  \Lambda_{\mathrm{d}}
  \Lambda_{\mathrm{q}}^{2}+
  \frac{21493836}{5}~
  \Lambda_{\mathrm{d}}
  \Lambda_{\mathrm{o}}^{2},
  \label{eq1-scaling-3rd}
  \\
  {\mathfrak{F}}_{\mathrm{q}}
  &=&
  -12
  \Lambda_{\mathrm{d}}
  \Lambda_{\mathrm{q}}-
  \frac{1458}{5}~
  \Lambda_{\mathrm{q}}
  \Lambda_{\mathrm{o}}+
  \frac{945}{8}~
  \Lambda_{\mathrm{q}}
  \Lambda_{\mathrm{d}}^{2}+
  \nonumber \\ && +
  17472~
  \Lambda_{\mathrm{q}}^{3}+
  \frac{15380348}{5}~
  \Lambda_{\mathrm{q}}
  \Lambda_{\mathrm{o}}^{2},
  \label{eq2-scaling-3rd}
  \\
  {\mathfrak{F}}_{\mathrm{o}}
  &=&
  -18
  \Lambda_{\mathrm{d}}
  \Lambda_{\mathrm{o}}-
  \frac{64}{9}~
  \Lambda_{\mathrm{q}}^{2}+
  \frac{306}{5}~
  \Lambda_{\mathrm{o}}^{2}+
  \frac{1365}{8}~
  \Lambda_{\mathrm{o}}
  \Lambda_{\mathrm{d}}^{2}+
  \nonumber \\ && +
  12096~
  \Lambda_{\mathrm{o}}
  \Lambda_{\mathrm{q}}^{2}+
  \frac{16769916}{5}~
  \Lambda_{\mathrm{o}}^{3}.
  \label{eq3-scaling-3rd}
\end{eqnarray}
  \label{subeqs-scaling-3rd}
\end{subequations}
The symmetry of the scaling equations (\ref{scaling-3rd}) should
be noted:
${\mathfrak{F}}_{\mathrm{d}}$ and ${\mathfrak{F}}_{\mathrm{o}}$
are even with respect to the inversion transformation
$\Lambda_{\mathrm{q}}\to-\Lambda_{\mathrm{q}}$, whereas
${\mathfrak{F}}_{\mathrm{q}}$ is odd. Therefore we can safely
conclude that the scaling equations (\ref{scaling-3rd}) are
invariant with respect to the inversion
$\Lambda_{\mathrm{q}}\to-\Lambda_{\mathrm{q}}$.
The fixed points of the scaling equations (\ref{scaling-3rd}) are
found from the conditions, ${\mathfrak{F}}_{\mathrm{d,q,o}}=0$.
Numerical solution of the last set of equations yields seven fixed
points in 3D parameter space, $P_{n}=(\Lambda_{\mathrm{d}}^{(n)},
\Lambda_{\mathrm{q}}^{(n)},\Lambda_{\mathrm{o}}^{(n)})$,
$n=1,2,\ldots7$:
\begin{subequations}
\begin{eqnarray}
&&  P_{1}  \mbox{=}
  \big(
      0.0285714, 0, 0
  \big),
  \label{fixed-point-1}
  \\
 && P_{2}  \mbox{=}
  \big(
      0.0193713, \mbox{-}0.00192056, \mbox{-}0.000158648
  \big),
  \label{fixed-point-2}
  \\
 && P_{3}  \mbox{=}
  \big(
      0.0193713, 0.00192056, \mbox{-}0.000158648
  \big),
  \label{fixed-point-3}
  \\
&&  P_{4}  \mbox{=}
  \big(
      0.0147126, \mbox{-}0.00101842, 0.00026056
  \big),
  \label{fixed-point-4}
  \\
&&  P_{5}  \mbox{=}
  \big(
      0.0140075, 0, \mbox{-}0.000264616
  \big),
  \label{fixed-point-5}
  \\
 &&   P_{6}  \mbox{=}
  \big(
      0.0140587, 0, 0.000246764
  \big),
  \label{fixed-point-6}
  \\
 && P_{7} \mbox{=}
  \big(
      0.0147126, 0.00101842, 0.00026056
  \big).
  \label{fixed-point-7}
\end{eqnarray}
  \label{subeqs-fixed-points}
\end{subequations}
There is one more fixed point, $P_0=(0,0,0)$, but it is
unstable, see scaling equations (\ref{subeqs-scaling-2nd}).

The scaling pattern of the parameters $\Lambda_{\beta}$
($\beta={\mathrm{d,q,o}}$) depends on the initial values of
the parameters. The initial values of $\Lambda_{\beta}$ are
given by eq. (\ref{subeqs-Lambda-initial})
[see also eq. (\ref{Jd-Jq-Jo-res})].
They consist of the dimensionless parameter $\lambda \rho_0$
which is calculated from a microscopic model of interaction of
a Yb atom in the $^{1}$S$_{0}$ state with an Yb atom in
the $^{3}$P$_{2}$ state, see eq. (\ref{exchange-strength-def}).
It is seen that $\Lambda_{\mathrm{d}}$ is positive, whereas
$\Lambda_{\mathrm{q}}$ and $\Lambda_{\mathrm{o}}$
are negative.

To proceed further, it is necessary to 
carry out stability analysis and study the scaling of
$\Lambda_{\mathrm{d}}$, $\Lambda_{\mathrm{q}}$ and
$\Lambda_{\mathrm{o}}$ near the fixed points
$P_{n}=(\Lambda_{\mathrm{d}}^{(n)},\Lambda_{\mathrm{q}}^{(n)},%
\Lambda_{\mathrm{o}}^{(n)})$.
For this purpose we introduce the variables $x_{\mathrm{d}}$,
$x_{\mathrm{q}}$ and $x_{\mathrm{o}}$,
\begin{eqnarray*}
  \Lambda_{\mathrm{d}}
  &=&
  \Lambda_{\mathrm{d}}^{(n)}+
  x_{\mathrm{d}},
  \\
  \Lambda_{\mathrm{q}}
  &=&
  \Lambda_{\mathrm{q}}^{(n)}+
  x_{\mathrm{q}},
  \\
  \Lambda_{\mathrm{o}}
  &=&
  \Lambda_{\mathrm{o}}^{(n)}+
  x_{\mathrm{o}},
\end{eqnarray*}
and assume that $x_{\beta}$ [$\beta={\mathrm{d,q,o}}$] are small. Expanding
the functions ${\mathfrak{F}}_{\beta}$, eq.
(\ref{subeqs-scaling-3rd}), in $x_{\beta}$ to first (linear) order
we get,
\begin{eqnarray*}
  {\mathfrak{F}}_{\beta}
  \big(
      \Lambda_{\mathrm{d}},~
      \Lambda_{\mathrm{q}},~
      \Lambda_{\mathrm{o}}
  \big)
  &=&
  \sum_{\beta'={\mathrm{d,q,o}}}
  A_{\beta,\beta'}~
  x_{\beta'}+
  O\big(x^2\big),
\end{eqnarray*}
where
\begin{eqnarray*}
  A_{\beta,\beta'} &=&
  \bigg(
       \frac{\partial {\mathfrak{F}}_{\beta}}
            {\partial \Lambda_{\beta'}}
  \bigg)_{P_n},
\end{eqnarray*}
the derivative is taken at the fixed point $P_n$.
Thereby we get a set of linear differential equations for
$x_{\beta}$,
\begin{subequations}
\begin{eqnarray}
  \frac{\partial x_{\mathrm{d}}}{\partial \ln D}
  &=&
  A_{\mathrm{d,d}}~
  x_{\mathrm{d}}+
  A_{\mathrm{d,q}}~
  x_{\mathrm{q}}+
  A_{\mathrm{d,o}}~
  x_{\mathrm{o}},
  \label{eq1-fixed-point-stability}
  \\
  \frac{\partial x_{\mathrm{q}}}{\partial \ln D}
  &=&
  A_{\mathrm{q,d}}~
  x_{\mathrm{d}}+
  A_{\mathrm{q,q}}~
  x_{\mathrm{q}}+
  A_{\mathrm{q,o}}~
  x_{\mathrm{o}},
  \label{eq2-fixed-point-stability}
  \\
  \frac{\partial x_{\mathrm{o}}}{\partial \ln D}
  &=&
  A_{\mathrm{o,d}}~
  x_{\mathrm{d}}+
  A_{\mathrm{o,q}}~
  x_{\mathrm{q}}+
  A_{\mathrm{o,o}}~
  x_{\mathrm{o}}.
  \label{eq3-fixed-point-stability}
\end{eqnarray}
  \label{subeqs-fixed-point-stability}
\end{subequations}
The solution of the set of equations
(\ref{subeqs-fixed-point-stability}) is of the form,
\begin{eqnarray*}
  x_{\beta} &\propto&
  D^{\gamma},
\end{eqnarray*}
where the Lyapunov exponent $\gamma$ is an eigenvalue of the set
of equations (\ref{subeqs-fixed-point-stability}). The set of
three linear equations has, as a rule, three eigenvalues. A fixed
point $P_n$ is stable when all $x_{\beta}$ tend to zero as
$D$ tends to zero. This occurs when all $\gamma$'s are positive.
Accordingly, we now  write down the numerical values of the
triples ($\gamma_1,\gamma_2,\gamma_3$) for each one of the fixed
points $P_1$ -- $P_7$ in its turn and determine its stability
(s=stable, u=unstable).
\\
\ \\
$P_1:  (\gamma_1,\gamma_2,\gamma_3)$=%
$(0.0285714, $-$0.246429, $-$0.375) ~~~ \Rightarrow ~ \mbox{u}.$
\\
$P_2: (\gamma_1,\gamma_2,\gamma_3)$=%
$(0.341287,$-$0.228946, 0.163813) ~ \Rightarrow ~ \mbox{u}.$
\\
$P_3: (\gamma_1,\gamma_2,\gamma_3)$=%
$(0.341287,$-$0.228946, 0.163813) ~ \Rightarrow ~ \mbox{u}.$
\\
$P_4: (\gamma_1,\gamma_2,\gamma_3)$=%
$(0.44974, 0.320668, 0.0632014) ~ \Rightarrow ~~ \mbox{s}.$
\\
$P_5: (\gamma_1,\gamma_2,\gamma_3)$=%
$(0.434777, 0.14764, 0.14764) ~~~~ \Rightarrow ~~ \mbox{s}.$
\\
$P_6: (\gamma_1,\gamma_2,\gamma_3)$=%
$(0.40609, 0.271873, $-$0.0300051) ~ \Rightarrow ~ \mbox{u}.$
\\
$P_7:(\gamma_1,\gamma_2,\gamma_3)$=%
$(0.44974, 0.320668, 0.0632014) ~ \Rightarrow ~~ \mbox{s}.$
\\
\noindent
Note that $P_1$ is the NB fixed point, that in
this case is unstable. Accordingly, only $P_4$, $P_5$ and $P_7$ are
stable. Elucidation of these three stable fixed points such that
the corresponding fixed point Hamiltonians display non-Fermi
liquid behavior (see section \ref{sec-ground-state}) is one of
the central results of the present work.

\section{Analysis of the Scaling Equations}
  \label{sec-analysis}
\noindent
Analysis of
the flow pattern for the system of three coupled non-linear scaling
equations constructed above is rather rich and complicated. To some extent, in
Figs.~ \ref{Fig-d-neg-FP-num}  and  \ref{Fig-ScEqs-3rd-d-o-num},
we work out the analogue of the celebrated Anderson-Yuval
equations (derived for the anisotropic Kondo effect), as adapted
for the present model. Unlike the former case, however, where
the fixed points are at infinity, the present analysis leads to
the occurrence of finite fixed points.

Let us inspect the scaling equations (\ref{scaling-3rd}) in some details.
Note that when $\Lambda_{\mathrm{q}}=\Lambda_{\mathrm{o}}=0$,
then the s-d model is recovered as a special case, and scaling of
$\Lambda_{\mathrm{d}}$ depends on the sign of
$\Lambda_{\mathrm{d}}^{(0)}$.  For $\Lambda_{\mathrm{d}}^{(0)}>0$
$\Lambda_{\mathrm{d}}(D)$ flows towards
the Nosi\`er-Blandin fixed point $P_1$, eq. (\ref{fixed-point-1})
indicating an over-screened Kondo effect. When $\Lambda_{\mathrm{d}}^{(0)}<0$,
$\Lambda_{\mathrm{d}}(D)$ flows towards zero
which means that there is no Kondo effect.
When $\Lambda_{\mathrm{q}}^{(0)}$ and/or
$\Lambda_{\mathrm{o}}^{(0)}$ are non-zero,
the scenario is more complicated. In order to analyze
scaling, we consider the following equations,
\begin{subequations}
\begin{eqnarray}
  \frac{\partial \Lambda_{\mathrm{q}}}
       {\partial \Lambda_{\mathrm{d}}}
  &=&
  \frac{{\mathfrak{F}}_{\mathrm{q}}
        \big(
            \Lambda_{\mathrm{d}},
            \Lambda_{\mathrm{q}},
            \Lambda_{\mathrm{o}}
        \big)}
       {{\mathfrak{F}}_{\mathrm{d}}
        \big(
            \Lambda_{\mathrm{d}},
            \Lambda_{\mathrm{q}},
            \Lambda_{\mathrm{o}}
        \big)},
  \label{scaling-eq-q-vs-d}
  \\
  \frac{\partial \Lambda_{\mathrm{o}}}
       {\partial \Lambda_{\mathrm{d}}}
  &=&
  \frac{{\mathfrak{F}}_{\mathrm{o}}
        \big(
            \Lambda_{\mathrm{d}},
            \Lambda_{\mathrm{q}},
            \Lambda_{\mathrm{o}}
        \big)}
       {{\mathfrak{F}}_{\mathrm{d}}
        \big(
            \Lambda_{\mathrm{d}},
            \Lambda_{\mathrm{q}},
            \Lambda_{\mathrm{o}}
        \big)},
  \label{scaling-eq-o-vs-d}
\end{eqnarray}
  \label{subeqs-scaling-eqs-q-o-vs-d}
\end{subequations}
where ${\mathfrak{F}}_{\mathrm{d,q,o}}$ are given by
eq. (\ref{subeqs-scaling-3rd}). Solving the set of equations
(\ref{subeqs-scaling-eqs-q-o-vs-d}), we get
$\Lambda_{\mathrm{q}}$ and $\Lambda_{\mathrm{o}}$
as functions of $\Lambda_{\mathrm{d}}$.
Numerical solution of the set of equations is illustrated in
Figs. \ref{Fig-d-neg-FP-num} and \ref{Fig-ScEqs-3rd-d-o-num},
and implies that there are sveral scaling regimes as follows:
\begin{itemize}
\item All the running coupling constants $\Lambda_{\beta}$ flow to zero as
      $D \to 0$, hence there is no Kondo effect.

\item The running coupling constants $\Lambda_{\beta}$ flow to one
      of the stable fixed points. In this case, Kondo effect realizes.
\end{itemize}
At this stage it should be determined for which
values of the coupling constants $\Lambda_{\beta}^{(0)}$ there is Kondo effect,
and for which ones there is not. Our numerical analysis shows
that when $\Lambda_{\mathrm{d}}>0$,  Kondo effect
 {\it always exists} (see Fig. \ref{Fig-ScEqs-3rd-d-o-num}).
Therefore it is left to investigate the case
$\Lambda_{\mathrm{d}}^{(0)}<0$. The result of our numerical
calculations for this case is shown in Fig. \ref{Fig-d-neg-FP-num}.

We are now in a position to analyze the different scaling regimes displayed in this figure.
\begin{figure}[htb]
\centering
  \includegraphics[width=60 mm,angle=0]
   {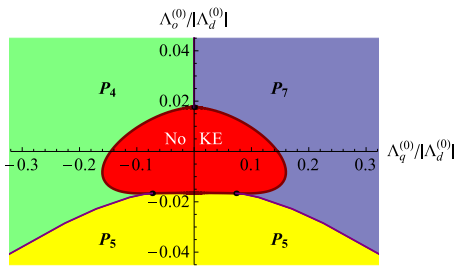}
 \caption{\footnotesize
   ({\color{blue}Color online})
   Scaling of $\Lambda_{\mathrm{d}}$,
   $\Lambda_{\mathrm{q}}$ and $\Lambda_{\mathrm{o}}$
   for $\Lambda_{\mathrm{d}}^{(0)}=-0.00105$ and different
   values of $\Lambda_{\mathrm{q}}^{(0)}$ and
   $\Lambda_{\mathrm{o}}^{(0)}$.
   The red area: all $\Lambda$'s frow to zero and there is no Kondo effect.
   Green area: $\Lambda$'s flow to the fixed point $P_4$.
   Yellow area: $\Lambda$'s flow to the fixed point $P_5$.
   Blue area: $\Lambda$'s flow to the fixed point $P_7$.}
 \label{Fig-d-neg-FP-num}
\end{figure}
When the effective bandwidth decreases, the coupling
$\Lambda_{\mathrm{d}}$ increases from its negative initial value
and tends to 0. At this stage, it is important to determine
whether $|\Lambda_{\mathrm{q}}(D)|$ and
$|\Lambda_{\mathrm{o}}(D)|$ decrease faster or slower
than $|\Lambda_{\mathrm{d}}(D)|$. In other words,
we should consider the dimensionless parameters
${\mathcal{K}}_{\beta}$ ($\beta=$d,q,o), defined as,
$$
  {\mathcal{K}}_{\beta} ~=~
  \frac{\partial \ln |\Lambda_{\beta}(D_0)|}
       {\partial \ln D_0}.
$$
(For $\Lambda_{\mathrm{d}}<0$, all ${\mathcal{K}}_{\beta}$ are
positive).
When ${\mathcal{K}}_{\mathrm{d}}<{\mathcal{K}}_{\mathrm{q,o}}$,
the couplings $\Lambda_{\mathrm{q,o}}$ vanish faster
than $\Lambda_{\mathrm{d}}$. As a result, the Kondo Hamiltonian
renormalizes towards the s-d model Hamiltonian with ferromagnetic
coupling $\Lambda_{\mathrm{d}}(D)$. This coupling flows towards
zero when $D$ vanishes. This is the case when $\Lambda_{\beta}$
are in the red area in Fig. \ref{Fig-d-neg-FP-num} (see also dark
red arrowed curves in Fig. \ref{Fig-ScEqs-3rd-d-o-num}).

\begin{figure}[htb]
\centering
  \includegraphics[width=60 mm,angle=0]
   {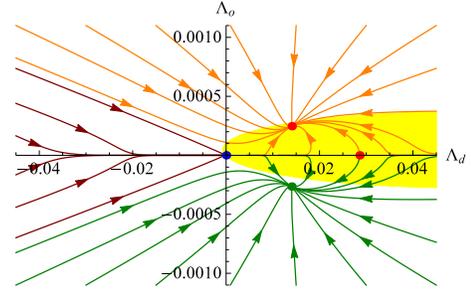}
 \caption{\footnotesize
   ({\color{blue}Color online})
   Scaling of $\Lambda_{\mathrm{d}}$ and
   $\Lambda_{\mathrm{o}}$ for $\Lambda_{\mathrm{q}}=0$.
   Dark red arrowed lines: the couplings rescale to zero.
   Green lines: the couplings renormalize towards the stable fixed
   point $P_5$, eq. (\ref{fixed-point-5}).
   Orange lines: the couplings renormalize towards the saddle
   fixed point $P_6$, eq. (\ref{fixed-point-6}).
   When $\Lambda_{\mathrm{q}}$ is very small but not zero,
   the couplings renormalize towards the fixed points
   $P_4$ or $P_7$, eqs. (\ref{fixed-point-4}) or (\ref{fixed-point-7})
   depending on whether the initial value of $\Lambda_{\mathrm{q}}$
   is negative or positive.}
 \label{Fig-ScEqs-3rd-d-o-num}
\end{figure}

When ${\mathcal{K}}_{\mathrm{q}}<{\mathcal{K}}_{\mathrm{d}}$
and/or ${\mathcal{K}}_{\mathrm{o}}<{\mathcal{K}}_{\mathrm{d}}$,
then $\Lambda_{\mathrm{d}}$ vanishes when $\Lambda_{\mathrm{q}}$
an/or $\Lambda_{\mathrm{o}}$ assume finite values. At this point,
$\Lambda_{\beta}$ continues to flow [see eqs. (\ref{scaling-3rd})
and (\ref{subeqs-scaling-3rd})]. $\Lambda_{\mathrm{d}}$,
for example, changes its sign and the couplings $\Lambda_{\beta}$
flow towards one of the fixed points, $P_4$, $P_5$ or $P_7$
(green, yellow and blue areas in Fig. \ref{Fig-d-neg-FP-num}).
Note that quadrupole and octupole
interaction give rise to exotic property of the Kondo effect:
{\textit{The effective dipole
coupling $\Lambda_{\mathrm{d}}(T)$ as a function of temperature
turns from ferromagnetic at high temperature to antiferromagnetic
at low temperature}}. This property is shown in Fig. \ref{Fig-ScEqs-3rd-d-o-num},
see orange and green arrowed curves.

It should be noted that when $\Lambda_{\mathrm{q}}=0$,
the function ${\mathfrak{F}}_{\mathrm{q}}=0$ [see
eq. (\ref{eq2-scaling-3rd})]. Therefore, when $\Lambda_{\mathrm{q}}^{(0)}=0$,
then $\Lambda_{\mathrm{q}}(D)=0$ for any $D<D_0$.
Consider renormalization of $\Lambda_{\mathrm{d}}$
and $\Lambda_{\mathrm{o}}$ in the plane
$\Lambda_{\mathrm{q}}=0$. Numerical solution
of eq. (\ref{scaling-eq-o-vs-d}) for $\Lambda_{\mathrm{q}}(0)=0$
is displayed in Fig. \ref{Fig-ScEqs-3rd-d-o-num}. It is seen
that the couplings flow to one of the fixed points,
$P_0$, $P_5$ or $P_6$, eq. (\ref{subeqs-fixed-points}).
In order to check stability of the solution, we apply
the Lyapunov method for stability. For this purpose,
we consider the scaling equation for $\Lambda_{\mathrm{q}}(D)$,
\begin{eqnarray*}
  \frac{\partial \Lambda_{\mathrm{q}}}
       {\partial \ln D}
  &=&
  {\mathfrak{F}}_{\mathrm{q}}
  \big(
      \Lambda_{\mathrm{d}},
      \Lambda_{\mathrm{q}},
      \Lambda_{\mathrm{o}}
  \big),
\end{eqnarray*}
with infinitesimal initial condition $\Lambda_{\mathrm{q}}^{(0)}$.
Keeping  just the linear  power of $\Lambda_{\mathrm{q}}$ on the right hand side
of the last equation we may write,
\begin{eqnarray}
  \frac{\partial \Lambda_{\mathrm{q}}}
       {\partial \ln D}
  &=&
  \Lambda_{\mathrm{q}}~
  {\mathcal{A}}_{\mathrm{q}}
  \big(
      \Lambda_{\mathrm{d}},
      \Lambda_{\mathrm{o}}
  \big),
  \label{eq-for-q-linear}
\end{eqnarray}
where
$$
  {\mathcal{A}}_{\mathrm{q}}
  \big(
      \Lambda_{\mathrm{d}},
      \Lambda_{\mathrm{o}}
  \big)
  ~=~
  \lim_{\Lambda_{\mathrm{q}}\to0}
  \Bigg(
       \frac{\partial
             {\mathfrak{F}}_{\mathrm{q}}
             \big(
                 \Lambda_{\mathrm{d}},
                 \Lambda_{\mathrm{q}},
                 \Lambda_{\mathrm{o}}
             \big)}
            {\partial \Lambda_{\mathrm{q}}}
  \Bigg).
$$
The solution of eq. (\ref{eq-for-q-linear}) is,
\begin{eqnarray}
  \frac{\Lambda_{\mathrm{q}}(D)}
       {\Lambda_{\mathrm{q}}^{(0)}}
  =
  \exp
  \Bigg\{
       -\int\limits_{D}^{D_0}
       {\mathcal{A}}_{\mathrm{q}}
       \Big(
           \Lambda_{\mathrm{d}}(D'),
           \Lambda_{\mathrm{o}}(D')
       \Big)~
       \frac{d D'}{D'}
  \Bigg\}.
  \label{solution-q-linear}
\end{eqnarray}
Note that when $D$ vanishes, $\Lambda_{\mathrm{d}}$ and
$\Lambda_{\mathrm{o}}$ flow to one of the fixed points
(\ref{subeqs-fixed-points}), where ${\mathcal{A}}_{\mathrm{q}}$
takes a finite value. In this case the integral on the right hand
side of eq. (\ref{solution-q-linear}) diverges. Thus, we conclude
that when ${\mathcal{A}}_{\mathrm{q}}$ is positive along all
the scaling trajectories of $\Lambda_{\mathrm{d}}(D)$ and
$\Lambda_{\mathrm{o}}(D)$, then $\Lambda_{\mathrm{q}}(D)$ flows
to zero, and the solution displayed in Fig.
\ref{Fig-ScEqs-3rd-d-o-num} is stable. When
${\mathcal{A}}_{\mathrm{q}}$ is negative, then
$\Lambda_{\mathrm{q}}(D)$ flows away from zero, and the solution
displayed in Fig. \ref{Fig-ScEqs-3rd-d-o-num} is unstable.
The interval of $\Lambda_{\mathrm{d}}$ and $\Lambda_{\mathrm{o}}$
where the solution displayed in Fig. \ref{Fig-ScEqs-3rd-d-o-num}
is unstable is marked by yellow.

Finally, we just state our result pertaining to scaling of
the couplings satisfying the initial conditions
(\ref{subeqs-Lambda-initial}). Numerical analysis shows
that for any positive $\lambda$, the couplings flow towards
the fixed point $P_5$.

\section{The Strong Coupling Regime}
  \label{sec-ground-state}

It is expected that the physics of an over-screened Kondo effect is exposed 
mainly in the strong coupling regime.  Limiting one to the weak coupling regime 
turns it difficult to determine whether
the pertinent Kondo physics at the stable points is that of
over-screening or under screening.
Thus, after the candidates for stable fixed points are identified,
it is necessary to elucidate the ground-state wave functions
at these points. The reason is at least two-fold. First, it is
required in order to evaluate physical observables at low temperatures 
$T<T_K$.  Second,
it is essential  to determine whether the strong coupling fixed
point is unstable, so that according to NB analysis, there is
a stable finite fixed point, and over-screening does occur.
This task is carried out below, using variational wave functions.
It is then found that the nature of the system (whether there is
or there is no over-screening) depends on the initial values of
the bare constants $\lambda_{\mathrm{d}}$,
$\lambda_{\mathrm{q}}$ and $\lambda_{\mathrm{o}}$.
Similar (albeit simpler) situation is encountered in
the two-channel Kondo effect based on the $s-d$ Hamiltonian.

The variational method is
an appropriate tool for that purpose,
as it is not based on
perturbation theory. It can be shown that the minimal energy
can be reached when the number of atoms over the fully occupied
Fermi sphere is ${\mathcal{N}}=I+\frac{1}{2}$.
For $I=\frac{5}{2}$, ${\mathcal{N}}=3$. Such a three particle
wave function is encoded by the spin $S$ of the three atoms.
The maximal value of the spin is $S=\frac{9}{2}$. This value
is obtained as follows: according to the Pauli principle,
the magnetic quantum numbers satisfy the inequalities
${i}_{1}\neq{i}_{2}$, ${i}_{2}\neq{i}_{3}$ and
${i}_{3}\neq{i}_{1}$. Therefore the maximal magnetic quantum
number of the three atoms is $S=\frac{9}{2}$.

A simple form of a variational wave function is,
\begin{eqnarray}
  \big|
      \Sigma,m
  \big\rangle
  &=&
  \sum_{f,\{i\}_3,\{n\}_3}
  C^{\Sigma,m}_{S,s;F,f}~
  \psi_{\Sigma}(n_1)~
  \psi_{\Sigma}(n_2)
  \psi_{\Sigma}(n_3)
  \times \nonumber \\ && \times
  C^{S,s}_{I,i_1;I,i_2;I,i_3}~
  c_{n_1,i_1}^{\dag}~
  c_{n_2,i_2}^{\dag}~
  c_{n_3,i_3}^{\dag}~
  \big|
      f;\Omega
  \big\rangle,
  \label{WF-var}
\end{eqnarray}
where $|f;\Omega\rangle$ describes an impurity with magnetic
quantum number $f$ and a Fermi sea at Fermi energy $\epsilon_F$.
The Clebsch-Gordan coefficients, $C^{\Sigma,m}_{S,s;F,f}$}
 entail the restriction $\Sigma =S+F,S+F-1,S+F-2,\ldots,S-F$
on the total angular momentum of the four atom system (three itinerant and one impurity).
The symbols $C^{S,s}_{I,i_1;I,i_2;I,i_3}$ are the so called three particle
Clebsch-Gordan coefficients. Here we use the fact that
$H_{\mathrm{d}}$, $H_{\mathrm{q}}$ and $H_{\mathrm{o}}$ commute
with each other (see Appendix \ref{append-eigenstates-dd-qq-oo}
for details), and therefore the four-atomic orbital angular
momentum $\Sigma $ is a good quantum number.

In order to find the components $\psi_{\Sigma}(n)$, we write down the Schr\"odinger
equation,
$$
  H
  \big| \Sigma,m\big\rangle
  ~=~
  \veps
  \big| \Sigma,m\big\rangle,
$$
and that yields the algebraic eigenvalue problem,
\begin{eqnarray}
  \big(
      \veps_n-
      \veps
  \big)
  \psi_{\Sigma}(n)+
  g_{\Sigma}
  \sum_{n'}
  \Theta(\veps_{n'}-\epsilon_F)
  \psi_{\Sigma}(n')
  = 0.
  \label{eq-Schrodinger-1atom}
\end{eqnarray}
Here
\begin{eqnarray}
  g_{\Sigma} &=&
  \Big\{
      \lambda_{\mathrm{d}}~
      {\mathcal{D}}_{\Sigma}+
      \lambda_{\mathrm{q}}~
      {\mathcal{Q}}_{\Sigma}+
      \lambda_{\mathrm{o}}~
      {\mathcal{O}}_{\Sigma}
  \Big\}~
  \frac{k_F}{a_{\g}^{2}},
  \label{gL-def}
\end{eqnarray}
where
\begin{eqnarray*}
  {\mathcal{D}}_{\Sigma} &=&
  \frac{1}{2}~
  \Big\{
      \Sigma(\Sigma +1)-
      F(F+1)-
      I(I+1)
  \Big\},
  \nonumber
  \\
  {\mathcal{Q}}_{\Sigma} &=&
  4~
  {\mathcal{D}}_{\Sigma}^{2}+
  2~
  {\mathcal{D}}_{\Sigma}-
  \frac{4}{3}~
  F(F+1)~
  S(S+1),
  \nonumber
  \\
  {\mathcal{O}}_{\Sigma} &=&
  36
  {\mathcal{D}}_{\Sigma} ^{3}+
  72
  {\mathcal{D}}_{\Sigma}^{2}+
  12
  {\mathcal{D}}_{\Sigma}-
  \nonumber \\ && -
  \frac{12}{5}~
  \big(3S(S+1)-1\big)~
  \big(2F(F+1)-1\big)~
  {\mathcal{D}}_{\Sigma}-
  \nonumber \\ && -
  18~
  S(S+1)~
  F(F+1).
\end{eqnarray*}
The solution of eq. (\ref{eq-Schrodinger-1atom}) is,
\begin{eqnarray*}
  \psi_{\Sigma}(n) &=&
  \frac{A_{\Sigma}}{\veps_{\Sigma}-\veps_{n}},
\end{eqnarray*}
where $A_{\Sigma}$ is a normalization constant.
The energy $\veps_{\Sigma}$ can be found from
the equation,
\begin{eqnarray}
  g_{\Sigma}
  \sum_{n}
  \frac{\Theta(\veps_{n}-\epsilon_F)}
       {\veps_{\Sigma}-\veps_{n}}+
  1 &=& 0.
  \label{eq-for-veps-1atom}
\end{eqnarray}
We are interested in the energies $\veps_{\Sigma}$ which are below
the Fermi energy $\epsilon_F$. This is the case when $g_{\Sigma}<0$.
Introducing the density of states, we can write,
\begin{eqnarray*}
  \veps_{\Sigma}-\epsilon_F &=&
  D_0
  \exp\bigg(-\frac{1}{|g_L|~\rho_0}\bigg).
\end{eqnarray*}

The energy of the ground state is found as
$$
  \veps_{\mathrm{gs}} ~=~
  \min_{\Sigma}\veps_{\Sigma}.
$$
Thus, the problem of finding the ground state reduces
to that of finding a minimum of $g_{\Sigma}$.
In order to check whether the magnetic impurity is over-screened
or under-screened, we consider the operator
\begin{eqnarray}
  \big(
      {\boldsymbol \Sigma}
      \cdot
      {\mathbf{F}}
  \big)
  =
  \frac{1}{2}~
  \big\{
      \Sigma(\Sigma +1)+
      F(F+1)-
      S(S+1)
  \big\}.
  \label{L-dot-F}
\end{eqnarray}
When $({\boldsymbol \Sigma}\cdot{\mathbf{F}})<0$,
there is over-screened Kondo effect. Using eq. (\ref{L-dot-F}),
the inequality becomes,
\begin{eqnarray}
  \Sigma(\Sigma +1)+
  F(F+1)-
  S(S+1)
  ~<~ 0.
  \label{cond-over-scr}
\end{eqnarray}
For $S=\frac{9}{2}$ and $F=\frac{3}{2}$, the inequality
(\ref{cond-over-scr}) is fulfilled whenever $\Sigma =3$ or $4$.
Thus, when the condition
(\ref{cond-over-scr}) is fulfilled, there is an over-screened Kondo effect with
non Fermi liquid ground state. Note that this is not a necessary condition:
When the inequality (\ref{cond-over-scr}) is not satisfied, we cannot determine
the nature of the ground state.

We now apply our analysis for elucidating the nature of the stable
fixed points $P_4$, $P_5$ and $P_{7}$. For the fixed point $P_4$,
the ground state corresponds to the energy level with quantum number 
$\Sigma =3$, and
therefore there is an over-screened Kondo effect. Similarly, for the fixed
point $P_5$, the ground state corresponds to the energy level
with with quantum number 
$\Sigma =4$, and therefore there is over-screened Kondo effect.
Finally, for the fixed point $P_7$, the ground state corresponds
to the energy level with with quantum number 
$\Sigma =5$, and therefore we cannot conclude
whether the impurity is over-screened or under-screened.
As an example, when the initial values of the couplings
$\lambda_{\mathrm{d}}$, $\lambda_{\mathrm{q}}$ and
$\lambda_{\mathrm{o}}$ are given by eq.
(\ref{subeqs-Lambda-initial}), the Kondo Hamiltonian flows toward
the fixed point $P_5$, and therefore we conclude that there is
an over-screened Kondo effect. In order to substantiate this statement, 
the exchange interaction between the ``dressed'' impurity
atom and the Fermi sea should be considered. 
For this purpose, it is assumed that the temperature
is low enough, so that the ``dressed'' impurity is in its ground state
described by the wave functions (\ref{WF-var}), 
and the following representation of the identity operator is employed,
\begin{eqnarray*}
  \sum_{n,i,m}
  c_{n,i}^{\dag}~
  \big|
      \Sigma,m;{\mathrm{g}}
  \big\rangle
  \big\langle
      \Sigma,m;{\mathrm{g}}
  \big|
  c_{n,i} &+&
  \nonumber \\ +
  \sum_{n,i,m}
  c_{n,i}~
  \big|
      \Sigma,m;{\mathrm{g}}
  \big\rangle
  \big\langle
      \Sigma,m;{\mathrm{g}}
  \big|
  c_{n,i}^{\dag}
  &=& 1,
\end{eqnarray*}
where $| \Sigma,m;{\mathrm{g}}\rangle$ describes the degenerate Fermi
sea of the itinerate atoms and the ``dressed'' impurity.
Then the exchange interaction of the ``dressed'' impurity with
the itinerant atoms is
\begin{eqnarray*}
  \tilde{H}_{K} &=&
  \sum_{n,n'}
  \sum_{i,i'}
  \sum_{m,m'}
  \Big\{
      \big\langle
          \Sigma,m;{\mathrm{g}}
      \big|
          c_{n',i'}
          H_K
          c_{n,i}^{\dag}
      \big|
          \Sigma,m';{\mathrm{g}}
      \big\rangle-
  \nonumber \\ &-&
      \big\langle
          \Sigma,m;{\mathrm{g}}
      \big|
          c_{n,i}^{\dag}
          H_K
          c_{n',i'}
      \big|
          \Sigma,m';{\mathrm{g}}
      \big\rangle
  \Big\}~
  Y^{m,m'}
  c_{n',i'}^{\dag}
  c_{n,i},
\end{eqnarray*}
where $Y^{m,m'}=|{\Sigma,m}\rangle\langle{\Sigma,m'}|$ are 
Hubbard operators.
Taking into account eqs. (\ref{HK-def}) and (\ref{WF-var}), we may
write
\begin{eqnarray}
  \tilde{H}_{K} &=&
  \frac{\tilde\lambda}{a_{\mathrm{g}}^{2}}
  \sum_{n,n'}
  \sum_{i,i'}
  \sum_{m,m'}
  \sum_{j}
  \sum_{\{i\}_{3}}
  \sum_{\{i'\}_{3}}
  Y^{m,m'}
  c_{n',i'}^{\dag}
  c_{n,i}
  \times \nonumber \\ && \times
  C^{\Sigma,m}_{I,i_1;I,i_2;I,i_3;I,i;J,j}~
  C^{\Sigma,m'}_{I,i'_1;I,i'_2;I,i'_3;I,i';J,j},
  \label{HK-strong}
\end{eqnarray}
where $C^{\Sigma,m}_{I,i_1;I,i_2;I,i_3;I,i;J,j}$ are
the Clebsch-Gordan coefficients, $\{i\}_{3}=\{i_1,i_2,i_3\}$ and
$\{i'\}_{3}=\{i'_1,i'_2,i'_3\}$. The coupling
$\tilde\lambda\sim{T}_{K}$ is positive.

The Hamiltonian (\ref{HK-strong}) is decomposed into a sum of
multipole interactions, similar to the Hamiltonian
(\ref{HK-Hd-Hq-Ho}). However, because of high spins that are involved [the dressed
impurity has the spin $\Sigma =4$ and the itinerant atoms have spin
$I=\frac{5}{2}$], the exchange Hamiltonian
consists of dipole, quadrupole, octupole, 16-pole and 32-pole
interactions. Derivation of the scaling equations for
this Hamiltonian is much more cumbersome than the derivation of
the scaling equations for the bare Hamiltonian (\ref{HK-Hd-Hq-Ho}).
Therefore we will be content with a qualitative picture pertaining to
the Hamiltonian (\ref{HK-strong}): The fact that 
$\tilde\lambda>0$, implies that $\tilde{H}_{K}$ displays an
antiferromagnetic exchange interaction. This is a typical situation leading to
 over-screening Kondo effect, where the exchange interaction
between the dressed impurity and the Fermi sea is
anti-ferromagnetic \cite{NB-JPh-80,Hewson-book}. It can be shown
that in the framework of second order poor man's scaling
technique, the anti-ferromagnetic coupling flows towards
$\tilde\lambda\to\infty$, and therefore the weak coupling
fixed point $\lambda
       \to\infty$ is unstable. Thus, the weak coupling
fixed points $\lambda=0$ and $\lambda \to\infty$  are unstable, 
implying that there is at least
one stable strong coupling fixed point with finite $\lambda$'s
which describes a non Fermi phase \cite{NB-JPh-80}.

\ \\
\section{Entropy, Specific Heat and Magnetic Susceptibility}
  \label{sec-suscept}
\noindent
We are now in a position to calculate a few 
experimentally relevant physical quantities. 
A possible candidate for elucidating
the special features of the multipolar 
Kondo effect is
the temperature
dependence of numerous thermodynamic quantities.
Here we compute
the impurity contribution to the specific heat, the entropy, and the magnetic susceptibility,  
 and compare some of our results with
those obtained within the standard Kondo effect based on the $s-d$ Hamiltonian.

The formalism developed so far enables us to carry out these
calculations in the weak coupling regime
$T>T_K$,
wherein it is expected that the general form of the thermodynamic
quantities is dominated by logarithmic functions of
$\frac{D}{T}$. Whereas for dipolar exchange interaction
(governed by the $s-d$ Hamiltonian), the derivation is quite
standard, the derivation and handling of the spin algebra in
the present case of multipolar exchange interactions ( carried out below) is more
involved. It is found that for the magnetic susceptibility,
the temperature dependencies in  the standard and multipolar Kondo
effect are quite close to each other but for the specific heat and
entropy the differences are quite sizeable. We are tempted to expect
that in the strong coupling regime, the dependencies will be
qualitatively and quantitatively distinct.

\subsection{Entropy and Specific Heat}

The impurity contributions to the
entropy $S_{\mathrm{imp}}$ and the specific heat $C_{\mathrm{imp}}$
are given by, 
\begin{eqnarray}
  S_{\mathrm{imp}} &=&
  -k_{\mathrm{B}}~
  \frac{\partial \big(T \ln Z_{\mathrm{imp}}\big)}
       {\partial T},
  \label{entropy-def}
  \\
  C_{\mathrm{imp}} &=&
  T~
  \frac{\partial S_{\mathrm{imp}}}
       {\partial T}.
  \label{Cv-def}
\end{eqnarray}
Here $Z_{\mathrm{imp}}$ is the partition function of the impurity
\cite{Hewson-book},
\begin{eqnarray}
  Z_{\mathrm{imp}} &=&
  \frac{Z}{Z_{\mathrm{c}}},
  \label{Z-imp-def}
\end{eqnarray}
where $Z$ is the partition function of the total system and
$Z_{\mathrm{c}}$ is the partition function of the itinerant
atoms without impurity,
\begin{eqnarray*}
  Z ~=
  {\mathrm{Tr}}
  \big(
      e^{-\beta H}
  \big),
  \ \ \ \ \
  Z_{\mathrm{c}} =
  {\mathrm{Tr}}
  \big(
      e^{-\beta H_{\mathrm{c}}}
  \big).
\end{eqnarray*}

Perturbation calculations up to forth order in the coupling constants yield
the entropy (\ref{entropy-def}),
\begin{eqnarray}
  S_{\mathrm{imp}} =
  k_{\mathrm{B}}~
  \Big\{
      \ln\big(2F+1\big)+
      Z_{3}+
      Z_{4}~
      \ln\big(\beta D\big)
  \Big\},
  \label{entropy-2nd-3rd}
\end{eqnarray}
where
\begin{eqnarray}
  &&
  Z_3 =
  -\frac{2\pi^2}{3}~
  \bigg\{
       \frac{525}{2}~
       \Lambda_{\mathrm{d}}^{3}-
       \frac{18213545952}{5}~
       \Lambda_{\mathrm{o}}^{3}+
  \nonumber \\ && ~~~~~ +
       80640~
       \Lambda_{\mathrm{q}}^{2}
       \Lambda_{\mathrm{d}}+
  \nonumber \\ && ~~~~~ +
       23514624~
       \Lambda_{\mathrm{q}}^{2}
       \Lambda_{\mathrm{o}}+
       39680928~
       \Lambda_{\mathrm{o}}^{2}
       \Lambda_{\mathrm{d}}
  \bigg\},
  \label{Z3-res}
  \\
  &&
  Z_4 =
  -\pi^2~
  \bigg\{
       525~
       \Lambda_{\mathrm{d}}^{4}+
       \frac{656474112}{5}~
       \Lambda_{\mathrm{q}}^{4}+
  \nonumber \\ && ~~~~~ +
       \frac{50025111034752}{25}~
       \Lambda_{\mathrm{o}}^{4}+
       1537536~
       \Lambda_{\mathrm{d}}^{2}~
       \Lambda_{\mathrm{q}}^{2}+
  \nonumber \\ && ~~~~~ +
       1009659168~
       \Lambda_{\mathrm{d}}^{2}~
       \Lambda_{\mathrm{o}}^{2}+
       1065996288~
       \Lambda_{\mathrm{d}}~
       \Lambda_{\mathrm{q}}^{2}~
       \Lambda_{\mathrm{o}}-
  \nonumber \\ && ~~~~~ -
       \frac{671877472896}{5}~
       \Lambda_{\mathrm{d}}~
       \Lambda_{\mathrm{o}}^{3}-
  \nonumber \\ && ~~~~~ -
       \frac{767799502848}{25}~
       \Lambda_{\mathrm{q}}^{2}~
       \Lambda_{\mathrm{o}}^{2}
  \bigg\}.
  \label{Z4-res}
\end{eqnarray}
This result should be supported by the condition imposing 
the invariance of the entropy under
the poor mans scaling transformation \cite{Hewson-book}, implying
\begin{eqnarray}
  \frac{\partial}{\partial \ln D}~
  \Bigg\{
       Z_{\mathrm{imp}}^{(3)}
       \Big(
           \Lambda_{\mathrm{d}},
           \Lambda_{\mathrm{q}},
           \Lambda_{\mathrm{o}}
       \Big)+
  \nonumber \\ +
       {\mathcal{Z}}_{4}
       \Big(
           \Lambda_{\mathrm{d}},
           \Lambda_{\mathrm{q}},
           \Lambda_{\mathrm{o}}
       \Big)~
       \ln\bigg(\frac{D}{k_{\mathrm{B}}T}\bigg)
  \Bigg\}
  &=& 0.
  \label{entropy-scaling}
\end{eqnarray}
Within the accuracy of this equation, when differentiating
the second term, any implicit dependence on $D$
through the couplings $\Lambda_{\mathrm{d,q,o}}$ is neglected.
The renormalization procedure should proceed until the
bandwidth $D$ is reduced to the temperature $T$. At this
point, the fourth order perturbation theory contribution vanishes and
the entropy takes the final form,
\begin{eqnarray}
  S_{\mathrm{imp}} &=&
  k_{\mathrm{B}}~
  \Bigg\{
       \ln\big(2F+1\big)+
  \nonumber \\ && +
       Z_3
       \Big(
           \Lambda_{\mathrm{d}}(T),~
           \Lambda_{\mathrm{q}}(T),~
           \Lambda_{\mathrm{o}}(T)
       \Big)
  \Bigg\},
  \label{S-imp-res}
\end{eqnarray}
where $Z_3$ (as a function of $\Lambda_{\mathrm{d}}$,
$\Lambda_{\mathrm{q}}$ and $\Lambda_{\mathrm{o}}$) is given by
eq. (\ref{Z3-res}), whereas $\Lambda_{\mathrm{d,q,o}}(T)$ are
solution of the scaling equations (\ref{subeqs-scaling-2nd}).

The above results for the entropy 
pave the way for calculating the specific heat (\ref{Cv-def}),
\begin{eqnarray*}
  C_{\mathrm{imp}} &=&
  k_{\mathrm{B}}~
  \frac{\partial}{\partial \ln T}~
  \bigg[
       Z_3
       \Big(
           \Lambda_{\mathrm{d}}(T),~
           \Lambda_{\mathrm{q}}(T),~
           \Lambda_{\mathrm{o}}(T)
       \Big)
  \bigg].
\end{eqnarray*}
Taking into account the scaling equation (\ref{entropy-scaling}),
we get
\begin{eqnarray}
  C_{\mathrm{imp}} &=&
  -k_{\mathrm{B}}~
  Z_4
  \Big(
      \Lambda_{\mathrm{d}}(T),~
      \Lambda_{\mathrm{q}}(T),~
      \Lambda_{\mathrm{o}}(T)
  \Big),
  \label{Cv-res}
\end{eqnarray}
where $Z_4$ (as a function of $\Lambda_{\mathrm{d}}$,
$\Lambda_{\mathrm{q}}$ and $\Lambda_{\mathrm{o}}$) is given by
eq. (\ref{Z4-res}), whereas $\Lambda_{\mathrm{d,q,o}}(T)$ are
solution of the scaling equations (\ref{subeqs-scaling-2nd}).

\begin{figure}[htb]
\centering
\subfigure[]
  {\includegraphics[width=55 mm,angle=0]
   {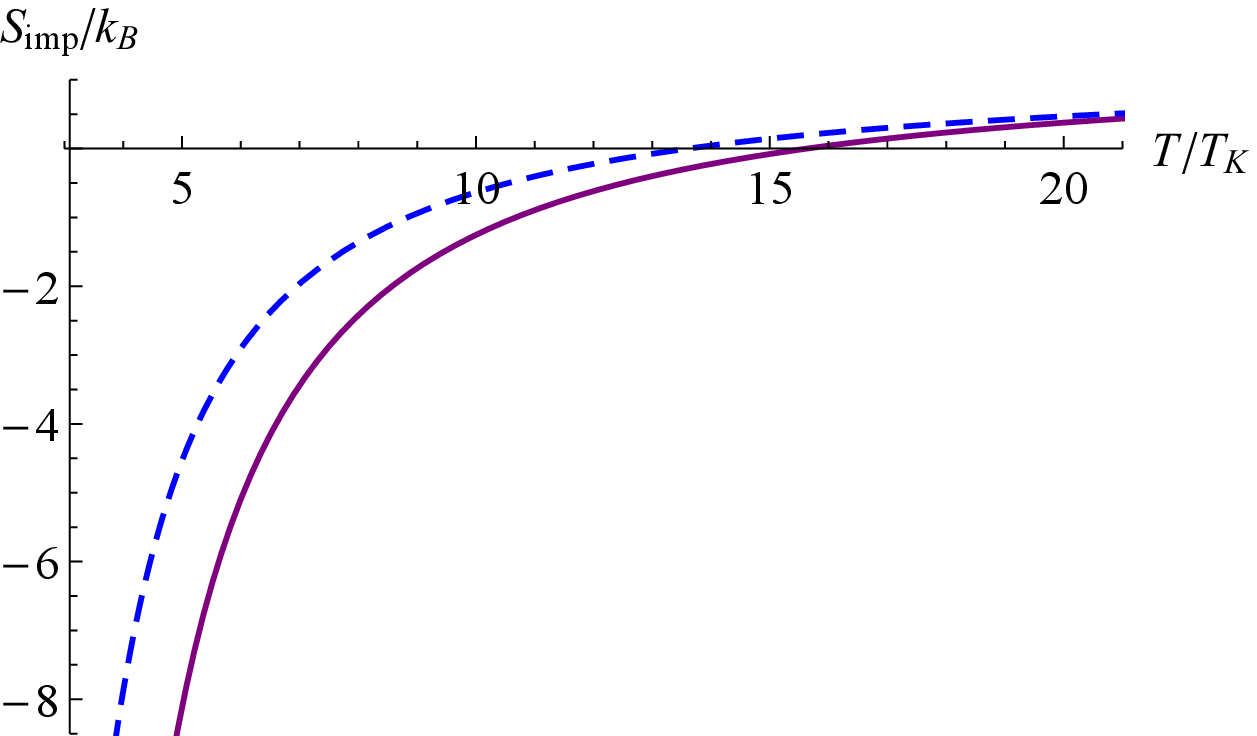}
   \label{Fig-Simp}}
\subfigure[]
  {\includegraphics[width=55 mm,angle=0]
   {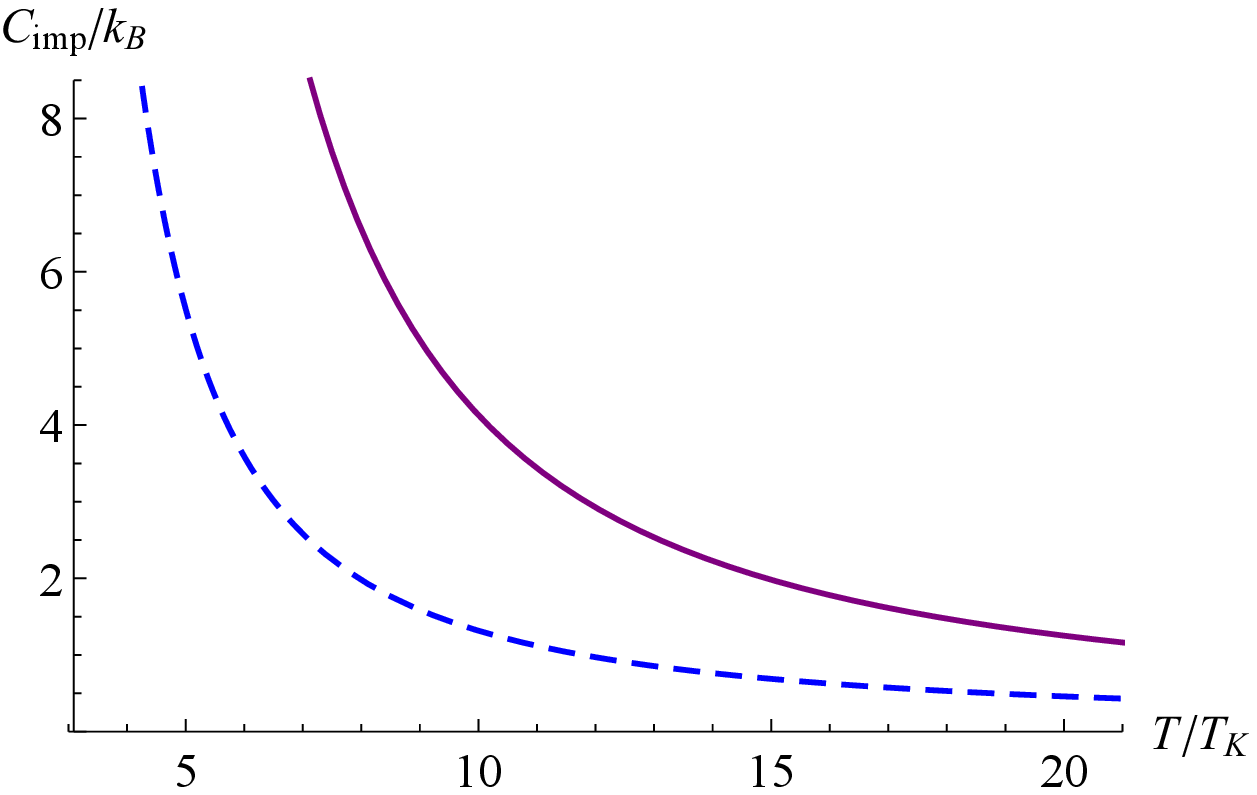}
   \label{Fig-Cimp}}
 \caption{\footnotesize
   ({\color{blue}Color online}) Results for $\lambda \rho_0=0.5$:
   {\underline{Solid curves}}:
   The entropy (\ref{S-imp-res}) [panel (a)] and
   the specific heat (\ref{Cv-res}) [panel (b)]
   for the $^{173}$Yb(\1S0) -- $^{173}$Yb(\3P2) system
  governed by the multipolar Kondo Hamiltonian.
   {\underline{Dashed curves}}:
   The entropy (\ref{S-Hewson}) [panel (a)] and
   the specific heat (\ref{Cv-Hewson}) [panel (b)]
   for the $^{171}$Yb(\1S0) -- $^{171}$Yb(\3P2) system governed 
   by the $s-d$ Kondo Hamiltonian.}
 \label{Fig-Simp-Cimp}
\end{figure}
The entropy (\ref{S-imp-res}) and the specific heat (\ref{Cv-res})
of the impurity are shown in Fig. \ref{Fig-Simp-Cimp} for
$\lambda \rho_0=0.5$ [solid curves in panels (a) and (b)].
Fig. \ref{Fig-Simp} illustrates decreasing of the entropy due to
the Kondo interaction. The entropy of the isolated impurity atom
is $\ln(2F+1)$. Fig. \ref{Fig-Cimp} demonstrates a monotonic behaviour of
the specific heat and the entropy in the weak coupling regime.

For comparison with the standard Kondo effect , consider
the system consisting of $^{171}$Yb(\1S0)
itinerant atoms and $^{171}$Yb(\3P2) localized impurities.
In this case, the orbital angular momentum of the itinerant atoms is
$I=\frac{1}{2}$, whereas the angular momentum
 of the localized
impurity atoms is $F=\frac{3}{2}$ [the electronic orbital moment
is $J=2$, and the nuclear spin is $I=\frac{1}{2}$].
The entropy and the specific heat in this case are given by
eqs. (3.4) and (3.5) in Ref. \cite{Hewson-book}. Taking into
account the invariance of the entropy and the specific heat under
the poor man's scaling, we get the following expressions,
\begin{eqnarray}
  S_{s-d} &=&
  k_{\mathrm{B}}
  \Big\{
      \ln(2F+1)-
  \nonumber \\ && -
      \frac{\pi^2}{3}~
      F(F+1)~
      \Lambda_{s-d}^{3}(T)
  \Big\},
  \label{S-Hewson}
  \\
  C_{s-d} &=&
  k_{\mathrm{B}}~
  \pi^2~
  F(F+1)~
  \Lambda_{s-d}^{4}(T),
  \label{Cv-Hewson}
\end{eqnarray}
where
\begin{eqnarray}
  \Lambda_{s-d}(T) &=&
  \frac{1}{\ln(T/T_K)}.
  \label{Lambda-sd-vs-T}
\end{eqnarray}

The entropy (\ref{S-Hewson}) and the specific heat
(\ref{Cv-Hewson}) for the $^{171}$Yb(\1S0) -- $^{171}$Yb(\3P2)
system is shown in Fig. \ref{Fig-Simp-Cimp} [dashed curves in
panels (a) and (b)].

\subsection{Magnetic Susceptibility}

In order to derive an expression for the magnetic susceptibility
of the atomic gas with the Kondo impurity, we note that
the itinerant atoms are in the electronic spin-singlet state,
whereas the impurity is in the electronic spin-triplet state.
Therefore interaction of itinerant atoms with the magnetic field
is proportional to the nuclear magneton $\mu_{\mathrm{n}}$,
whereas the interaction of the impurity with the magnetic field
is proportional to the Bohr magneton $\mu_B$. The interaction of
the itinerant atoms and the impurity with the magnetic field
${\mathbf{B}}=B{\mathbf{e}}_{z}$ is described by the Hamiltonian,
\begin{eqnarray}
  H_B &=&
  -g_{\mathrm{Yb}}
  \mu_{\mathrm{n}}
  \sum_{i,i',n}
  \big(
      {\mathbf{B}}
      \cdot
      {\mathbf{I}}_{i,i'}
  \big)~
  c_{n,i}^{\dag}
  c_{n,i'}-
  \nonumber \\ &&
  -g
  \mu_B
  \sum_{f}
  \big(
      {\mathbf{B}}
      \cdot
      {\mathbf{F}}_{f,f'}
  \big)~
  X^{f,f'},
  \label{HB-impurity}
\end{eqnarray}
where $g_{\mathrm{Yb}}=-0.2592$ is the nuclear g-factor
of $^{173}$Yb \cite{nuclear-g-factor},
$g$ is electronic g-factor of Yb atom in the $^{3}$P$_{2}$
state,
\begin{eqnarray}
  g ~=~
  \frac{3J(J+1)+S(S+1)-\ell(\ell+1)}{2J(J+1)}
  ~=~
  \frac{3}{2},
  \label{g-factor-3P2}
\end{eqnarray}
where for the $^{3}$P$_{2}$ configuration, $J=2$ and
$\ell=S=1$
[in this section, $S$ and $\ell$ denote the electronic
spin and orbital angular moment of the Yb(\3P2) atom].
Then the impurity magnetization
${\mathbf{M}}_{\mathrm{imp}}=M_{\mathrm{imp}}{\mathbf{e}}_{z}$
can be written as \cite{Hewson-book},
\begin{eqnarray}
  M_{\mathrm{imp}} =
  g
  \mu_B~
  \big\langle
      \hat{F}^{z}
  \big\rangle+
  g_{\mathrm{Yb}}
  \mu_{\mathrm{n}}~
  \Big\{
      \big\langle
          \hat{I}^{z}
      \big\rangle-
      \big\langle
          \hat{I}^{z}
      \big\rangle_{0}
  \Big\},
  \label{magnetiz-def}
\end{eqnarray}
where $\langle\cdots\rangle$ indicates a thermal average with respect
to the total Hamiltonian $H+H_B$, and $\langle\cdots\rangle_{0}$
with respect to $H_0+H_B$,
\begin{eqnarray*}
  \big\langle
      {\mathcal{O}}
  \big\rangle
  &=&
  \frac{{\mathrm{tr}}
        \big(e^{-\beta(H+H_B)}~{\mathcal{O}}\big)}
       {{\mathrm{tr}}~e^{-\beta(H+H_B)}},
  \\
  \big\langle
      {\mathcal{O}}
  \big\rangle_{0}
  &=&
  \frac{{\mathrm{tr}}
        \big(e^{-\beta(H_0+H_B)}~{\mathcal{O}}\big)}
       {{\mathrm{tr}}~e^{-\beta(H_0+H_B)}}.
\end{eqnarray*}
Here $H$ and $H_0$ are given by eq. (\ref{H-tot-def}).

The magnetic interaction described by the Hamiltonian
(\ref{HB-impurity}) has a standard form of a scalar product of
the external magnetic field and the magnetic dipole angular momentum
operators of the impurity and itinerant atoms. It reflects
the fact that (usually), only the dipole moment contributes to
the linear magnetization of atoms. However, somewhat unexpectedly,
the Kondo Hamiltonian (\ref{HK-Hd-Hq-Ho}) gives rise to nontrivial
contributions of the quadrupole and octupole magnetic moments to
the linear magnetization of the system. This requires an analysis
that is distinct from the one employed in the standard treatment of magnetic
susceptibility as applied to the $s-d$ Hamiltonian. Here
 we derive the magnetic susceptibility of the multipolar Kondo Hamiltonian 
(as a function of temperature), in the weak coupling regime, ${T}\gg{T}_{K}$.

\subsection{Contributions to $M_{\mathrm{imp}}$ due to $H_K$}
  \label{subsection-HK-sec-magnet}
  \vspace{-0.1in}
First, let us recall the expression for the magnetization of
an isolated atom. To linear order in the magnetic field $B$,
the magnetization of a single $^{173}$Yb atom in the $^{3}$P$_{2}$
state with $F=\frac{3}{2}$ is,
\begin{eqnarray}
  M_{\mathrm{imp}}^{(0)}
  &=&
  \frac{F(F+1)}{3T}~
  \big(g\mu_B\big)^2
  B.
  \label{magnet-imp-isolat-linear}
\end{eqnarray}
Next, consider the contributions to $M_{\mathrm{imp}}$ due to $H_K$,
$$
  \delta{M}_{\mathrm{imp}}
  ~=~
  M_{\mathrm{imp}}-
  M_{\mathrm{imp}}^{(0)}.
$$
By definition, this contribution is given by,
\begin{eqnarray}
  \delta{M}_{\mathrm{imp}} &=&
  g
  \mu_B~
  \Big\{
      \big\langle
          \hat{F}^{z}
      \big\rangle-
      \big\langle
          \hat{F}^{z}
      \big\rangle_{0}
  \Big\}+
  \nonumber \\ &+&
  g_{\mathrm{Yb}}
  \mu_{\mathrm{n}}~
  \Big\{
      \big\langle
          \hat{I}^{z}
      \big\rangle-
      \big\langle
          \hat{I}^{z}
      \big\rangle_{0}
  \Big\}.
  \label{magnetiz-HK-def}
\end{eqnarray}
Assuming that the couplings $\lambda$'s are small and expanding
$\delta{M}_{\mathrm{imp}}$ with powers of $H_K$ yield,
\begin{eqnarray}
  \delta{M}_{\mathrm{imp}} &=&
  \sum_{n=1}^{\infty}
  \delta{M}_{\mathrm{imp}}^{(n)},
  \label{magnetiz-series}
\end{eqnarray}
where $\delta{M}_{\mathrm{imp}}^{(n)}$ is proportional to
$\lambda_{\beta}^{n}$.
Below we will calculate $\delta{M}_{\mathrm{imp}}^{(1)}$  and
$\delta{M}_{\mathrm{imp}}^{(2)}$.

\subsubsection{Corrections linear with $\lambda$'s }
  \label{subsubsec-linear-subsec-HK-sec-magnet}

The correction $\delta{M}_{\mathrm{imp}}^{(1)}$ can be written
as,
\begin{eqnarray}
  \delta{M}_{\mathrm{imp}}^{(1)} &=&
  \sum_{\beta}
  \Big\{
      \delta{M}_{\mathrm{f};\beta}+
      \delta{M}_{\mathrm{i};\beta}
  \Big\}.
  \label{magnet-1st-fd+fq+fo+id+iq+io}
\end{eqnarray}
Here
\begin{eqnarray}
  \delta{M}_{{\mathrm{f}};\beta}
  &=&
  -g
  \mu_B
  \int\limits_{0}^{\beta}
  \Big\langle
      \hat{F}^{z}
      H_{\beta}(\tau)
  \Big\rangle_{0}~
  d\tau,
  \label{magnet-f-beta-def}
  \\
  \delta{M}_{{\mathrm{i}};\beta}
  &=&
  -g_{\mathrm{Yb}}
  \mu_{\mathrm{n}}
  \int\limits_{0}^{\beta}
  \Big\langle
      \hat{I}^{z}
      H_{\beta}(\tau)
  \Big\rangle_{0}~
  d\tau.
  \label{magnet-i-beta-def}
\end{eqnarray}
The expectation values
$\langle{F}^{\vec{\alpha}_{\beta}}\rangle$,
$\langle{I}^{\vec{\alpha}_{\beta}}\rangle$,
$\langle{F}^{z}{F}^{\vec{\alpha}_{\beta}}\rangle$ and
$\langle{I}^{z}{I}^{\vec{\alpha}_{\beta}}\rangle$
[where $\hat{F}^{\vec{\alpha}_{\beta}}$ or
$\hat{I}^{\vec{\alpha}_{\beta}}$ are dipole ($\beta={\mathrm{d}}$),
quadrupole ($\beta={\mathrm{q}}$) and octupole ($\beta={\mathrm{o}}$)
tensors for a localized impurity or itinerant atoms,
$\alpha$'s are the Cartesian indices]
are calculated in Appendix \ref{append-average-D-Q-O}.
Then $\delta{M}_{\mathrm{imp}}^{(1)}$ takes the form,
\begin{eqnarray}
  \delta{M}_{\mathrm{imp}}^{(1)} &=&
  -\frac{175B}{4T}~
  g
  \mu_B~
  g_{\mathrm{Yb}}
  \mu_{\mathrm{n}}~
  \Lambda_{\mathrm{d}}.
  \label{delta-M-linear-res}
\end{eqnarray}
Note that the factor $\frac{175}{4}$ comes from,
$$
  \frac{2}{9}~
  F(F+1)~
  I(I+1)
  (2I+1)
  ~=~
  \frac{175}{4}.
$$
If instead of itinerant atoms with spin $I=\frac{5}{2}$, we use
atoms with spin $s=\frac{1}{2}$, the last expression turns out to be,
$$
  \frac{2}{9}~
  F(F+1)~
  s(s+1)
  (2s+1)
  ~=~
  \frac{F(F+1)}{3},
$$
which agrees with eq. (3.2) from Ref. \cite{Hewson-book}.
\subsubsection{Corrections quadratic with $\lambda$'s }
  \label{subsubsec-quad-subsec-HK-sec-magnet}

Calculating the second order correction,
$\delta{M}_{\mathrm{imp}}^{(2)}$, we get
$\delta{M}_{\mathrm{imp}}$ up to $\lambda^2$,
\begin{eqnarray}
  \delta{M}_{\mathrm{imp}} &=&
  -M_{\mathrm{imp}}^{(0)}~
  N(I)~
  \frac{g \mu_B}
       {g_{\mathrm{Yb}} \mu_{\mathrm{n}}}
  \times \nonumber \\ &\times&
  \Bigg\{
       \Lambda_{\mathrm{d}}-
       {\mathfrak{F}}_{\mathrm{d}}^{(2)}~
       \ln
       \bigg(
            \frac{D}{T}
       \bigg)
  \Bigg\},
  \label{delta-M-second-res}
\end{eqnarray}
where $M_{\mathrm{imp}}^{(0)}$ is given by
eq. (\ref{magnet-imp-isolat-linear}),
$N(I)$ is given by eq. (\ref{Ns}), and
\begin{eqnarray}
  &&
  {\mathcal{N}}_{I} ~=~
  \frac{2}{3}~
  I
  \big(
      I+1
  \big)
  \big(
      2I+1
  \big),
  \label{NI-def}
  \\
  &&
  {\mathfrak{F}}_{\mathrm{d}}^{(2)} =
  -\Lambda_{\mathrm{d}}^{2}-
  \frac{9216}{25}~
  \Lambda_{\mathrm{q}}^{2}-
  \frac{1469664}{25}~
  \Lambda_{\mathrm{o}}^{2}.
  \label{Fd-2nd-def}
\end{eqnarray}
${\mathfrak{F}}_{\mathrm{d}}^{(2)}$ is obtained
from eq. (\ref{eq1-scaling-3rd}) neglecting the terms of
order $\lambda^3$.
The condition imposing the invariance of the magnetization
under the poor manâ scaling transformation is
\begin{eqnarray}
  \frac{\partial}{\partial \ln D}~
  \Bigg\{
       \Lambda_{\mathrm{d}}-
       {\mathfrak{F}}_{\mathrm{d}}^{(2)}
       \ln
       \bigg(
            \frac{D}{T}
       \bigg)
  \Bigg\}
  &=& 0.
  \label{dMimp=0}
\end{eqnarray}
Within the accuracy of this equation, when differentiating
the second term, we should neglect any implicit dependence
on $D$ through the couplings $\Lambda_{\beta}$.
The renormalization procedure should proceed until the bandwidth
$D$ is reduced to the temperature $T$. At this point, the second
order of the perturbation theory vanishes and the magnetization
takes the form,
\begin{eqnarray}
  M_{\mathrm{imp}} &=&
  M_{\mathrm{imp}}^{(0)}~
  \bigg\{
       1-
       \frac{g_{\mathrm{Yb}}
             \mu_{\mathrm{n}}}
            {g \mu_B}~
       N(I)~
       \Lambda_{\mathrm{d}}(T)
  \bigg\},
  \label{M-imp-scaling}
\end{eqnarray}
where $N(I)$ is given by eq. (\ref{Ns}),
$\Lambda_{\mathrm{d}}(T)$ is the solution of the second order
scaling equation (\ref{subeqs-scaling-2nd}).

It is useful to write the magnetic susceptibility
$\chi_{\mathrm{imp}}=\partial{M}_{\mathrm{imp}}/\partial{B}$ as,
\begin{eqnarray}
  \chi_{\mathrm{imp}}(T) &=&
  \chi_{\mathrm{imp}}^{(0)}+
  \delta \chi_{\mathrm{imp}}(T),
  \label{chi=chi0+dchi}
\end{eqnarray}
where $\chi_{\mathrm{imp}}^{(0)}$ is the susceptibility of the isolated
impurity atom, and $\delta\chi_{\mathrm{imp}}(T)$ is correction to
the susceptibility due to the Kondo interaction. Explicitly,
\begin{eqnarray}
  &&
  \chi_{\mathrm{imp}}^{\mathrm{(0)}} ~=~
  \frac{\chi_{0}}{3}~
  \frac{T_K}{T}~
  F(F+1),
  \label{chi-isolated-def}
  \\
  &&
  \delta \chi_{\mathrm{imp}} =
  X_{\mathrm{imp}}~
  N(I)~
  \Lambda_{\mathrm{d}},
  \label{d-chi-def}
\end{eqnarray}
where
\begin{eqnarray*}
  \chi_0 =
  \frac{\big(g \mu_B\big)^{2}}{T_K},
  \ \ \ \ \
  X_{\mathrm{imp}} =
  -\frac{g_{\mathrm{Yb}} \mu_{\mathrm{n}}}
       {g \mu_B}~
  \chi_{\mathrm{imp}}^{(0)}.
\end{eqnarray*}

\begin{figure}[htb]
\centering
  \includegraphics[width=60 mm,angle=0]
   {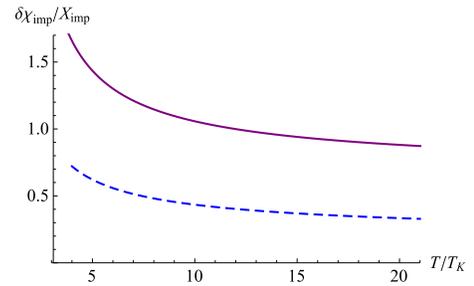}
 \caption{\footnotesize
   ({\color{blue}Color online})
   Solid curve: The ratio
   $\delta\chi_{\mathrm{imp}}/X_{\mathrm{imp}}$ (\ref{d-chi-def})
   as a function of $T$ for $\lambda
       {a_{\g}^{2}}\rho_0=0.5$. The values 
   of the parameters are
   $\Lambda_{\mathrm{d}}^{(0)}=0.0247619$,
   $\Lambda_{\mathrm{q}}^{(0)}=-0.0005952$,
   $\Lambda_{\mathrm{o}}^{(0)}=-0.0002646$.
   Dashed curve: The ratio
   $\delta\chi_{s-d}/X_{\mathrm{imp}}$
   (\ref{suscept-sd-model}) for the $^{171}$Yb(\1S0) --
   $^{171}$Yb(\3P2) system.}

 \label{Fig-suscept-2nd-num}
\end{figure}

The ratio
$\delta\chi_{\mathrm{imp}}/X_{\mathrm{imp}}$ as a function of
temperature is shown in Fig. \ref{Fig-suscept-2nd-num}, solid
curve. It should be noted that
$X_{\mathrm{imp}}/\chi_{\mathrm{imp}}^{\mathrm{(0)}}\ll1$
since the ratio,
$$
  \frac{g_{\mathrm{Yb}}~\mu_{\mathrm{n}}}
       {g~\mu_B}
  ~=~ -9.411\cdot10^{-5},
$$
is small. When $T$ approaches the Kondo temperature,
$\delta\chi_{\mathrm{imp}}$ diverges as $1/\ln(T/T_K)$
which indicates breaking down of the underlying perturbation
theory.

For comparison, consider a system consisting of $^{171}$Yb(\1S0)
itinerant atoms and a $^{171}$Yb(\3P2) atom as localized impurity.
The atomic orbital angular momentum is assumed to be $F=\frac{3}{2}$,
therefore the susceptibility of the isolated impurity is given by
eq. (\ref{chi-isolated-def}). Magnetic susceptibility of
the $^{171}$Yb(\3P2) atom interacting with the $^{171}$Yb(\1S0)
atoms [atomic orbital moment is $I=\frac{1}{2}$] is
\cite{Hewson-book},
\begin{eqnarray}
  &&
  \chi_{s-d} =
  \chi_{\mathrm{imp}}^{(0)}+
  \delta\chi_{s-d},
  \nonumber
  \\
  &&
  \delta\chi_{s-d} =
  X_{\mathrm{imp}}~
  \Lambda_{s-d}(T),
  \label{suscept-sd-model}
\end{eqnarray}
where $\Lambda_{s-d}(T)$ is given by eq.
(\ref{Lambda-sd-vs-T}). The ratio
$\delta\chi_{s-d}/X_{\mathrm{imp}}$, eq.
(\ref{suscept-sd-model}), is shown in Fig.
\ref{Fig-suscept-2nd-num}, dashed curve.
\subsection{Experimental Feasibility} 
Having developed the theoretical framework for calculating
 entropy, specific heat and magnetization, a few words on the
 experimental feasibility of measuring thermodynamic 
 quantities (specific for the pertinent system), are in order.  It is
worth mentioning that some of these thermodynamic observables
have been successfully measured in one-component Bose
gases~\cite{navon11}, two-component Fermi
gases~\cite{Nascimbne10}, and SU($N$) fermions trapped in optical
lattices~\cite{Hofrichter16} either using \textit{in-situ} local
probe of the inhomogeneous atomic density~\cite{Qi09,Nascimbne10}
or performing a spin transport measurements~\cite{Valtolina16}.
In particular, the magnetic susceptibility of the two-component
Fermi gas is determined in various ways, including
(1) the relative spin fluctuation measurement~\cite{sanner11},
(2) the sum-rule approach for the spin-dipole mode
frequency~\cite{Valtolina16} or (3) the direct measurement of
susceptibility from the inhomogeneous density profile of
the spin-imbalanced atomic gas~\cite{lee13}. Similar measurements
should be feasible in a  $^1$S$_0$-$^3$P$_2$ ytterbium mixture.
Indeed, the spin-dependent trapping potential available in
the ytterbium mixture allows one to induce spin-selective
transport and consequently monitor the spin-dipole mode of
the system. In a similar manner, the heat capacity of
the two-component gas can be determined from the local density of
the atomic gas~\cite{Ku13}.

\section{Conclusion}
  \label{sec-conclusion}

Let us then briefly summarize our results. Our main arena
concerns the Kondo physics in an ultracold Fermi gas of
$^{173}$Yb($^{1}$S$_0$) atoms (in their electronic ground-state)
in which a few  $^{173}$Yb($^{3}$P$_2$) atoms (in a long lived
excited state) are trapped in a specially designed optical
potential. The main objectives are:
1) To explore the feasibility of experimental realization;
2) To calculate the exchange interaction between the itinerant
$^{173}$Yb($^{1}$S$_0$) and $^{173}$Yb($^{3}$P$_2$) atoms and to
verify that it is an antiferromagnetic exchange;
3) To construct the Kondo Hamiltonian and to identify its
underlying symmetry;
4) To carry out the corresponding poor-man scaling, to identify
the stable fixed points and to determine whether some of them
display non-Fermi liquid behaviour;
5) To calculate some experimentally accessible observable in such
a system.

As far as objective 1) is concerned, we have considered
a mixture of $^{1}$S$_0$ and $^{3}$P$_2$ ytterbium fermions
that can be readily prepared in contemporary experiments,
in which a state-dependent optical potential employes a strong
$^3$P$_2$-$^3$S$_1$ transition and tightly confines $^{3}$P$_2$
atoms while leaves the ground-state $^{1}$S$_0$ atoms itinerant.
By properly choosing the wavelength of the optical potential,
we have shown that the spontaneous light scattering can be
sufficiently reduced to observe a many-body effect.
The localized and itinerant atoms can be independently
detected with the combination of an optical pumping and a blast.
Finally a $^{1}$S$_0$-$^{3}$P$_2$ mixture of ytterbium atoms
displays a magnetic Feshbach resonance by which the interaction
strength between localized and itinerant atoms  can be further
controlled \cite{Feshbach-prl-13}.
Such novel features may open a new route to
investigate the Kondo effect with tuneable atom-atom interactions
in this system.
Calculating the exchange interaction proceeds along similar lines
as in our previous paper\cite{IK-TK-YA-GBJ-PRB-16}.

The main difficulty is encountered in achieving goals (3) and (4).
It is required to write down the Kondo Hamiltonian in terms of
multipole expansion, since otherwise, the RG procedure is
inapplicable. This requires a technically tedious procedure
related to the pertinent spin algebra. Moreover, identifying
the corresponding fixed points requires calculations of RG
diagrams to third order in the exchange constant, which turn out
to be rather involved. Details of the calculations are explained in
the Appendices.

Having overcame these technical difficulties,
we have found seven fixed points for $\lambda_{\mathrm{d}}$
and $\lambda_{\mathrm{q}}$ and $\lambda_{\mathrm{o}}$.
Three of them, $P_4$, $P_5$ and $P_7$ [eqs. (\ref{fixed-point-4}),
(\ref{fixed-point-5}) and (\ref{fixed-point-7})]
are stable, and the other fixed points are unstable. The fixed
points found here are distinct from the NB non Fermi
liquid fixed point described in our previous paper
\cite{KKAK-prb-15}, in which we studied the Kondo
physics in a mixture of $^{23}$Na and $^{6}$Li atoms.
In the present work, the NB non Fermi liquid fixed point
corresponds to $P_1$ in the list (\ref{fixed-point-4}), that is
found to be {\it unstable}.

The remaining task, that is, elucidating the Kondo physics
{\it in the strong coupling regime} for the new stable fixed
points $P_4$, $P_5$ and $P_7$ (identified in this work) is beyond
the scope of our present study. It is perceived that the standard
techniques that are applied to the dipolar Kondo effect such as
Bethe Ansatz and conformal field theory might work also in this
case albeit with non-trivial modifications.

\subsection*{Acknowledgement}
 Y.A, I.K and T.K acknowledge many years of
 collaboration and  discussions pertaining to the Kondo Physics with
their colleague and their close  friend \\
$~~~~~~~~~~~~~ $ \underline {\bf Konstantin Abramovich Kikoin.} \\
His sudden death left us shocked and wordless.\\
\ \\
The authors thank S. Zhang for useful discussion.
G.B.J acknowledges financial support from the Hong Kong Research
Grants Council (Project No. 26300014/16300215/16311516) and from
the Croucher Foundation. The research of Y.A is partially supported by
grant 400/12 of the Israel Science Foundation.

\appendix

\section{Trapping of the Yb Atoms by the Optical Potential}
  \label{append-trap}

Since the Yb(\3P2) atom has the electronic
orbital moment $J=2$, the polarizability
$\hat\alpha_{\e}(\omega)$ is a $5\times5$ matrix.
We introduce the matrices $\hat\alpha_{\e}^{x}(\omega)$,
$\hat\alpha_{\e}^{y}(\omega)$ and
$\hat\alpha_{\e}^{z}(\omega)$, for the electric field collinear
to the axes $x$, $y$ and $z$, respectively.
Explicitly, they are
\begin{eqnarray}
  &&
  \hat\alpha_{\mathrm{e}}^{z} ~=~
  {\mathrm{diag}}
  \Big(
      \alpha_2,~
      \alpha_1,~
      \alpha_0,~
      \alpha_1,~
      \alpha_2
  \Big),
  \nonumber
  \\
  &&
  \hat\alpha_{\mathrm{e}}^{\beta} ~=~
  \hat{\mathcal{U}}_{\beta}~
  \hat\alpha_{\mathrm{e}}^{z}~
  \hat{\mathcal{U}}_{\beta}^{\dag},
  \label{alpha-xyz-def}
\end{eqnarray}
where $\beta=x,y$ is a Cartesian index, the spin-rotation
matrices $\hat{\mathcal{U}}_{\beta}$ are
\begin{eqnarray*}
  \hat{\mathcal{U}}_{x} ~=~
  e^{i \pi \hat{J}^{x}/2},
  \ \ \ \ \
  \hat{\mathcal{U}}_{y} ~=~
  e^{-i \pi \hat{J}^{y}/2}.
\end{eqnarray*}

We consider the optical potential generated by standing
electromagnetic wave in the directions of the axes $x$,$y$
and $z$ with the double-magic wavelength $\lambda_0$
\cite{Khramov-PRL-14},
\begin{eqnarray}
  \lambda_0 ~=~
  546~{\text{nm}}.
  \label{lambda-0}
\end{eqnarray}
The polarizability $\hat\alpha_{F}^{\beta}(\lambda_0)=%
\alpha_{\e}(\lambda_0)\hat{J}^{0}$
[the Cartesian index $\beta$ indicates the direction of
the electric field] is proportional to the identity matrix
$\hat{J}^{0}$. Explicitly, $\alpha_{\g}(\lambda_0)$
and $\alpha_{\e}(\lambda_0)$, the polarizability of
the Yb(\1S0) and Yb(\3P2) atoms are
\cite{Khramov-PRL-14}
\begin{eqnarray}
  \alpha_{\e}(\lambda_0)
  ~=~
  250~{\mathrm{a.u.}},
  \ \ \
  \alpha_{\g}(\lambda_0)
  ~=~
  200~{\mathrm{a.u.}}
  \label{polar-high}
\end{eqnarray}

\subsection{Optical Potential}
  \label{subsec-potential}

\begin{figure}[htb]
\centering
  \includegraphics[width=60 mm,angle=0]
   {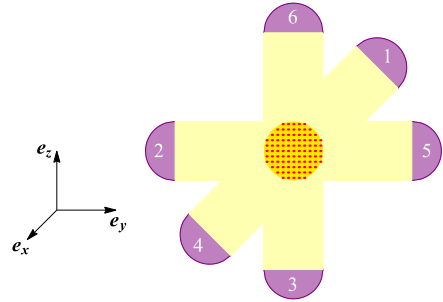}
 \caption{\footnotesize
   {\color{blue}(Color online)}
   Three pairs of lasers labelled by the numbers
   1 -- 6 generating trapping optical potential
   [golden yellow disk with red dots].}
 \label{Fig-lasers}
\end{figure}

We consider possibility of formation of the short-
and long-wavelength potentials the light of the double-magic
wavelength $\lambda_0$. The optical potential is
generated by three pairs of lasers, as illustrated in
Fig. \ref{Fig-lasers}. The light of the first, second or
third pair of lasers propagates parallel and antiparallel
to the axes $x$, $y$ or $z$. The optical potential is,
\begin{eqnarray}
  V_{\nu}(\mbfr) &=&
  -\alpha_{\nu}(\omega)~
  \lim_{{\mathcal{T}}\to\infty}
  \frac{1}{{\mathcal{T}}}
  \int\limits_{0}^{{\mathcal{T}}}
  \big|
      \mbfE(\mbfr,t)
  \big|^{2}
  dt,
  \label{V-def}
\end{eqnarray}
where $\nu=\g$ or $\e$ for the Yb(\1S0) or Yb(\3P2) atoms.
The electric field $\mbfE(\mbfr)$ is,
\begin{eqnarray}
  \mbfE(\mbfr,t) &=&
  \sum_{\beta=1}^{6}
  \mbfE_{\beta}(\mbfr,t).
  \label{E=E1+E2+E3}
\end{eqnarray}
Here
\begin{eqnarray}
  \mbfE_{\beta}(\mbfr,t) &=&
  \mbfE_{\beta}^{(0)}
  \cos\big(\mbfk_{\beta} \mbfr-\omega_0 t\big)
  \times \nonumber \\ && \times
  \exp
  \bigg(
       -\frac{x_{a_{\beta}}^{2}-
              x_{b_{\beta}}^{2}}
             {2 L^2}
  \bigg),
  \label{Ej-def}
\end{eqnarray}
where $\mbfE_{\beta}^{(0)}=E_u\mbfe_{a_{\beta}}$,
$\mbfE_{\beta+3}^{(0)}=E_v\mbfe_{a_{\beta}}$,
$\beta=1,2,3$. 
The indices $a_{\beta}=a_{\beta+3}$ and
$b_{\beta}=b_{\beta+3}$ are
\begin{eqnarray}
  &&
  a_1 = 2,
  \ \ \
  a_2 = 3,
  \ \ \
  a_3 = 1,
  \nonumber
  \\
  &&
  b_1 = 3,
  \ \ \
  b_2 = 1,
  \ \ \
  b_3 = 2.
  \label{j-aj-bj-def}
\end{eqnarray}
$\mbfe_1$, $\mbfe_2$ and $\mbfe_3$ are unit vectors
parallel to the axes $x$, $y$ and $z$.
The amplitudes $E_u$ and $E_v$ are real and satisfy
the inequalities
$$
  E_u ~\gg~
  E_v ~>~
  0.
$$
The wave vectors of the light are
$\mbfk_{\beta}=-\mbfk_{\beta+3}=k_0\mbfe_{\beta}$,
where $\beta=1,2,3$,
$k_0=2\pi_0/\lambda_0$ is the wavenumber of
the light. $\omega_0=k_0c$ is the frequency of the light.
The waist radius $\sqrt{2}L$ satisfies the inequality
$$
  k_0 L ~\gg~ 1.
$$

Explicitly, the optical potential is
\begin{eqnarray}
  V_{\nu}(\mbfr) &=&
  V_{\nu}^{\mathrm{(slow)}}(\mbfr)+
  V_{\nu}^{\mathrm{(fast)}}(\mbfr).
  \label{V=Vslow+Vfast}
\end{eqnarray}
Here
\begin{eqnarray}
  V_{\nu}^{\mathrm{(slow)}}(\mbfr) &=&
  \sum_{\beta=1}^{3}
  V_{\beta,\nu}^{\mathrm{(slow)}}(\mbfr),
  \label{V-slow}
  \\
  V_{\nu}^{\mathrm{(fast)}}(\mbfr) &=&
  \sum_{\beta=1}^{3}
  V_{\beta,\nu}^{\mathrm{(fast)}}(\mbfr),
  \label{V-fast}
\end{eqnarray}
where
\begin{eqnarray}
  V_{\beta,\nu}^{\mathrm{(slow)}}(\mbfr)
  &=&
  -V_{0,\nu}~
  \exp
  \bigg(
       -\frac{x_{a_{\beta}}^{2}+
              x_{b_{\beta}}^{2}}
             {L^2}
  \bigg),
  \label{V-j-1D-slow}
  \\
  V_{\beta,\nu}^{\mathrm{(fast)}}(\mbfr)
  &=&
  -V_{1,\nu}~
  \cos\big(2k_0 x_{\beta}\big)
  \times \nonumber \\ && \times
  \exp
  \bigg(
       -\frac{x_{a_{\beta}}^{2}+
              x_{b_{\beta}}^{2}}
             {L^2}
  \bigg).
  \label{V-j-1D-fast}
\end{eqnarray}
The strengths $V_{0,\nu}$ and $V_{1,\nu}$ are 
\begin{eqnarray}
  V_{0,\nu} &=&
  \frac{\alpha_{\nu}(\omega_0)}{2}~
  \Big\{
      E_{u}^{2}+
      E_{v}^{2}
  \Big\},
  \nonumber
  \\
  V_{1,\nu} &=&
  \alpha_{\nu}(\omega_0)~
  E_u
  E_v.
  \label{V0-V1-def-append}
\end{eqnarray}
We assume that $E_u$ and $E_v$ are real and positive and
$\alpha_{\nu}(\lambda_0)>0$, and
therefore $V_{0,\nu}>0$ and $V_{1,\nu}>0$.
Here $\beta$ is a Cartesian index,
the indices $a_{\beta}$ and $b_{\beta}$ are given by
eq. (\ref{j-aj-bj-def}).

The optical potential (\ref{V=Vslow+Vfast}) is
illustrated in Fig. \ref{Fig-V-Gauss-cos} for
$L=10\lambda_0$ and $E_v=0.05E_u$.
Here the solid blue and green
curves show $V_{\nu}(x,0,0)$ and
$V_{\nu}(x,\lambda_0/4,\lambda_0/4)$.
For comparison, the dashed red curve
illustrates $V_{\nu}^{\mathrm{(slow)}}(x,0,0)$,
eq. (\ref{V-slow}). Note that we take here
$L=10\lambda_0$ just for better illustration.
Real values of $L$ are larger than
$10^2\lambda_0$.

In the following discussions,
we assume that the Yb(\3P2) atoms are trapped
by the fast oscillating potential (\ref{V-fast}) and
are localized near the stable equilibrium points
$\mbfr=(n_1\lambda_0/2,n_2\lambda_0/2,n_3\lambda_0/2)$,
where $n_1$, $n_2$ and $n_3$ are integers.
From the other side, we assume that the density
of the Yb(\1S0) atoms is such that the Fermi energy
$\epsilon_F$ [measured from the bottom of
the potential well] satisfies the inequality
$\epsilon_F\gg{V}_{1,\g}$, and therefore the atoms
with energy close to $\epsilon_F$ can be considered
as itinerant: their motion is restricted by the potential
$V_{\g}^{\mathrm{(slow)}}(\mbfr)$, eq. (\ref{V-slow}).

\subsection{Wave Function and Energy of the Trapped
  ${\text{Yb}}$(\3P2) Atoms}
  \label{subsec-disp-3P2}

The density of the Yb(\3P2) atoms are low, so that
all the atoms are localized by the fast oscillating
potential (\ref{V-fast}). When the energy level
of the atom is deep enough, we can derive
the wave function and the energy level in
harmonic approximation. Consider, for example,
the atom trapped near the stable equilibrium
point $\mbfr=(0,0,0)$. When the radius of localization
of the atom is small with respect to $\lambda_0/4$,
the optical potential (\ref{V-fast}) can be approximated
as,
\begin{eqnarray}
  V_{\e}^{\mathrm{(fast)}}(\mbfr)
  &\approx&
  -V_{1,\e}+
  2 V_{1,\e}~
  \big(k_0 r\big)^{2},
  \label{V-e-fast-harmonic}
\end{eqnarray}
whereas $V_{\e}^{\mathrm{(slow)}}(\mbfr)$ is almost
constant for $r<\lambda_0/4$, where $r=|\mbfr|$.
The wave function of the atom trapped by the harmonic
potential (\ref{V-e-fast-harmonic}) is given by eq. (\ref{WF-trapped-3P2}).

The energy $\veps_{\mathrm{imp}}$ measured
from the bottom of the well is,
\begin{eqnarray}
  \veps_{\mathrm{imp}} &=&
  \frac{3}{2}~
  \hbar \Omega_{\e}.
  \label{veps-imp-append}
\end{eqnarray}

\subsection{Wave Functions and Energy Levels of the Trapped
  ${\text{Yb}}$(\1S0) Atoms}
  \label{sec-disp-1S0}

When the energy $\epsilon$ of the trapped Yb(\1S0) atom
[measured from the bottom of the potential well]
satisfies the inequality $\epsilon\gg{V}_{1,\g}$,
we can approximate the potential (\ref{V=Vslow+Vfast})
as
\begin{eqnarray}
  V_{\g}(\mbfr) &\approx&
  V_{\g}^{\mathrm{(slow)}}(\mbfr).
  \label{V-g-approx-slow}
\end{eqnarray}
Moreover, when the energy level is deep enough,
we can approximate $V_{\g}^{\mathrm{(slow)}}(\mbfr)$
as,
\begin{eqnarray}
  V_{\g}^{\mathrm{(slow)}}(\mbfr) &\approx&
  -3 V_{0,\g}+
  2 V_{0,\g}~
  \frac{r^2}{L^2}.
  \label{V-g-slow-harmonic}
\end{eqnarray}
Quantum states of atoms in isotropic potential
are described by the radial quantum number
$n$ [$n=0,1,2,\ldots$],
the angular momentum $l$ [$l=0,1,2,\ldots$] and
projection $m$ of the angular moment on the axis
$z$ [$m=-l,-l+1,\ldots,l$]. Due to the centrifugal
barrier, only the atoms with $l=0$ can approach
the impurity and be involved in the exchange
interaction with it. The wave functions of the atoms
with $l=0$ trapped by the harmonic potential
(\ref{V-g-slow-harmonic}) are,
\begin{eqnarray}
  \Psi_{n}(\mbfr) &=&
  \frac{{\mathcal{N}}_{n}}{\sqrt{4 \pi}}~
  L_{n}^{(\frac{1}{2})}
  \bigg(\frac{r^2}{a_{\g}^{2}}\bigg)~
  \exp
  \bigg(
       -\frac{r^2}{2a_{\g}^{2}}
  \bigg),
  \label{WF-trapped-1S0}
\end{eqnarray}
where $L_{n}^{(\frac{1}{2})}(\varrho)$ are generalized
Laguerre polynomials. The normalization factor is
\begin{eqnarray*}
  {\mathcal{N}}_{n} &=&
  \bigg(
       \frac{2}{\pi a_{\g}^{6}}
  \bigg)^{1/4}~
  \sqrt{\frac{2^{n+2}~n!}{(2n+1)!!}}.
\end{eqnarray*}
The harmonic length $a_{\g}$ and frequency
$\Omega_{\g}$ are defined as
\begin{eqnarray}
  \frac{a_{\g}}{L} ~=~
  \bigg(
       \frac{{\mathcal{E}}_{L}}{2 V_{0,\g}}
  \bigg)^{1/4},
  \ \ \ \ \
  \hbar \Omega_{\g} ~=~
  2
  \sqrt{2 {\mathcal{E}}_{L} V_{0,\g}},
  \label{a-omega-g-def-append}
\end{eqnarray}
where ${\mathcal{E}}_{L}$ is defined as,
\begin{eqnarray}
  {\mathcal{E}}_{L} &=&
  \frac{\hbar^2}{2 M L^2}.
  \label{EL-def-append}
\end{eqnarray}

The energy levels of the states with $l=0$ are,
\begin{eqnarray}
  \veps_{n} &=&
  \hbar \Omega_{\g}~
  \bigg(
       2 n+\frac{3}{2}
  \bigg).
  \label{veps-itinerant-append}
\end{eqnarray}

In what following we assume that
\begin{eqnarray}
  \Omega_{\e}
  ~\gg~
  \Omega_{\g}.
  \label{inequality-append}
\end{eqnarray}
Within this framework, the spectrum is nearly continuous
and the ytterbium atoms in the ground-state form a
Fermi gas. The Fermi energy $\epsilon_F$ is such that
$\epsilon_F\gg\hbar\Omega_{\g}$, hence the Fermi gas
is 3D.

\section{Multipole Operators}
  \label{append-multipole}
In subsections \ref{subsec-notations} and \ref{subsec-momentum}
we introduced multipole operators:
A $2^n$ pole operator is an expression involving $n$ spin
operators, with appropriate coefficients. These operators,
explicitly calculated in this section, are the building blocks of
the exchange interaction of the multipolar Kondo Hamiltonian to be
introduced in the next section.
\subsection{Notations}
  \label{subsec-notations}
In this subsection we introduce the definitions and expressions
for the $2^n$ poles required for the representation of the Kondo
Hamiltonian in terms of multipole expansion. These $2^n$ poles
result from the spin content of the underlying atomic system.

We consider exchange interaction of itinerant atoms (which are
$^{173}$Yb atoms in the ground $^{1}$S$_{0}$ state with atomic
spin $I=\frac{5}{2}$),  and localized impurities (which are
the same $^{173}$Yb atoms in the long lived excited $^{3}$P$_{2}$
state with atomic spin $F=\frac{3}{2}$). Note that $I$ is
contributed solely from the nuclear spin while $F$ is the sum of
electronic and nuclear spins. An atom with total angular momentum
$F=\frac{3}{2}$ has nontrivial dipole, quadrupole and octupole
magnetic momenta. They are denoted here as
$$
  \hat{F}^{\alpha},
  \ \ \ \ \
  \hat{F}^{\alpha,\alpha'},
  \ \ \ \ \
  \hat{F}^{\alpha,\alpha',\alpha''},
$$
where $\alpha$, $\alpha'$ and $\alpha''$ are Cartesian indices.
An atom with total angular momentum $I=\frac{5}{2}$ has nontrivial
dipole, quadrupole, octupole, 16-pole and 32-pole magnetic momenta,
denoted here as
\begin{eqnarray*}
  &&
  \hat{I}^{\alpha_1},
  \ \ \ \ \ \
  \hat{I}^{\alpha_1,\alpha_2},
  \ \ \ \ \ \
  \hat{I}^{\alpha_1,\alpha_2,\alpha_3},
  \\
  &&
  \hat{I}^{\alpha_1,\alpha_2,\alpha_3,\alpha_4},
  \ \ \ \
  \hat{I}^{\alpha_1,\alpha_2,\alpha_3,\alpha_4,\alpha_5}.
\end{eqnarray*}
When an expression applies for both itinerant atoms and
impurities, we use the notations $\hat{S}^{\alpha}$,
$\hat{S}^{\alpha,\alpha''}$, $\hat{S}^{\alpha,\alpha',\alpha''}$
for the dipole, quadrupole and octupole angular momenta. Here $\hat{S}$
denotes the operators $\hat{F}$ or $\hat{I}$.


\subsection{Explicit expressions for $2^n$-Pole Momenta}
  \label{subsec-momentum}

The magnetic dipole operator is collinear with the vector of
its spin (more precisely its total angular momentum) operator.
When a particle has spin $S$, the vector $\mbfS$ of the spin
matrices (generators of the $2S+1$-dimensional representation
of the SU(2) group) are,
\begin{eqnarray}
  S^{z}_{s,s'} &=&
  s~
  \delta_{s,s'},\nonumber
  \\
  S^{+}_{s,s'} &=&
  {\mathcal{L}}(S,s)~
  \delta_{s,s'+1}, \nonumber
  \\
  S^{-}_{s,s'} &=&
  {\mathcal{L}}(S,s')~
  \delta_{s',s+1},
  \label{Fdipole-def}
   \label{subeqs-dipole}
\end{eqnarray}
where $s,s'$ are magnetic quantum numbers such that $|s|\leq{S}$,
and
\begin{eqnarray*}
  {\mathcal{L}}(S,s) &=&
  \sqrt{(S+s)(S-s+1)}.
\end{eqnarray*}
Next, the quadrupole moment operators are represented by symmetric
traceless matrices $S^{\alpha,\alpha'}$ ($\alpha,\alpha'=x,y,z$
are Cartesian indices) defined as,
\begin{eqnarray}
  \hat{S}^{\alpha,\alpha'} &=&
  \Big\{
      \hat{S}^{\alpha},~
      \hat{S}^{\alpha'}
  \Big\}-
  \frac{2}{3}~
  \hat{\mbfS}^2~
  \delta^{\alpha,\alpha'},
  \label{quadrupole-def}
\end{eqnarray}
where
$$
  \Big\{
      \hat{S}^{\alpha},~
      \hat{S}^{\alpha'}
  \Big\}
  ~=~
  \hat{S}^{\alpha}
  \hat{S}^{\alpha'}+
 \hat{S}^{\alpha'}
  \hat{S}^{\alpha}.
$$
The quadrupole operators satisfy the following equalities,
\begin{eqnarray}
 \hat{S}^{\alpha,\alpha'}
  &=&
  \hat{S}^{\alpha',\alpha},
  \ \ \ \ \
  \sum_{\alpha}
  \hat{S}^{\alpha,\alpha}
  = 0.
  \label{trace-quadrupole}
\end{eqnarray}
Continuing this analysis, the octupole moment operators are
represented by matrices,
\begin{eqnarray}
  \hat{S}^{\alpha,\alpha',\alpha''}
  &=&
  \Big\{
      \hat{S}^{\alpha},~
      \hat{S}^{\alpha'},~
      \hat{S}^{\alpha''}
  \Big\}-
  \frac{1}{5}~
  \Big(
      3~
      \hat\mbfS^2-
      1
  \Big)
  \times \nonumber \\ &\times&
  \sum_{\alpha_1,\alpha'_1,\alpha''_1}
  P^{\alpha,\alpha',\alpha''}_{\alpha_1,\alpha'_1,\alpha''_1}
  \delta^{\alpha_1,\alpha'_1}~
  \hat{S}^{\alpha''_1}.
  \label{octupole-def}
\end{eqnarray}
Here the symbol $P^{\alpha,\alpha',\alpha''}%
_{\alpha_1,\alpha'_1,\alpha''_1}$
denotes permutation of the indices
$\alpha$, $\alpha'$, $\alpha''$,
\begin{eqnarray}
  P^{\alpha,\alpha',\alpha''}_{\alpha_1,\alpha'_1,\alpha''_1}
  &=&
  \delta_{\alpha,\alpha_1}~
  P^{\alpha',\alpha''}_{\alpha'_1,\alpha''_1}+
  \delta_{\alpha,\alpha'_1}~
  P^{\alpha',\alpha''}_{\alpha''_1,\alpha_1}+
  \nonumber \\ && +
  \delta_{\alpha,\alpha''_1}~
  P^{\alpha',\alpha''}_{\alpha_1,\alpha'_1},
  \label{permutation-3-def}
\end{eqnarray}
the symbol $P^{\alpha,\alpha'}_{\alpha_1,\alpha'_1}$
denotes permutation of the indices $\alpha$, $\alpha'$,
\begin{eqnarray}
  P^{\alpha,\alpha'}_{\alpha_1,\alpha'_1}
  &=&
  \delta_{\alpha,\alpha_1}
  \delta_{\alpha',\alpha'_1}+
  \delta_{\alpha,\alpha'_1}
  \delta_{\alpha',\alpha_1}.
  \label{permutation-2-def}
\end{eqnarray}
The symbol
$\{\hat{S}^{\alpha},\hat{S}^{\alpha'},\hat{S}^{\alpha''}\}$
is fully symmetric product of $\hat{S}^{\alpha}$,
$\hat{S}^{\alpha'}$ and $\hat{S}^{\alpha''}$,
$$
  \big\{
      \hat{S}^{\alpha_1},~
      \hat{S}^{\alpha_2},~
      \hat{S}^{\alpha_3}
  \big\}
  ~=~
  \sum_{\{\alpha'\}_{3}}
  P^{\alpha_1,\alpha_2,\alpha_3}_{\alpha'_1,\alpha'_2,\alpha'_3}~
  \hat{S}^{\alpha'_1}
  \hat{S}^{\alpha'_2}
  \hat{S}^{\alpha'_3},
$$
where $\{\alpha'\}_{3}=\{\alpha'_1,\alpha'_2,\alpha'_3\}$.
The octupole operators are symmetric with all the indices,
$$
  \hat{S}^{\alpha,\alpha',\alpha''}
  ~=~
  \hat{S}^{\alpha',\alpha,\alpha''}
  ~=~
  \hat{S}^{\alpha,\alpha'',\alpha'}.
$$
Moreover, they are constructed in such a way that the trace over
any two indices vanishes that is,
\begin{eqnarray*}
  \sum_{\alpha'}
  \hat{S}^{\alpha,\alpha',\alpha'}
  ~=~ 0.
\end{eqnarray*}

\section{Exchange Interaction}
  \label{append-exchange}
\noindent
\underline{Appendices \ref{append-exchange},
\ref{append-derivation-2nd-PMS}, \ref{apppend-derivation-dE-2nd},
\ref{apppend-derivation-3rd-PMS} main points:}
The second order correction terms are defined in Eqs.~(\ref{dH2-def},
\ref{dH2-alpha-alpha}), while the third ordered correction terms
are defined in Eqs.~(\ref{dH3-beta-beta-beta}). The formidable
task of evaluating these terms is carried out in these subsections.

When an impurity atom is localized at the origin of coordinates and
an itinerant atom is placed at position $\mbfR$ so that they are separated
by $R=|\mbfR|$, there is an exchange interaction between them.
The interaction Hamiltonian is,
\begin{eqnarray}
  {\mathcal{H}}_{\mathrm{exch}}(R) &=&
  \sum_{f,f'}
  \sum_{i,i'}
  V_{f,f';i,i'}(R)
  \times \nonumber \\ && \times
  X^{f,f'}~
  \hat\psi_{i'}^{\dag}(\mbfR)
  \hat\psi_{i}(\mbfR),
  \label{H-exch-def}
\end{eqnarray}
where $X^{f,f'}=|{f}\rangle\langle{f}'|$ are Hubbard operators
of the localized impurity, $\hat\psi_{i}(\mbfR)$ and
$\hat\psi_{i}^{\dag}(\mbfR)$ are annihilation and creation
operators of itinerant atoms at position $\mbfR$ with
the nuclear magnetic quantum number $i$.
The rate $V_{f,f';i,i'}(R)$ is,
\begin{eqnarray}
  V_{f,f';i,i'}(R) =
  \frac{t_{\mathrm{s}}(R)~
        t_{\mathrm{p}}(R)}
       {3~\Delta\epsilon}
  \sum_{j}
  C_{J,j;I,i}^{F,f}
  C_{J,j;I,i'}^{F,f'}.
  \label{g(R)-def}
\end{eqnarray}
Here $t_{\mathrm{s}}(R)$ and $t_{\mathrm{p}}(R)$ are given by
eq. (\ref{tunneling-rate}),
$$
  \Delta\epsilon=
  \veps_{\mathrm{ion}}+
  \veps_{\mathrm{ea}}+
  \epsilon_{\mathrm{g}}-
  \epsilon_{\mathrm{x}}
  ~=~
  4.1104~{\text{eV}},
$$
where $\veps_{\mathrm{ion}}=6.2542$~{eV} is the ionization energy
\cite{e-ion-Yb-78},
$\veps_{\mathrm{ea}}=0.3$~{eV} is the electron affinity
\cite{e-affin-Yb-04} and
$\epsilon_{\mathrm{x}}-\epsilon_{\mathrm{g}}=2.4438$~{eV} is
the excitation energy of the $^{3}$P$_{2}$ state
\cite{e-ion-Yb-78}.

Substituting eq. (\ref{g(R)-def}) into eq. (\ref{H-exch-def}), we get
\begin{eqnarray}
  {\mathcal{H}}_{\mathrm{exch}}(R) &=&
  g(R)
  \sum_{j}
  \sum_{f,f'}
  \sum_{i,i'}
  C_{J,j;I,i}^{F,f}
  C_{J,j;I,i'}^{F,f'}
  \times \nonumber \\ && \times
  X^{f,f'}~
  \hat\psi_{i'}^{\dag}(\mbfR)
  \hat\psi_{i}(\mbfR),
  \label{H-exch-res}
\end{eqnarray}
where
\begin{eqnarray}
  g(R)
  &=&
  \frac{t_{\mathrm{s}}(R)~
        t_{\mathrm{p}}(R)}
       {3~\Delta\epsilon}.
  \label{g(R)-def-append}
\end{eqnarray}

\section{Derivation of $\delta{H}_{\beta,\beta'}^{(2)}$,
  Eq. (\ref{dH2-alpha-alpha})}
  \label{append-derivation-2nd-PMS}

Here we consider in turn the various multipole contributions
to $\delta{H}_{\beta,\beta'}^{(2)}$ with
$\beta,\beta'={\mathrm{d}},{\mathrm{q}},{\mathrm{o}}$.

\noindent
{\textbf{1}}. {\underline{\textbf{Dipole-dipole contribution}}}:
The correction $\delta{H}_{\mathrm{d,d}}^{(2)}$
[eq. (\ref{dH2-alpha-alpha})] is,
\begin{eqnarray}
  \delta{H}^{(2)}_{\mathrm{d,d}} &=&
  -\frac{\lambda_{\mathrm{d}}^{2}}{D}
  \sum_{\alpha,\alpha'}
  \sum_{f,f',f''}
  \sum_{i,i',i''}
  \sum_{n,n',n''}
  F^{\alpha}_{f,f''}
  F^{\alpha'}_{f'',f'}
  \times \nonumber \\ && \times
  X^{f,f'}
  c_{n,i}^{\dag}
  c_{n',i'}
  \Big(
      I^{\alpha}_{i,i''}
      I^{\alpha'}_{i'',i'}
      \big\langle
          c_{n'',i''}
          c_{n'',i''}^{\dag}
      \big\rangle-
  \nonumber \\ && -
      I^{\alpha'}_{i,i''}
      I^{\alpha}_{i'',i'}
      \big\langle
          c_{n'',i''}^{\dag}
          c_{n'',i''}
      \big\rangle
  \Big).
  \label{dH2-dd-def}
\end{eqnarray}
Here the energy of atoms with harmonic quantum numbers
$n,n'$ belong to the reduced energy band, whereas the energy
of atoms with the harmonic quantum number $n''$ is located
near the edge of the energy band such that
$$
  \big|\veps_{n}\big|,~
  \big|\veps_{n'}\big|~<~ D',
  \ \ \ \ \
  D' ~<~
  \big|\veps_{n''}\big|
  ~<~ D,
$$
where $D>D'=D-\delta{D}$. When $D'\gg{T}$ ($T$ is the temperature
of the gas), we can write
\begin{eqnarray*}
  \big\langle
      c_{n'',i''}
      c_{n'',i''}^{\dag}
  \big\rangle
  &=&
  \Theta(\veps_{n''}-D)
  \Theta(-D'-\veps_{n''}),
  \\
  \big\langle
      c_{n'',i''}^{\dag}
      c_{n'',i''}
  \big\rangle
  &=&
  \Theta(\veps_{n''}-D')
  \Theta(D-\veps_{n''}),
\end{eqnarray*}
where $\Theta(\epsilon)$ is the Heaviside theta function equal
to 1 for $\epsilon>0$, 0 for $\epsilon<0$ and $\frac{1}{2}$
for $\epsilon=0$.
Thus,  the correction $H_{\mathrm{d,d}}^{(2)}$ can be written as,
\begin{eqnarray}
  \delta{H}^{(2)}_{\mathrm{d,d}} &=&
  -\frac{\lambda_{\mathrm{d}}^{2}\rho_0\delta{D}}{2D}
  \sum_{\alpha,\alpha'}
  \sum_{f,f'}
  \sum_{i,i'}
  \sum_{n.n'}
  \Big[
      \hat{F}^{\alpha},
      \hat{F}^{\alpha'}
  \Big]_{f,f'}
  \times \nonumber \\ && \times
  \Big[
      \hat{I}^{\alpha},~
      \hat{I}^{\alpha'}
  \Big]_{i,i'}
  X^{f,f'}~
  c_{n,i}^{\dag}
  c_{n',i'},
  \label{dH2-dd-a}
\end{eqnarray}
where
$$
  \big[
      \hat{A},
      \hat{B}
  \big]
  ~=~
  \hat{A}\hat{B}-\hat{B}\hat{A},
$$
denotes the commutator of the matrices $\hat{A}$ and $\hat{B}$.
$\rho_0$ is the density of states of itinerant atoms.
Taking into account the property of the Levi-Civita symbols,
\begin{eqnarray}
  \sum_{\alpha,\alpha'}
  \epsilon^{\alpha,\alpha',\alpha_1}
  \epsilon^{\alpha,\alpha',\alpha'_1}
  &=&
  2~
  \delta^{\alpha_1,\alpha'_1},
  \label{LCS-prop}
\end{eqnarray}
we get,
\begin{eqnarray}
  \delta{H}^{(2)}_{\mathrm{d,d}} &=&
  -\frac{\lambda_{\mathrm{d}}^{2}\rho_0\delta{D}}{D}
  \sum_{\alpha}
  \sum_{f,f'}
  \sum_{i,i'}
  \sum_{n,n'}
  F^{\alpha}_{f,f'}~
  I^{\alpha}_{i,i'}
  \times \nonumber \\ && \times
  X^{f,f'}~
  c_{n,i}^{\dag}
  c_{n',i'}.
  \label{dH2-dd-res}
\end{eqnarray}

\noindent
{\textbf{2}}. {\underline{\textbf{Dipole-quadrupole contribution}}}:
The correction $\delta{H}_{\mathrm{d,q}}^{(2)}$
[eq. (\ref{dH2-alpha-alpha})] is,
\begin{eqnarray}
  \delta{H}^{(2)}_{\mathrm{d,q}} &=&
  -\frac{\lambda_{\mathrm{d}}
         \lambda_{\mathrm{q}}
         \rho_0\delta{D}}{2D}
  \sum_{\alpha_1,\alpha_2,\alpha'_2}
  \sum_{f,f'}
  \sum_{i,i'}
  \sum_{n,n'}
  \times \nonumber \\ && \times
  \Big[
      \hat{F}^{\alpha_1},
      \hat{F}^{\alpha_2,\alpha'_2}
  \Big]_{f,f'}
  \Big[
      \hat{I}^{\alpha_1},~
      \hat{I}^{\alpha_2,\alpha'_2}
  \Big]_{i,i'}
  \times \nonumber \\ && \times
  X^{f,f'}~
  c_{n,i}^{\dag}
  c_{n',i'},
  \label{dH2-dq-a}
\end{eqnarray}
where $\rho_0$ is the density of states (\ref{DOS-def}) of
itinerant atoms.
Taking into account the property (\ref{LCS-prop}) of
the Levi-Civita symbols, we get
\begin{eqnarray}
  \delta{H}^{(2)}_{\mathrm{d,q}} &=&
  -\frac{12
        \lambda_{\mathrm{d}}
        \lambda_{\mathrm{q}}
        \rho_0\delta{D}}{D}
  \sum_{\alpha,\alpha'}
  \sum_{f,f'}
  \sum_{i,i'}
  \sum_{n,n'}
  F^{\alpha,\alpha'}_{f,f'}~
  I^{\alpha,\alpha'}_{i,i'}
  \times \nonumber \\ && \times
  X^{f,f'}~
  c_{n,i}^{\dag}
  c_{n',i'}.
  \label{dH2-dq-res}
\end{eqnarray}

{\textbf{3}}. {\underline{\textbf{Dipole-octupole contribution}}}:
The correction $\delta{H}_{\mathrm{d,o}}^{(2)}$
[eq. (\ref{dH2-alpha-alpha})] is,
\begin{eqnarray}
  \delta{H}^{(2)}_{\mathrm{d,o}} &=&
  -\frac{\lambda_{\mathrm{d}}
         \lambda_{\mathrm{o}}
         \rho_0\delta{D}}{2D}
  \sum_{\alpha_1}
  \sum_{\alpha_2,\alpha'_2,\alpha''_2}
  \sum_{f,f'}
  \sum_{i,i'}
  \sum_{n,n'}
  \nonumber \\ && \times
  \Big[
      \hat{F}^{\alpha_1},
      \hat{F}^{\alpha_2,\alpha'_2,\alpha''_2}
  \Big]_{f,f'}
  \Big[
      \hat{I}^{\alpha_1},~
      \hat{I}^{\alpha_2,\alpha'_2,\alpha''_2}
  \Big]_{i,i'}
  \times \nonumber \\ && \times
  X^{f,f'}~
  c_{n,i}^{\dag}
  c_{n',i'},
  \label{dH2-do-a}
\end{eqnarray}
where $\rho_0$ is the density of states (\ref{DOS-def}) of
itinerant atoms.
Taking into account the property
(\ref{LCS-prop}) of the Levi-Civita symbols, we get
\begin{eqnarray}
  \delta{H}^{(2)}_{\mathrm{d,o}} &=&
  -\frac{18
        \lambda_{\mathrm{d}}
        \lambda_{\mathrm{o}}
        \rho_0\delta{D}}{D}
  \sum_{\alpha,\alpha',\alpha''}
  \sum_{f,f'}
  \sum_{i,i'}
  \sum_{n,n'}
  \nonumber \\ && \times
  F^{\alpha,\alpha',\alpha''}_{f,f'}~
  I^{\alpha,\alpha',\alpha''}_{i,i'}
  X^{f,f'}~
  c_{n,i}^{\dag}
  c_{n',i'}.
  \label{dH2-do-res}
\end{eqnarray}

{\textbf{4}}. {\underline{\textbf{Quadrupole-quadrupole contribution}}}:
The correction $\delta{H}_{\mathrm{q,q}}^{(2)}$
[eq. (\ref{dH2-alpha-alpha})] is,
\begin{eqnarray}
  \delta{H}^{(2)}_{\mathrm{q,q}} &=&
  -\frac{\lambda_{\mathrm{q}}^{2}\rho_0\delta{D}}{2D}
  \sum_{\alpha_1,\alpha'_1}
  \sum_{\alpha_2,\alpha'_2}
  \sum_{f,f'}
  \sum_{i,i'}
  \sum_{n,n'}
  \nonumber \\ && \times
  \Big[
      \hat{F}^{\alpha_1,\alpha'_1},
      \hat{F}^{\alpha_2,\alpha'_2}
  \Big]_{f,f'}
  \Big[
      \hat{I}^{\alpha_1,\alpha'_1},~
      \hat{I}^{\alpha_2,\alpha'_2}
  \Big]_{i,i'}
  \times \nonumber \\ && \times
  X^{f,f'}
  c_{n,i}^{\dag}
  c_{n',i'},
  \label{dH2-qq-a}
\end{eqnarray}
where $\rho_0$ is the density of states (\ref{DOS-def}) of
itinerant atoms.
Taking into
account the property (\ref{LCS-prop}) of the Levi-Civita
symbols, we get
\begin{eqnarray}
  \delta{H}^{(2)}_{\mathrm{q,q}} &=&
  -\frac{64\lambda_{\mathrm{q}}^{2}\rho_0\delta{D}}{9D}
  \sum_{\alpha,\alpha',\alpha''}
  \sum_{f,f'}
  \sum_{i,i'}
  \sum_{n,n'}
  \nonumber \\ && \times
  F^{\alpha,\alpha',\alpha''}_{f,f'}~
  I^{\alpha,\alpha',\alpha''}_{i,i'}~
  X^{f,f'}~
  c_{n,i}^{\dag}
  c_{n',i'}-
  \nonumber \\ &-&
  \frac{9216\lambda_{\mathrm{q}}^{2}\rho_0\delta{D}}{25D}
  \sum_{\alpha}
  \sum_{f,f'}
  \sum_{i,i'}
  \sum_{n,n'}
  \nonumber \\ && \times
  F^{\alpha}_{f,f'}~
  I^{\alpha}_{i,i'}~
  X^{f,f'}~
  c_{n,i}^{\dag}
  c_{n',i'}.
  \label{dH2-qq-res}
\end{eqnarray}

{\textbf{5}}. {\underline{\textbf{Quadrupole-octupole contribution}}}:
The correction $\delta{H}_{\mathrm{q,o}}^{(2)}$
[eq. (\ref{dH2-alpha-alpha})] is,
\begin{eqnarray}
  \delta{H}^{(2)}_{\mathrm{q,o}} &=&
  -\frac{\lambda_{\mathrm{q}}
         \lambda_{\mathrm{o}}
         \rho_0\delta{D}}{2D}
  \sum_{\alpha_1,\alpha'_1}
  \sum_{\alpha_2,\alpha'_2,\alpha''_2}
  \sum_{f,f'}
  \sum_{i,i'}
  \sum_{n,n'}
  \nonumber \\ && \times
  \Big[
      \hat{F}^{\alpha_1,\alpha'_1},
      \hat{F}^{\alpha_2,\alpha'_2,\alpha''_2}
  \Big]_{f,f'}
  \times \nonumber \\ && \times
  \Big[
      \hat{I}^{\alpha_1,\alpha'_1},~
      \hat{I}^{\alpha_2,\alpha'_2,\alpha''_2}
  \Big]_{i,i'}
  \times \nonumber \\ && \times
  X^{f,f'}
  c_{n,i}^{\dag}
  c_{n',i'},
  \label{dH2-qo-a}
\end{eqnarray}
where $\rho_0$ is the density of states (\ref{DOS-def}) of
itinerant atoms.
Then eq. (\ref{dH2-qo-a}) takes
the form,
\begin{eqnarray}
  \delta{H}^{(2)}_{\mathrm{q,o}} &=&
  -\frac{1458
        \lambda_{\mathrm{q}}
        \lambda_{\mathrm{o}}}
       {5 D}
  \sum_{\alpha,\alpha'}
  \sum_{f,f'}
  \sum_{i,i'}
  \sum_{n,n'}
  F^{\alpha,\alpha'}_{f,f'}
  I^{\alpha',\alpha}_{i',i}
  \times \nonumber \\ && \times
  X^{f,f'}~
  c_{n',i'}^{\dag}
  c_{n,i}.
  \label{dH2-qo-res}
\end{eqnarray}

{\textbf{6}}. {\underline{\textbf{Octupole-octupole contribution}}}:
The correction $\delta{H}_{\mathrm{o,o}}^{(2)}$
[eq. (\ref{dH2-alpha-alpha})] is,
\begin{eqnarray}
  \delta{H}^{(2)}_{\mathrm{o,o}} &=&
  -\frac{\lambda_{\mathrm{q}}\lambda_{\mathrm{o}}\rho_0\delta{D}}{2D}
  \sum_{\alpha_1,\alpha'_1,\alpha''_1}
  \sum_{\alpha_2,\alpha'_2,\alpha''_2}
  \sum_{f,f'}
  \sum_{i,i'}
  \sum_{n,n'}
  \nonumber \\ && \times
  \Big[
      \hat{F}^{\alpha_1,\alpha'_1,\alpha''_1},
      \hat{F}^{\alpha_2,\alpha'_2,\alpha''_2}
  \Big]_{f,f'}
  \times \nonumber \\ && \times
  \Big[
      \hat{I}^{\alpha_1,\alpha'_1,\alpha''_1},~
      \hat{I}^{\alpha_2,\alpha'_2,\alpha''_2}
  \Big]_{i,i'}
  \times \nonumber \\ && \times
  X^{f,f'}
  c_{n,i}^{\dag}
  c_{n',i'},
  \label{dH2-oo-a}
\end{eqnarray}
where $\rho_0$ is the density of states (\ref{DOS-def}) of
itinerant atoms.
Then eq. (\ref{dH2-oo-a}) takes
the form,
\begin{eqnarray}
  \delta{H}^{(2)}_{\mathrm{o,o}} &=&
  -\frac{1469664
        \lambda_{\mathrm{o}}^{2}\rho_0 \delta D}
       {25 D}
  \sum_{\alpha}
  \sum_{f,f'}
  \sum_{i,i'}
  \sum_{n,n'}
  F^{\alpha}_{f,f'}
  I^{\alpha}_{i',i}
  \times \nonumber \\ && \times
  X^{f,f'}~
  c_{n',i'}^{\dag}
  c_{n,i}+
  \nonumber \\ && +
  \frac{306\lambda_{\mathrm{o}}^{2}\rho_0 \delta D}{5 D}
  \sum_{\alpha,\alpha',\alpha''}
  \sum_{f,f'}
  \sum_{i,i'}
  \sum_{n,n'}
  \times \nonumber \\ && \times
  F^{\alpha,\alpha',\alpha''}_{f,f'}
  I^{\alpha'',\alpha',\alpha}_{i',i}
  X^{f,f'}~
  c_{n',i'}^{\dag}
  c_{n,i}.
  \label{dH2-oo-res}
\end{eqnarray}

\section{Derivation of $\delta{E}$,
  Eq. (\ref{dE-res})}
  \label{apppend-derivation-dE-2nd}

The second order correction to the energy is  illustrated by
the diagrams displayed in Fig. \ref{Fig-2nd-energy-diagrams}.
is decomposed
into its multipole components as,
\begin{eqnarray}
  \delta{E} &=&
  \delta{E}_{\mathrm{d}}+
  \delta{E}_{\mathrm{q}}+
  \delta{E}_{\mathrm{o}}.
  \label{dE=dEd+dEq+dEo}
\end{eqnarray}
Here $\delta{E}_{\mathrm{d}}$, $\delta{E}_{\mathrm{q}}$ and
$\delta{E}_{\mathrm{o}}$ are dipole, quadrupole and octupole
contributions to $\delta{E}$, given explicitly as,
\begin{subequations}
\begin{eqnarray}
  \delta{E}_{\mathrm{d}} &=&
  -\lambda_{\mathrm{d}}^{2}~
  \frac{E}{D^2}
  \sum_{f,f'}
  X^{f,f'}
  \sum_{\alpha_1,\alpha_2}
  \Big(
      \hat{F}^{\alpha_1}
      \hat{F}^{\alpha_2}
  \Big)_{f,f'}
  \times \nonumber \\ && \times
  {\mathrm{Tr}}
  \Big(
      \hat{I}^{\alpha_1}
      \hat{I}^{\alpha_2}
  \Big)~
  2 \rho_0^2 D \delta D,
  \label{dE-d-def}
  \\
  \delta{E}_{\mathrm{q}} &=&
  -\lambda_{\mathrm{q}}^{2}~
  \frac{E}{D^2}
  \sum_{f,f'}
  X^{f,f'}
  \times \nonumber \\ && \times
  \sum_{\alpha_1,\alpha'_1}
  \sum_{\alpha_2,\alpha'_2}
  \Big(
      \hat{F}^{\alpha_1,\alpha'_1}
      \hat{F}^{\alpha_2,\alpha'_2}
  \Big)_{f,f'}
  \times \nonumber \\ && \times
  {\mathrm{Tr}}
  \Big(
      \hat{I}^{\alpha_1,\alpha'_1}
      \hat{I}^{\alpha_2,\alpha'_2}
  \Big)~
  2 \rho_0^2 D \delta D,
  \label{dE-q-def}
  \\
  \delta{E}_{\mathrm{o}} &=&
  -\lambda_{\mathrm{o}}^{2}~
  \frac{E}{D^2}
  \sum_{f,f'}
  X^{f,f'}
  \sum_{\alpha_1,\alpha'_1,\alpha''_1}
  \sum_{\alpha_2,\alpha'_2,\alpha''_2}
  \nonumber \\ && \times
  \Big(
      \hat{F}^{\alpha_1,\alpha'_1,\alpha''_1}
      \hat{F}^{\alpha_2,\alpha'_2,\alpha''_2}
  \Big)_{f,f'}
  \times \nonumber \\ && \times
  {\mathrm{Tr}}
  \Big(
      \hat{I}^{\alpha_1,\alpha'_1,\alpha''_1}
      \hat{I}^{\alpha_2,\alpha'_2,\alpha''_2}
  \Big)~
  2 \rho_0^2 D \delta D,
  \label{dE-o-def}
\end{eqnarray}
  \label{subeqs-dE-2nd}
\end{subequations}
We consider dipole, quadrupole and octupole contributions to
$\delta{E}$ in turn.

\noindent
{\textbf{1. Dipole contribution}}: The trace of the product of two spin
matrices is,
\begin{eqnarray}
  \sum_{i_1,i_2}
  I^{\alpha_1}_{i_1,i_2}
  I^{\alpha_2}_{i_2,i_1}
  &=&
  \frac{1}{3}~
  I(I+1)(2I+1)~
  \delta^{\alpha_1,\alpha_2}.
  \label{trace-d-d}
\end{eqnarray}
Using eq. (\ref{trace-d-d}), we can write
\begin{eqnarray}
  \sum_{\alpha}
  \sum_{f_1}
  F^{\alpha}_{f,f_1}
  F^{\alpha}_{f_1,f'}
  &=&
  F(F+1)~
  \delta_{f,f'}.
  \label{dE-d-F-Casimir}
\end{eqnarray}
Finally, the dipole contribution to the self energy is,
\begin{eqnarray*}
  \delta{E}_{\mathrm{d}} =
  -\frac{2\delta{D}}{3D}~
  E~
  \Lambda_{\mathrm{d}}^{2}~
  F(F+1)
  I(I+1)(2I+1).
\end{eqnarray*}
Taking into account that $F=\frac{3}{2}$ and $I=\frac{5}{2}$, we get
\begin{eqnarray}
  \delta{E}_{\mathrm{d}} &=&
  -\frac{525}{4}~
  \frac{\delta{D}}{D}~
  E~
  \Lambda_{\mathrm{d}}^{2}.
  \label{dE-d-res}
\end{eqnarray}
\noindent
{\textbf{2. Quadrupole contribution}}: The trace of the product of two
quadrupole matrices is,
\begin{eqnarray}
  &&
  \sum_{i_1,i_2}
  I^{\alpha_1,\alpha'_1}_{i_1,i_2}
  I^{\alpha_2,\alpha'_2}_{i_2,i_1}
  ~=~
  -\frac{112}{3}~
  \Big\{
      2~
      \delta_{\alpha_1,\alpha'_1}~
      \delta_{\alpha_2,\alpha'_2}-
  \nonumber \\ && ~~~~~ ~~~~~ -
      3~
      \big[
          \delta_{\alpha_1,\alpha_2}~
          \delta_{\alpha'_1,\alpha'_2}+
          \delta_{\alpha_1,\alpha'_2}~
          \delta_{\alpha'_1,\alpha_2}
      \big]
  \Big\}.
  \label{trace-q-q}
\end{eqnarray}
Using eq. (\ref{trace-q-q}), we can write
\begin{eqnarray}
  &&
  \sum_{\alpha_1,\alpha'_1}
  \sum_{\alpha_2,\alpha'_2}
  \hat{F}^{\alpha_1,\alpha'_1}
  \hat{F}^{\alpha_2,\alpha'_2}~
  {\mathrm{Tr}}
  \big(
      \hat{I}^{\alpha_1,\alpha'_1}
      \hat{I}^{\alpha_2,\alpha'_2}
  \big)
  ~= \nonumber \\ && ~~~~~ ~~~~~ =~
  6720~
  \hat{1}_{4},
  \label{dE-q-F-Casimir}
\end{eqnarray}
where $\hat{1}_{4}$ is the $4\times4$ identity matrix.

Finally, quadrupole contribution to the self energy is,
\begin{eqnarray}
  \delta{E}_{\mathrm{q}} &=&
  -13440~
  \frac{\delta{D}}{D}~
  E~
  \Lambda_{\mathrm{q}}^{2}.
  \label{dE-q-res}
\end{eqnarray}
\noindent
{\textbf{3. Octupole contribution}}: The trace of the product of two
quadrupole matrices is,
\begin{eqnarray}
  &&
  \sum_{\alpha_1,\alpha'_1,\alpha''_1}
  \sum_{\alpha_2,\alpha'_2,\alpha''_2}
  \hat{F}^{\alpha_1,\alpha'_1,\alpha''_1}
  \hat{F}^{\alpha_2,\alpha'_2,\alpha''_2}
  \times \nonumber \\ && ~~~ \times
  {\mathrm{Tr}}
  \big(
      \hat{I}^{\alpha_1,\alpha'_1,\alpha''_1}
      \hat{I}^{\alpha_2,\alpha'_2,\alpha''_2}
  \big)
  ~=~
  1653372~
  \hat{1}_{4},
  \label{trace-o-o}
\end{eqnarray}
where $\hat{1}_{4}$ is the $4\times4$ identity matrix.

Finally, octupole contribution to the self energy is,
\begin{eqnarray}
  \delta{E}_{\mathrm{o}} &=&
  -3306744~
  \frac{\delta{D}}{D}~
  E~
  \Lambda_{\mathrm{q}}^{2}.
  \label{dE-o-res}
\end{eqnarray}
\noindent
{\textbf{Finally}}, the second order correction to the energy is,
\begin{eqnarray}
  \delta{E} &=&
  -\frac{\delta{D}}{D}~
  E~
  \bigg\{
       \frac{525}{4}~
       \Lambda_{\mathrm{d}}^{2}+
       13440~
       \Lambda_{\mathrm{q}}^{2}+
  \nonumber \\ && +~
       3306744~
       \Lambda_{\mathrm{o}}^{2}
  \bigg\}.
  \label{dE-res-append}
\end{eqnarray}

\section{Derivation of $\delta{H}^{(3)}_{\beta,\beta',\beta}$,
  Eq. (\ref{dH3-beta-beta-beta})}
  \label{apppend-derivation-3rd-PMS}

We consider $\delta{H}_{\beta',\beta,\beta'}^{(3)}$ for
$\beta,\beta'={\mathrm{d,q,o}}$, in turn.

{\textbf{1}}. {\underline{\textbf{Dipole-dipole contribution}}}:
The correction $\delta{H}_{\mathrm{d,d,d}}^{(3)}$ is,
\begin{eqnarray}
  \delta{H}_{\mathrm{d,d,d}}^{(3)} &=&
  \frac{2
        \lambda_{\mathrm{d}}^{3}
        \rho_{0}^{2}
        \delta{D}}
       {D}
  \sum_{\alpha_1,\alpha_2,\alpha_3}
  \sum_{f,f'}
  \Big(
      \hat{F}^{\alpha_1}
      \hat{F}^{\alpha_2}
      \hat{F}^{\alpha_3}
  \Big)_{f,f'}
  \nonumber \\ && \times
  X^{f,f'}
  \sum_{i,i'}
  \sum_{n,n'}
  I^{\alpha_2}_{i,i'}
  c_{n,i}^{\dag}
  c_{n',i'}
  \times \nonumber \\ && \times
  {\mathrm{Tr}}
  \Big(
      \hat{I}^{\alpha_1}
      \hat{I}^{\alpha_3}
  \Big).
  \label{dH-d-d-d}
\end{eqnarray}
In order to simplify the expression in the right hand side of eq. (\ref{dH-d-d-d}),
we use the following equalities,
\begin{eqnarray}
  &&
  {\mathrm{Tr}}
  \Big(
      \hat{I}^{\alpha_1}
      \hat{I}^{\alpha_3}
  \Big)
  ~=~
  \frac{1}{3}~
  I(I+1)(2I+1)
  ~= \nonumber \\ && ~~~~~ ~~~~~ ~~~~~ ~=~
  \frac{35}{2}~
  \delta_{\alpha_1,\alpha_3},
  \label{trace-d-d-3rd}
  \\
  &&
  \sum_{\alpha_1}
  \hat{F}^{\alpha_1}
  \hat{F}^{\alpha_2}
  \hat{F}^{\alpha_1}
  ~=~
  \Big(
      F(F+1)-1
  \Big)~
  \hat{F}^{\alpha_2}
  ~= \nonumber \\ && ~~~~~ ~~~~~ ~~~~~ ~=~
  \frac{11}{4}~
  \hat{F}^{\alpha_2}.
  \label{d-d-d}
\end{eqnarray}
Then the dipole-dipole contribution takes the form,
\begin{eqnarray}
  \delta{H}_{\mathrm{d,d,d}}^{(3)} &=&
  \frac{385}{4}~
  \frac{\lambda_{\mathrm{d}}^{3}
        \rho_{0}^{2}
        \delta{D}}
       {D}
  \sum_{\alpha}
  \sum_{f,f'}
  \sum_{i,i'}
  \sum_{n,n'}
  \nonumber \\ && \times
  F^{\alpha}_{f,f'}
  I^{\alpha}_{i,i'}
  X^{f,f'}
  c_{n,i}^{\dag}
  c_{n',i'}.
  \label{dH-d-d-d-res}
\end{eqnarray}

{\textbf{2}}. {\underline{\textbf{Dipole-quadrupole contribution}}}:
The correction $\delta{H}_{\mathrm{q,d,q}}^{(3)}$ is,
\begin{eqnarray}
  \delta{H}_{\mathrm{q,d,q}}^{(3)} &=&
  \frac{2
        \lambda_{\mathrm{d}}
        \lambda_{\mathrm{q}}^{2}
        \rho_{0}^{2}
        \delta{D}}
       {D}
  \sum_{\alpha_1,\alpha'_1}
  \sum_{\alpha_2}
  \sum_{\alpha_3,\alpha'_3}
  \sum_{f,f'}
  \nonumber \\ && \times
  \Big(
      \hat{F}^{\alpha_1,\alpha'_1}
      \hat{F}^{\alpha_2}
      \hat{F}^{\alpha_3,\alpha'_3}
  \Big)_{f,f'}
  \times \nonumber \\ && \times
  X^{f,f'}
  \sum_{i,i'}
  \sum_{n,n'}
  I^{\alpha_2}_{i,i'}
  c_{n,i}^{\dag}
  c_{n',i'}
  \times \nonumber \\ && \times
  {\mathrm{Tr}}
  \Big(
      \hat{I}^{\alpha_1,\alpha'_1}
      \hat{I}^{\alpha_3,\alpha'_3}
  \Big).
  \label{dH-q-d-q}
\end{eqnarray}
Using eqs. (\ref{trace-q-q}) and (\ref{dE-q-F-Casimir}), we can write
\begin{eqnarray*}
  &&
  \sum_{\alpha_1,\alpha'_1}
  \sum_{\alpha_3,\alpha'_3}
  \hat{F}^{\alpha_1,\alpha'_1}
  \hat{F}^{\alpha_2}
  \hat{F}^{\alpha_3,\alpha'_3}~
  {\mathrm{Tr}}
  \Big(
      \hat{I}^{\alpha_1,\alpha'_1}
      \hat{I}^{\alpha_3,\alpha'_3}
  \Big)
  ~= \nonumber \\ && ~~~~~ ~~~~~ =~
  1344~
  \hat{F}^{\alpha_2}.
\end{eqnarray*}
Then the dipole-quadrupole contribution can be written as,
\begin{eqnarray}
  \delta{H}_{\mathrm{q,d,q}}^{(3)} &=&
  2688~
  \frac{\lambda_{\mathrm{d}}
        \lambda_{\mathrm{q}}^{2}
        \rho_{0}^{2}
        \delta{D}}
       {D}
  \sum_{\alpha}
  \sum_{f,f'}
  \sum_{i,i'}
  \sum_{n,n'}
  \nonumber \\ && \times
  F^{\alpha}_{f,f'}
  I^{\alpha}_{i,i'}
  X^{f,f'}
  c_{n,i}^{\dag}
  c_{n',i'}.
  \label{dH-q-d-q-res}
\end{eqnarray}

{\textbf{3}}. {\underline{\textbf{Dipole-octupole contribution}}}:
The correction $\delta{H}_{\mathrm{o,d,o}}^{(3)}$ is,
\begin{eqnarray}
  \delta{H}_{\mathrm{o,d,o}}^{(3)} &=&
  \frac{2
        \lambda_{\mathrm{d}}
        \lambda_{\mathrm{o}}^{2}
        \rho_{0}^{2}
        \delta{D}}
       {D}
  \sum_{\alpha_1,\alpha'_1,\alpha''_1}
  \sum_{\alpha_2}
  \sum_{\alpha_3,\alpha'_3,\alpha''_3}
  \sum_{f,f'}
  \nonumber \\ && \times
  \Big(
      \hat{F}^{\alpha_1,\alpha'_1,\alpha''_1}
      \hat{F}^{\alpha_2}
      \hat{F}^{\alpha_3,\alpha'_3,\alpha''_3}
  \Big)_{f,f'}
  \times \nonumber \\ && \times
  X^{f,f'}
  \sum_{i,i'}
  \sum_{n,n'}
  I^{\alpha_2}_{i,i'}
  c_{n,i}^{\dag}
  c_{n',i'}
  \times \nonumber \\ && \times
  {\mathrm{Tr}}
  \Big(
      \hat{I}^{\alpha_1,\alpha'_1,\alpha''_1}
      \hat{I}^{\alpha_3,\alpha'_3,\alpha''_3}
  \Big).
  \label{dH-o-d-o}
\end{eqnarray}
Equation (\ref{dH-o-d-o}) can be simplified by using eq. (\ref{trace-o-o}),
\begin{eqnarray*}
  &&
  \sum_{\alpha_1,\alpha'_1,\alpha''_1}
  \sum_{\alpha_3,\alpha'_3,\alpha''_3}
  \hat{F}^{\alpha_1,\alpha'_1,\alpha''_1}
  \hat{F}^{\alpha_2}
  \hat{F}^{\alpha_3,\alpha'_3,\alpha''_3}
  \times \nonumber \\ && ~~~~~ ~~~~~ \times
  {\mathrm{Tr}}
  \Big(
      \hat{I}^{\alpha_1,\alpha'_1,\alpha''_1}
      \hat{I}^{\alpha_3,\alpha'_3,\alpha''_3}
  \Big) =
  \nonumber \\ && ~~~~~ ~~~~~ =~
  -\frac{4960116}{5}~
  \hat{F}^{\alpha_2}.
\end{eqnarray*}
Then the dipole-quadrupole contribution can be written as,
\begin{eqnarray}
  \delta{H}_{\mathrm{o,d,o}}^{(3)} &=&
  -\frac{9920232}{5}~
  \frac{\lambda_{\mathrm{d}}
        \lambda_{\mathrm{o}}^{2}
        \rho_{0}^{2}
        \delta{D}}
       {D}
  \sum_{\alpha}
  \sum_{f,f'}
  \sum_{i,i'}
  \sum_{n,n'}
  \nonumber \\ && \times
  F^{\alpha}_{f,f'}
  I^{\alpha}_{i,i'}
  X^{f,f'}
  c_{n,i}^{\dag}
  c_{n',i'}.
  \label{dH-o-d-o-res}
\end{eqnarray}

{\textbf{4}}. {\underline{\textbf{Quadrupole-dipole contribution}}}:
The correction $\delta{H}_{\mathrm{d,q,d}}^{(3)}$ is,
\begin{eqnarray}
  \delta{H}_{\mathrm{d,q,d}}^{(3)} &=&
  \frac{2
        \lambda_{\mathrm{q}}
        \lambda_{\mathrm{d}}^{2}
        \rho_{0}^{2}
        \delta{D}}
       {D}
  \sum_{\alpha_1}
  \sum_{\alpha_2,\alpha'_2}
  \sum_{\alpha_3}
  \sum_{f,f'}
  \nonumber \\ && \times
  \Big(
      \hat{F}^{\alpha_1}
      \hat{F}^{\alpha_2,\alpha'_2}
      \hat{F}^{\alpha_3}
  \Big)_{f,f'}
  \times \nonumber \\ && \times
  X^{f,f'}
  \sum_{i,i'}
  \sum_{n,n'}
  I^{\alpha_2,\alpha'_2}_{i,i'}
  c_{n,i}^{\dag}
  c_{n',i'}
  \times \nonumber \\ && \times
  {\mathrm{Tr}}
  \Big(
      \hat{I}^{\alpha_1}
      \hat{I}^{\alpha_3}
  \Big).
  \label{dH-d-q-d}
\end{eqnarray}
In order to simplify eq. (\ref{dH-d-q-d}), we use eq. (\ref{trace-d-d}) and
the following equality,
\begin{eqnarray*}
  \sum_{\alpha}
  \hat{F}^{\alpha}
  \hat{F}^{\alpha_2,\alpha'_2}
  \hat{F}^{\alpha}
  &=&
  \frac{3}{4}~
  \hat{F}^{\alpha_2,\alpha'_2}.
\end{eqnarray*}
Then the quadrupole-dipole contribution can be written as,
\begin{eqnarray}
  \delta{H}_{\mathrm{d,q,d}}^{(3)} &=&
  \frac{105}{4}~
  \frac{\lambda_{\mathrm{q}}
        \lambda_{\mathrm{d}}^{2}
        \rho_{0}^{2}
        \delta{D}}
       {D}
  \sum_{\alpha}
  \sum_{f,f'}
  \sum_{i,i'}
  \sum_{n,n'}
  \nonumber \\ && \times
  F^{\alpha}_{f,f'}
  I^{\alpha}_{i,i'}
  X^{f,f'}
  c_{n,i}^{\dag}
  c_{n',i'}.
  \label{dH-d-q-d-res}
\end{eqnarray}

{\textbf{5}}. {\underline{\textbf{Quadrupole-quadrupole contribution}}}:
The correction $\delta{H}_{\mathrm{q,q,q}}^{(3)}$ is,
\begin{eqnarray}
  \delta{H}_{\mathrm{q,q,q}}^{(3)} &=&
  \frac{2
        \lambda_{\mathrm{q}}^{3}
        \rho_{0}^{2}
        \delta{D}}
       {D}
  \sum_{\alpha_1,\alpha'_1}
  \sum_{\alpha_2,\alpha'_2}
  \sum_{\alpha_3,\alpha'_3}
  \sum_{f,f'}
  \nonumber \\ && \times
  \Big(
      \hat{F}^{\alpha_1,\alpha'_1}
      \hat{F}^{\alpha_2,\alpha'_2}
      \hat{F}^{\alpha_3,\alpha'_3}
  \Big)_{f,f'}
  \times \nonumber \\ && \times
  X^{f,f'}
  \sum_{i,i'}
  \sum_{n,n'}
  I^{\alpha_2,\alpha'_2}_{i,i'}
  c_{n,i}^{\dag}
  c_{n',i'}
  \times \nonumber \\ && \times
  {\mathrm{Tr}}
  \Big(
      \hat{I}^{\alpha_1,\alpha'_1}
      \hat{I}^{\alpha_3,\alpha'_3}
  \Big).
  \label{dH-q-q-q}
\end{eqnarray}
Using eqs. (\ref{trace-q-q}) and (\ref{dE-q-F-Casimir}), we can write
\begin{eqnarray*}
  &&
  \sum_{\alpha_1,\alpha'_1}
  \sum_{\alpha_3,\alpha'_3}
  \hat{F}^{\alpha_1,\alpha'_1}
  \hat{F}^{\alpha_2,\alpha'_2}
  \hat{F}^{\alpha_3,\alpha'_3}
  {\mathrm{Tr}}
  \Big(
      \hat{I}^{\alpha_1,\alpha'_1}
      \hat{I}^{\alpha_3,\alpha'_3}
  \Big)
  = \nonumber \\ && ~~~~~ ~~~~~ =
  -4032~
  \hat{F}^{\alpha_2,\alpha'_2}.
\end{eqnarray*}
Then $\delta{H}_{\mathrm{q,q,q}}^{(3)}$ takes the form,
\begin{eqnarray}
  \delta{H}_{\mathrm{q,q,q}}^{(3)} &=&
  -8064~
  \frac{\lambda_{\mathrm{q}}^{3}
        \rho_{0}^{2}
        \delta{D}}
       {D}
  \sum_{\alpha,\alpha'}
  \sum_{f,f'}
  \sum_{i,i'}
  \sum_{n,n'}
  \nonumber \\ && \times
  F^{\alpha,\alpha'}_{f,f'}
  I^{\alpha_2,\alpha'_2}_{i,i'}
  X^{f,f'}
  c_{n,i}^{\dag}
  c_{n',i'}.
  \label{dH-q-q-q-res}
\end{eqnarray}

{\textbf{6}}. {\underline{\textbf{Quadrupole-octupole contribution}}}:
The correction $\delta{H}_{\mathrm{o,q,o}}^{(3)}$ is,
\begin{eqnarray}
  \delta{H}_{\mathrm{o,q,o}}^{(3)} &=&
  \frac{2
        \lambda_{\mathrm{q}}
        \lambda_{\mathrm{o}}^{2}
        \rho_{0}^{2}
        \delta{D}}
       {D}
  \sum_{\alpha_1,\alpha'_1,\alpha''_1}
  \sum_{\alpha_2,\alpha'_2}
  \sum_{\alpha_3,\alpha'_3,\alpha''_3}
  \sum_{f,f'}
  \nonumber \\ && \times
  \Big(
      \hat{F}^{\alpha_1,\alpha'_1,\alpha''_1}
      \hat{F}^{\alpha_2,\alpha'_2}
      \hat{F}^{\alpha_3,\alpha'_3,\alpha''_3}
  \Big)_{f,f'}
  \times \nonumber \\ && \times
  X^{f,f'}
  \sum_{i,i'}
  \sum_{n,n'}
  I^{\alpha_2,\alpha'_2}_{i,i'}
  c_{n,i}^{\dag}
  c_{n',i'}
  \times \nonumber \\ && \times
  {\mathrm{Tr}}
  \Big(
      \hat{I}^{\alpha_1,\alpha'_1,\alpha''_1}
      \hat{I}^{\alpha_3,\alpha'_3,\alpha''_3}
  \Big).
  \label{dH-o-q-o}
\end{eqnarray}
In order to simplify eq. (\ref{dH-o-q-o}), we use the following equality,
\begin{eqnarray*}
  &&
  \sum_{\alpha_1,\alpha'_1,\alpha''_1}
  \sum_{\alpha_3,\alpha'_3,\alpha''_3}
  \hat{F}^{\alpha_1,\alpha'_1,\alpha''_1}
  \hat{F}^{\alpha_2,\alpha'_2}
  \hat{F}^{\alpha_3,\alpha'_3,\alpha''_3}
  \times \nonumber \\ && ~~~~~ ~~~~~ \times
  {\mathrm{Tr}}
  \Big(
      \hat{I}^{\alpha_1,\alpha'_1,\alpha''_1}
      \hat{I}^{\alpha_3,\alpha'_3,\alpha''_3}
  \Big)
  = \nonumber \\ && ~~~~~ ~~~~~ =
  \frac{1153372}{5}~
  \hat{F}^{\alpha_2,\alpha'_2}.
\end{eqnarray*}
Then eq. (\ref{dH-o-q-o}) takes the form,
\begin{eqnarray}
  \delta{H}_{\mathrm{o,q,o}}^{(3)} &=&
  \frac{2306744}{5}~
  \frac{\lambda_{\mathrm{q}}
        \lambda_{\mathrm{o}}^{2}
        \rho_{0}^{2}
        \delta{D}}
       {D}
  \sum_{\alpha,\alpha'}
  \sum_{f,f'}
  \sum_{i,i'}
  \sum_{n,n'}
  \nonumber \\ && \times
  F^{\alpha,\alpha'}_{f,f'}
  I^{\alpha,\alpha'}_{i,i'}
  X^{f,f'}
  c_{n,i}^{\dag}
  c_{n',i'}.
  \label{dH-o-q-o-res}
\end{eqnarray}

{\textbf{7}}. {\underline{\textbf{Octupole-dipole contribution}}}:
The correction $\delta{H}_{\mathrm{d,o,d}}^{(3)}$ is,
\begin{eqnarray}
  \delta{H}_{\mathrm{d,o,d}}^{(3)} &=&
  \frac{2
        \lambda_{\mathrm{o}}
        \lambda_{\mathrm{d}}^{2}
        \rho_{0}^{2}
        \delta{D}}
       {D}
  \sum_{\alpha_1}
  \sum_{\alpha_2,\alpha'_2,\alpha''_2}
  \sum_{\alpha_3}
  \sum_{f,f'}
  \nonumber \\ && \times
  \Big(
      \hat{F}^{\alpha_1}
      \hat{F}^{\alpha_2,\alpha'_2,\alpha''_2}
      \hat{F}^{\alpha_3}
  \Big)_{f,f'}
  \times \nonumber \\ && \times
  X^{f,f'}
  \sum_{i,i'}
  \sum_{n,n'}
  I^{\alpha_2,\alpha'_2,\alpha''_2}_{i,i'}
  c_{n,i}^{\dag}
  c_{n',i'}
  \times \nonumber \\ && \times
  {\mathrm{Tr}}
  \Big(
      \hat{I}^{\alpha_1}
      \hat{I}^{\alpha_3}
  \Big).
  \label{dH-d-o-d}
\end{eqnarray}
In order to simplify eq. (\ref{dH-d-o-d}), we use the following equality,
\begin{eqnarray*}
  &&
  \sum_{\alpha_1,\alpha_3}
  \hat{F}^{\alpha_1}
  \hat{F}^{\alpha_2,\alpha'_2,\alpha''_2}
  \hat{F}^{\alpha_3}~
  {\mathrm{Tr}}
  \Big(
      \hat{I}^{\alpha_1}
      \hat{I}^{\alpha_3}
  \Big)
  = \nonumber \\ && ~~~~~ ~~~~~ =
  -\frac{315}{8}~
  \hat{F}^{\alpha_2,\alpha'_2,\alpha''_2}.
\end{eqnarray*}
Then the octupole-dipole contribution can be written as,
\begin{eqnarray}
  \delta{H}_{\mathrm{d,o,d}}^{(3)} &=&
  -\frac{315}{4}~
  \frac{\lambda_{\mathrm{o}}
        \lambda_{\mathrm{d}}^{2}
        \rho_{0}^{2}
        \delta{D}}
       {D}
  \sum_{\alpha,\alpha',\alpha''}
  \sum_{f,f'}
  \sum_{i,i'}
  \sum_{n,n'}
  \nonumber \\ && \times
  F^{\alpha,\alpha',\alpha''}_{f,f'}
  I^{\alpha,\alpha',\alpha''}_{i,i'}
  X^{f,f'}
  c_{n,i}^{\dag}
  c_{n',i'}.
  \label{dH-d-o-d-res}
\end{eqnarray}

{\textbf{8}}. {\underline{\textbf{Octupole-quadrupole contribution}}}:
The correction $\delta{H}_{\mathrm{q,o,q}}^{(3)}$ is,
\begin{eqnarray}
  \delta{H}_{\mathrm{q,q,q}}^{(3)} &=&
  \frac{2
        \lambda_{\mathrm{o}}
        \lambda_{\mathrm{q}}^{2}
        \rho_{0}^{2}
        \delta{D}}
       {D}
  \sum_{\alpha_1,\alpha'_1}
  \sum_{\alpha_2,\alpha'_2,\alpha''_2}
  \sum_{\alpha_3,\alpha'_3}
  \sum_{f,f'}
  \nonumber \\ && \times
  \Big(
      \hat{F}^{\alpha_1,\alpha'_1}
      \hat{F}^{\alpha_2,\alpha'_2,\alpha''_2}
      \hat{F}^{\alpha_3,\alpha'_3}
  \Big)_{f,f'}
  \times \nonumber \\ && \times
  X^{f,f'}
  \sum_{i,i'}
  \sum_{n,n'}
  I^{\alpha_2,\alpha'_2,\alpha''_2}_{i,i'}
  c_{n,i}^{\dag}
  c_{n',i'}
  \times \nonumber \\ && \times
  {\mathrm{Tr}}
  \Big(
      \hat{I}^{\alpha_1,\alpha'_1}
      \hat{I}^{\alpha_3,\alpha'_3}
  \Big).
  \label{dH-q-o-q}
\end{eqnarray}
Using eqs. (\ref{trace-q-q}) and (\ref{dE-q-F-Casimir}), we can write
\begin{eqnarray*}
  &&
  \sum_{\alpha_1,\alpha'_1}
  \sum_{\alpha_3,\alpha'_3}
  \hat{F}^{\alpha_1,\alpha'_1}
  \hat{F}^{\alpha_2,\alpha'_2,\alpha''_2}
  \hat{F}^{\alpha_3,\alpha'_3}
  {\mathrm{Tr}}
  \Big(
      \hat{I}^{\alpha_1,\alpha'_1}
      \hat{I}^{\alpha_3,\alpha'_3}
  \Big)
  = \nonumber \\ && ~~~~~ ~~~~~ =
  1344~
  \hat{F}^{\alpha_2,\alpha'_2,\alpha''_2}.
\end{eqnarray*}
Then $\delta{H}_{\mathrm{q,o,q}}^{(3)}$ takes the form,
\begin{eqnarray}
  \delta{H}_{\mathrm{q,o,q}}^{(3)} &=&
  2688~
  \frac{\lambda_{\mathrm{o}}
        \lambda_{\mathrm{q}}^{2}
        \rho_{0}^{2}
        \delta{D}}
       {D}
  \sum_{\alpha,\alpha',\alpha''}
  \sum_{f,f'}
  \sum_{i,i'}
  \sum_{n,n'}
  \nonumber \\ && \times
  F^{\alpha,\alpha',\alpha''}_{f,f'}
  I^{\alpha,\alpha',\alpha''}_{i,i'}
  X^{f,f'}
  c_{n,i}^{\dag}
  c_{n',i'}.
  \label{dH-q-o-q-res}
\end{eqnarray}

{\textbf{9}}. {\underline{\textbf{Octupole-octupole contribution}}}:
The correction $\delta{H}_{\mathrm{o,o,o}}^{(3)}$ is,
\begin{eqnarray}
  \delta{H}_{\mathrm{o,o,o}}^{(3)} &=&
  \frac{2
        \lambda_{\mathrm{o}}^{3}
        \rho_{0}^{2}
        \delta{D}}
       {D}
  \sum_{\alpha_1,\alpha'_1,\alpha''_1}
  \sum_{\alpha_2,\alpha'_2,\alpha''_2}
  \sum_{\alpha_3,\alpha'_3,\alpha''_3}
  \sum_{f,f'}
  \nonumber \\ && \times
  \Big(
      \hat{F}^{\alpha_1,\alpha'_1,\alpha''_1}
      \hat{F}^{\alpha_2,\alpha'_2,\alpha''_2}
      \hat{F}^{\alpha_3,\alpha'_3,\alpha''_3}
  \Big)_{f,f'}
  \times \nonumber \\ && \times
  X^{f,f'}
  \sum_{i,i'}
  \sum_{n,n'}
  I^{\alpha_2,\alpha'_2,\alpha''_2}_{i,i'}
  c_{n,i}^{\dag}
  c_{n',i'}
  \times \nonumber \\ && \times
  {\mathrm{Tr}}
  \Big(
      \hat{I}^{\alpha_1,\alpha'_1,\alpha''_1}
      \hat{I}^{\alpha_3,\alpha'_3,\alpha''_3}
  \Big).
  \label{dH-o-o-o}
\end{eqnarray}
In order to simplify eq. (\ref{dH-o-o-o}), we use the following equality,
\begin{eqnarray*}
  &&
  \sum_{\alpha_1,\alpha'_1,\alpha''_1}
  \sum_{\alpha_3,\alpha'_3,\alpha''_3}
  \hat{F}^{\alpha_1,\alpha'_1,\alpha''_1}
  \hat{F}^{\alpha_2,\alpha'_2,\alpha''_2}
  \hat{F}^{\alpha_3,\alpha'_3,\alpha''_3}
  \times \nonumber \\ && ~~~~~ ~~~~~ \times
  {\mathrm{Tr}}
  \Big(
      \hat{I}^{\alpha_1,\alpha'_1,\alpha''_1}
      \hat{I}^{\alpha_3,\alpha'_3,\alpha''_3}
  \Big)
  = \nonumber \\ && ~~~~~ ~~~~~ =
  -\frac{236196}{5}~
  \hat{F}^{\alpha_2,\alpha'_2,\alpha''_2}.
\end{eqnarray*}
Then eq. (\ref{dH-o-o-o}) takes the form,
\begin{eqnarray}
  \delta{H}_{\mathrm{o,o,o}}^{(3)} &=&
  -\frac{472392}{5}~
  \frac{\lambda_{\mathrm{o}}^{3}
        \rho_{0}^{2}
        \delta{D}}
       {D}
  \sum_{\alpha,\alpha',\alpha''}
  \sum_{f,f'}
  \sum_{i,i'}
  \sum_{n,n'}
  \nonumber \\ && \times
  F^{\alpha,\alpha',\alpha''}_{f,f'}
  I^{\alpha,\alpha',\alpha''}_{i,i'}
  X^{f,f'}
  c_{n,i}^{\dag}
  c_{n',i'}.
  \label{dH-o-o-o-res}
\end{eqnarray}

\section{Eigenfunctions and Eigenenergies of the Dipole-Dipole,
  Quadrupole-Quadrupole and Octupole-Octupole Interactions}
  \label{append-eigenstates-dd-qq-oo}
\noindent
\underline{Appendix \ref{append-eigenstates-dd-qq-oo} main points}
The dipole-dipole, quadrupole-quadrupole and octopod-octopod
Hamiltonians are defined in Eqs.~(\ref{H-d-def-append},
\ref{H-q-def-append}, \ref{H-o-def-append}) respectively.
Their eigenfunctions and eigenenergies, required for
the calculations of the exchange constants, are explicitly
elucidated in this Appendix.

Consider four atoms, such that one atom (an ``impurity atom'') has
spin $F=\frac{3}{2}$ and the three other atoms (``itinerant
atoms'') have spin $I=\frac{5}{2}$. The interaction between
the itinerant atoms and the impurity is dipole-dipole,
quadrupole-quadrupole and octupole-octupole interactions. We derive
here eigenfunctions and corresponding eigenenergies of
the Hamiltonians of the dipole-dipole, quadrupole-quadrupole and
octupole-octupole interactions, in turn.

\subsection{Eigenfunctions of the Dipole-Dipole Interaction}

The Hamiltonian of the dipole-dipole interaction between the atoms
is,
\begin{eqnarray}
  {\mathcal{H}}_{\mathrm{d}} &=&
  \lambda_{\mathrm{d}}~
  \Big(
      \hat{\mathbf{F}}
      \cdot
      \hat{\mathbf{S}}
  \Big),
  \label{H-d-def-append}
\end{eqnarray}
where $\hat{\mathbf{F}}$ is a vector of the spin-$\frac{3}{2}$
operators, $\hat{\mathbf{S}}=\hat{\mathbf{I}}_1+%
\hat{\mathbf{I}}_2+\hat{\mathbf{I}}_3$, where $\hat{\mathbf{I}}_a$
($a=1,2,3$) is a vector of the spin-$\frac{5}{2}$ operators.

Note that ${\mathcal{H}}_{\mathrm{d}}$ can be expressed in terms of
the two-atonic spin $\hat{\mathbf{L}}$,
\begin{eqnarray}
  {\mathcal{H}}_{\mathrm{d}} &=&
  \frac{\lambda_{\mathrm{d}}}{2}~
  \Big(
      \hat{\mathbf{L}}^{2}-
      F(F+1)-
      S(S+1)
  \Big),
  \label{Hd-vs-F2-I2-L2-append}
\end{eqnarray}
where
$$
  \hat{\mathbf{L}} ~=~
  \hat{\mathbf{F}} +
  \hat{\mathbf{S}},
$$
Eq. (\ref{Hd-vs-F2-I2-L2-append}) shows that ${\mathcal{H}}_{\mathrm{d}}$
commutes with $\hat{\mathbf{L}}^{2}$, $\hat{\mathbf{F}}^{2}$ and
$\hat{\mathbf{S}}^{2}$. In addition, ${\mathcal{H}}_{\mathrm{d}}$
commutes with $\hat{\mathbf{L}}^{z}$, but neither with
$\hat{\mathbf{F}}^{z}$ nor $\hat{\mathbf{S}}^{z}$. The eigenfunctions
of ${\mathcal{H}}_{\mathrm{d}}$, $|L,L_z\rangle$, are defined as
\begin{eqnarray*}
  &&
  \hat{\mathbf{L}}^{2}~
  \big|
      L,L_z
  \big\rangle
  ~=~
  L(L+1)~
  \big|
      L,L_z
  \big\rangle,
  \\
  &&
  \hat{\mathbf{L}}^{z}~
  \big|
      L,L_z
  \big\rangle
  ~=~
  L_z~
  \big|
      L,L_z
  \big\rangle,
\end{eqnarray*}
where $L=3,4,5,6$.
The corresponding eigenvalues of ${\mathcal{H}}_{\mathrm{d}}$
are,
\begin{eqnarray}
  {\mathcal{E}}^{({\mathrm{d}})}_{L} =
  \lambda_{\mathrm{d}}~
  {\mathcal{D}}_{L},
  \label{En-d-res-append}
\end{eqnarray}
where
\begin{eqnarray}
  {\mathcal{D}}_{L}
  =
  \frac{1}{2}~
  \Big(
      L(L+1)-
      F(F+1)-
      S(S+1)
  \Big).
  \label{SL-def-append}
\end{eqnarray}
When $\lambda_{\mathrm{d}}>0$, the lowest energy level has maximal
value of $S$ and minimal value of $L$. For $I=\frac{5}{2}$, this is
the level $S=\frac{9}{2}$ and $L=3$ (see discussions in Sec.
\ref{sec-ground-state}).
Therefore, we conclude that the interaction is antiferromagnetic.

\subsection{Eigenfunctions of the Quadrupole-Quadrupole Interaction}

The Hamiltonian of the quadrupole-quadrupole interaction between
the atoms is,
\begin{eqnarray}
  {\mathcal{H}}_{\mathrm{q}} &=&
  \lambda_{\mathrm{q}}~
  \sum_{\alpha,\alpha'}
  \hat{F}^{\alpha,\alpha'}
  \hat{S}^{\alpha,\alpha'},
  \label{H-q-def-append}
\end{eqnarray}
where $\hat{F}^{\alpha,\alpha'}$ or $\hat{S}_{a}^{\alpha,\alpha'}$
is a quadrupole operator for the atom with spin-$\frac{3}{2}$
or spin-$\frac{9}{2}$,
\begin{eqnarray*}
  \hat{F}^{\alpha,\alpha'} &=&
  \hat{F}^{\alpha}
  \hat{F}^{\alpha'}+
  \hat{F}^{\alpha'}
  \hat{F}^{\alpha}-
  \frac{2}{3}~
  F(F+1)~
  \delta_{\alpha,\alpha'},
  \\
  \hat{S}^{\alpha,\alpha'} &=&
  \hat{S}^{\alpha}
  \hat{S}^{\alpha'}+
  \hat{S}^{\alpha'}
  \hat{S}^{\alpha}-
  \frac{2}{3}~
  S(S+1)~
  \delta_{\alpha,\alpha'}.
\end{eqnarray*}

${\mathcal{H}}_{\mathrm{q}}$ can be expressed in terms of
$(\hat{\mathbf{F}}\cdot\hat{\mathbf{S}})$ as,
\begin{eqnarray*}
  {\mathcal{H}}_{\mathrm{q}} &=&
  \lambda_{\mathrm{q}}~
  \Big\{
      4~
      \big(
          \hat{\mathbf{F}}
          \cdot
          \hat{\mathbf{S}}
      \big)^{2}+
      2~
      \big(
          \hat{\mathbf{F}}
          \cdot
          \hat{\mathbf{S}}
      \big)-
  \nonumber \\ && -
      \frac{4}{3}~
      F(F+1)~
      S(S+1)
  \Big\}.
\end{eqnarray*}

Eigenfunctions of ${\mathcal{H}}_{\mathrm{q}}$ are $|L,L_z\rangle$.
Corresponding eigenenergies are,
\begin{eqnarray}
  {\mathcal{E}}^{({\mathrm{q}})}_{L} &=&
  \lambda_{\mathrm{q}}~
  {\mathcal{Q}}_{L},
  \label{En-q-res-append}
\end{eqnarray}
where
\begin{eqnarray*}
  {\mathcal{Q}}_{L} =
  4~
  {\mathcal{D}}_{L}^{2}+
  2~
  {\mathcal{D}}_{L}-
  \frac{4}{3}~
  F(F+1)~
  S(S+1),
\end{eqnarray*}
${\mathcal{D}}_{L}$ is given by eq. (\ref{SL-def-append}).

The spin $L$ takes the values $L=3,4,5,6$. Corresponding energies
are,
\begin{eqnarray}
  {\mathcal{E}}^{({\mathrm{q}})}_{3}
  &=&
  132~
  \lambda_{\mathrm{q}},
  \nonumber
  \\
  {\mathcal{E}}^{({\mathrm{q}})}_{4}
  &=&
  -60~
  \lambda_{\mathrm{q}},
  \nonumber
  \\
  {\mathcal{E}}^{({\mathrm{q}})}_{5}
  &=&
  -120~
  \lambda_{\mathrm{q}},
  \label{En-qq-res-append}
  \\
  {\mathcal{E}}^{({\mathrm{q}})}_{6}
  &=&
  72~
  \lambda_{\mathrm{q}}.
  \nonumber
\end{eqnarray}
It is seen that when $\lambda_{\mathrm{q}}<0$, the lowest
energy level is the $L=3$ energy level. For
$\lambda_{\mathrm{q}}>0$, the lowest energy level is
the $L=5$ energy level.

\subsection{Eigenfunctions of the Octupole-Octupole Interaction}

The Hamiltonian of the octupole-octupole interaction between
the impurity and the itinerant atoms is,
\begin{eqnarray}
  {\mathcal{H}}_{\mathrm{o}} &=&
  \lambda_{\mathrm{o}}~
  \sum_{\alpha,\alpha',\alpha''}
  \hat{F}^{\alpha,\alpha',\alpha''}
  \hat{S}^{\alpha,\alpha',\alpha''},
  \label{H-o-def-append}
\end{eqnarray}
where $\hat{F}^{\alpha,\alpha',\alpha''}$ or
$\hat{S}^{\alpha,\alpha',\alpha''}$ is an octupole operator for
the atom with spin-$\frac{3}{2}$ or spin-$\frac{9}{2}$,
see eq. (\ref{octupole-def}).

Substituting eq. (\ref{octupole-def}) into eq. (\ref{H-o-def-append}),
we get
\begin{eqnarray*}
  {\mathcal{H}}_{\mathrm{o}} &=&
  \lambda_{\mathrm{o}}~
  \bigg\{
       36~
       \big(
           \hat{\mathbf{F}}
           \cdot
           \hat{\mathbf{I}}
       \big)^{3}+
       72~
       \big(
           \hat{\mathbf{F}}
           \cdot
           \hat{\mathbf{I}}
       \big)^{2}+
       12~
       \big(
           \hat{\mathbf{F}}
           \cdot
           \hat{\mathbf{I}}
       \big)-
  \nonumber \\ && -
       \frac{12}{5}~
       \big(
           3F(F+1)-1
       \big)
       \big(
           3S(S+1)-1
       \big)
       \big(
           \hat{\mathbf{F}}
           \cdot
           \hat{\mathbf{I}}
       \big)-
  \nonumber \\ && -
       18~
       F(F+1)~
       S(S+1)
  \bigg\}.
\end{eqnarray*}

The spin $L$ takes the values $L=3,4,5,6$. Corresponding energies
are,
\begin{eqnarray}
  {\mathcal{E}}^{({\mathrm{o}})}_{3}
  &=&
  -\frac{11088}{5}~
  \lambda_{\mathrm{o}},
  \nonumber
  \\
  {\mathcal{E}}^{({\mathrm{o}})}_{4}
  &=&
  \frac{22368}{5}~
  \lambda_{\mathrm{o}},
  \nonumber
  \\
  {\mathcal{E}}^{({\mathrm{o}})}_{5}
  &=&
  -\frac{14787}{5}~
  \lambda_{\mathrm{o}},
  \label{En-oo-res-append}
  \\
  {\mathcal{E}}^{({\mathrm{o}})}_{6}
  &=&
  2997~
  \lambda_{\mathrm{o}}.
  \nonumber
\end{eqnarray}
It is seen that when $\lambda_{\mathrm{o}}<0$, the lowest
energy level is the $L=4$ energy level.
When $\lambda_{\mathrm{o}}>0$, the lowest energy level is
the $L=5$ energy level.

\begin{figure}[htb]
\centering
  \includegraphics[width=30 mm,angle=0]
   {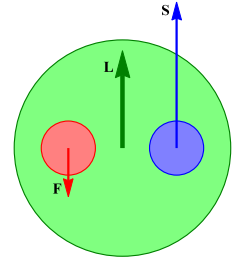}
 \caption{\footnotesize
   ({\color{blue}Color online})
   Impurity atom with spin $F$ (red disk),
   itinerant atoms with total spin $S$ (blue disk) creating
   a cloud screening the spin of the impurity. The spin of
   the ``dressed'' impurity is $L$.}
 \label{Fig-overscreening}
\end{figure}

Screening of the impurity spin by a cloud of itinerant atoms is
illustrated in Fig. \ref{Fig-overscreening}. Here the red disk
denotes the impurity atom with spin $F$, the blue disk denotes
a cloud of itinerant atoms with the total spin $S$. The green
arrow is a ``dressed'' spin of the impurity
${\mathbf{L}}={\mathbf{F}}+{\mathbf{S}}$.
When the lowest energy state is $|3,L_z\rangle$ or
$|4,L_z\rangle$, the ``dressed'' spin of the impurity $\mathbf{L}$
is antiparallel to  the ``bare'' spin $\mathbf{F}$ [see inequality
(\ref{cond-over-scr})], and therefore we deal with over-screened
Kondo effect.

\section{Yb(\3P2) Atom in Magnetic Field}
  \label{append-Yb-MF}
\begin{figure}[htb]
\centering \subfigure[]
  {\includegraphics[width=60 mm,angle=0]
   {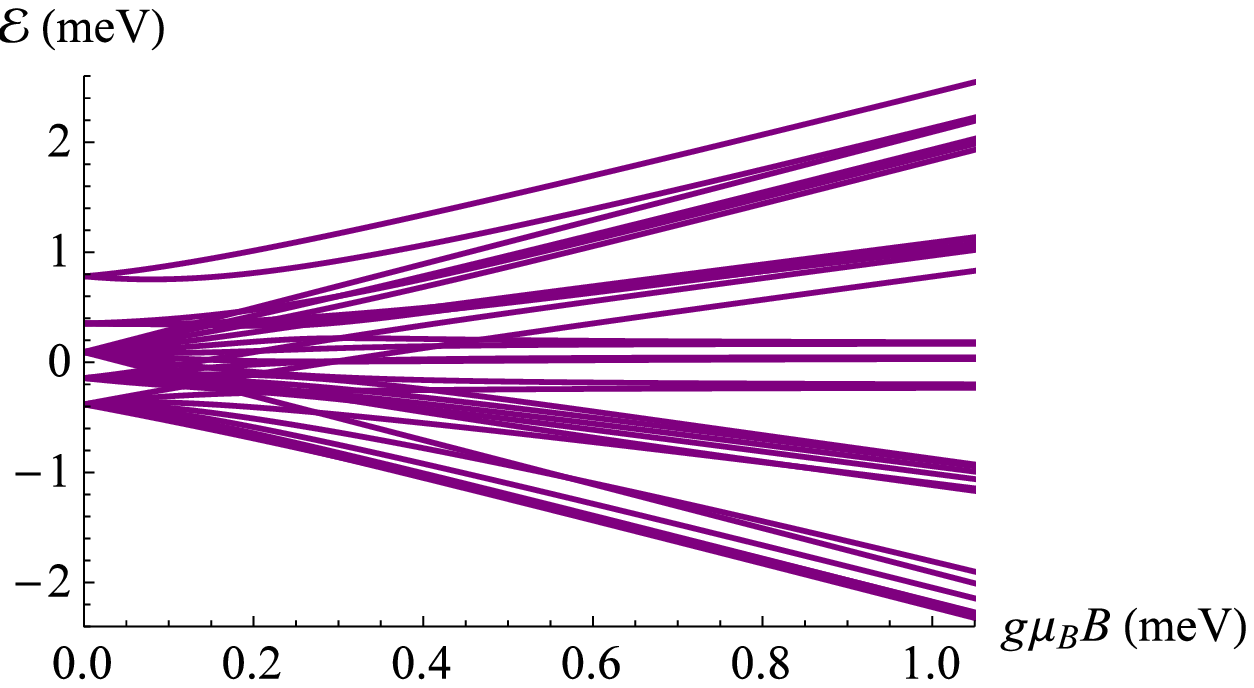}
   \label{Fig-Zeeman-gen}}
\centering \subfigure[]
  {\includegraphics[width=60 mm,angle=0]
   {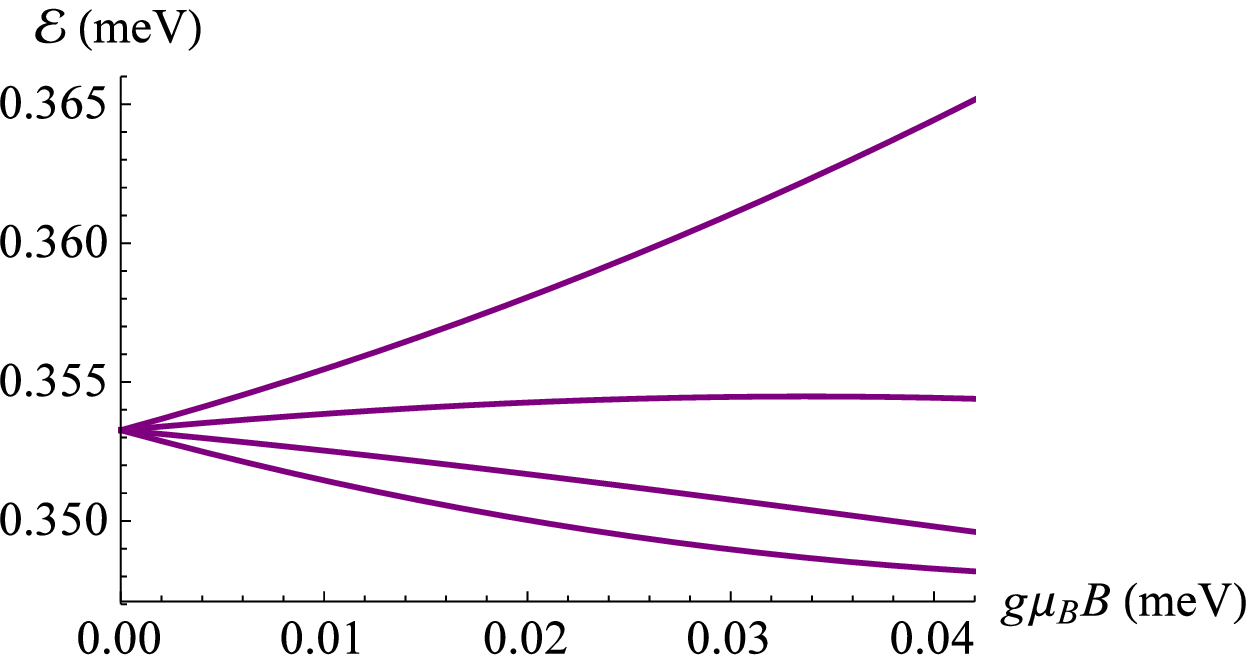}
   \label{Fig-Zeeman-small}}
 \caption{\footnotesize
   ({\color{blue}Color online})
   Energy spectrum of the $^{173}$Yb atom in the $^{3}$P$_{2}$
   quantum state [panel (a)].
   Zeeman splitting of the $F=\frac{3}{2}$ energy level by
   a weak magnetic field [panel (b)].}
 \label{Fig-Zeeman}
\end{figure}

  \noindent
\underline{Appendices \ref{append-Yb-MF} and
\ref {append-average-D-Q-O} main points:} Although we do not
subject our system to an external magnetic field (since it is
detrimental for the Kondo effect) we find it useful to employ our
detailed analysis of Yb atoms and inspect their properties under
an application of a weak magnetic field. In particular,
the multipole analysis worked out in this paper helps us to
elucidate the pattern of the dependence of energy levels on
the magnetic field, both for the ground-state $^1$S$_0$ and
the excited state $^3$P$_2$. This is shown in Figs.
\ref{Fig-Zeeman-gen}, and \ref{Fig-Zeeman-small}.

Consider an $^{3}$P$_{2}$ Yb atom in external magnetic field.
The Hamiltonian of the atom is,
\begin{eqnarray}
  {\mathcal{H}}_{\mathrm{at}} &=&
  {\mathcal{H}}_{\mathrm{at}}^{(0)}+
  {\mathcal{H}}_{B},
  \label{H=H0+HB-atom-append}
\end{eqnarray}
where ${\mathcal{H}}_{\mathrm{at}}^{(0)}$ is a Hamiltonian of
the isolated $^{173}$Yb atom in the $^{3}$P$_{2}$ state and
${\mathcal{H}}_{B}$ describes interaction of the atom with the magnetic field,
\begin{eqnarray}
  &&
  {\mathcal{H}}_{\mathrm{at}}^{(0)}
  =
  A_{\mathrm{d}}
  \sum_{\alpha}
  \hat{I}^{\alpha}
  \hat{J}^{\alpha}+
  A_{\mathrm{q}}
  \sum_{\alpha,\alpha'}
  \hat{I}^{\alpha,\alpha'}
  \hat{J}^{\alpha,\alpha'},
  \label{H0-atom-append}
  \\
  &&
  {\mathcal{H}}_{B}
  ~=
  -g
  \mu_B
  B~
  \hat{J}^{z}.
  \label{HB-atom-append}
\end{eqnarray}
Here $\hat{I}^{\alpha}$ and $\hat{I}^{\alpha,\alpha'}$ are spin
and quadrupole angular momentum operators for the nucleus, whereas
$J^{\alpha}$ and $J^{\alpha,\alpha'}$ are orbital angular momentum
and quadrupole angular momentum operators of the $^{3}$P$_{2}$
electronic configuration. $g$ is the electronic g-factor of the Yb
atom in the $^{3}$P$_{2}$ state, see eq. (\ref{g-factor-3P2}).

The constants $A_{\mathrm{d}}$ and $A_{\mathrm{q}}$ are
\cite{Yb-hyperfine-JPhysB99},
\begin{eqnarray}
  \frac{A_{\mathrm{d}}}{h} ~=~
  -738~{\mathrm{MHz}},
  \ \ \
  \frac{A_{\mathrm{q}}}{h} ~=~
  1312~{\mathrm{MHz}},
  \label{A-B-Yb-3P2}
\end{eqnarray}
where $h$ is the Planck constant, or
\begin{eqnarray*}
  A_{\mathrm{d}} ~=~
  -3.052~
  \mu{\text{eV}},
  \ \ \
  A_{\mathrm{d}} ~=~
  5.426~
  \mu{\text{eV}}.
\end{eqnarray*}

Taking into account definition (\ref{quadrupole-def}) for
the quadrupole angular momentum operators, we can write
the Hamiltonian (\ref{H0-atom-append}) in the form,
\begin{eqnarray}
  {\mathcal{H}}_{\mathrm{at}}^{(0)} &=&
  \big(
      A_{\mathrm{d}}-
      2~
      A_{\mathrm{q}}
  \big)~
  \big(
      \hat{\mathbf{I}}
      \cdot
      \hat{\mathbf{J}}
  \big)+
  4~
  A_{\mathrm{q}}~
  \big(
      \hat{\mathbf{I}}
      \cdot
      \hat{\mathbf{J}}
  \big)^{2}-
  \nonumber \\ && -
  \frac{4}{3}~
  A_{\mathrm{q}}
  I(I+1)
  J(J+1).
  \label{H0-atom-IJ-append}
\end{eqnarray}

Eq. (\ref{H0-atom-IJ-append}) shows that eigenfunctions of
${\mathcal{H}}_{\mathrm{at}}^{(0)}$ are also eigenfunctions
of the operators $\hat{\mathbf{F}}^{2}$ and $\hat{F}^{z}$
[where $\hat{\mathbf{F}}=\hat{\mathbf{I}}+\hat{\mathbf{J}}$
is the operator of the total atomic orbital momentum],
\begin{eqnarray}
  \big|
      F,f
  \big\rangle
  &=&
  \sum_{i,j}
  C^{F,f}_{I,i;J,j}
  \big|
      I,i;~
      J,j
  \big\rangle,
  \label{WF-L-lz-def-append}
\end{eqnarray}
where $i$, $j$ and $f$ are nuclear, electronic and total atomic
magnetic quantum numbers.
The wave functions $|I,i;J,j\rangle$ as eigenfunctions of
the operators $\hat{I}^{z}$ and $\hat{J}^{z}$,
\begin{eqnarray*}
  \hat{I}^{z}~
  \big|
      I,i;~
      J,j
  \big\rangle
  &=&
  i~
  \big|
      I,i;~
      J,j
  \big\rangle,
  \\
  \hat{J}^{z}~
  \big|
      I,i;~
      J,j
  \big\rangle
  &=&
  j~
  \big|
      I,i;~
      J,j
  \big\rangle.
\end{eqnarray*}

Corresponding eigenenergies ${\mathcal{E}}_{L}^{(0)}$ are
\begin{eqnarray*}
  {\mathcal{E}}_{\frac{1}{2}}^{(0)} &=&
  -7 A_{\mathrm{d}} + 140 A_{\mathrm{q}}
  ~=~
  781.0~\mu{\text{eV}},
  \\
  {\mathcal{E}}_{\frac{3}{2}}^{(0)} &=&
  -\frac{11}{2}~A_{\mathrm{d}} + 62 A_{\mathrm{q}}
  ~=~
  353.2~\mu{\text{eV}},
  \\
  {\mathcal{E}}_{\frac{5}{2}}^{(0)} &=&
  -3 A_{\mathrm{d}} - 28 A_{\mathrm{q}}
  ~=~
  -142.8~\mu{\text{eV}},
  \\
  {\mathcal{E}}_{\frac{7}{2}}^{(0)} &=&
  \frac{1}{2}~A_{\mathrm{d}}-70~A_{\mathrm{q}}
  ~=~
  -381.3~\mu{\text{eV}},
  \\
  {\mathcal{E}}_{\frac{9}{2}}^{(0)} &=&
  5~A_{\mathrm{d}}+20~A_{\mathrm{q}}
  ~=~
  93.2~\mu{\text{eV}}.
\end{eqnarray*}

The interaction Hamiltonian ${\mathcal{H}}_{B}$, eq. (\ref{HB-atom-append}),
commutes with the operator $\hat{F}^{z}$, but nt with $\hat{\mathbf{F}}^{2}$.
Therefore, eigenfunctions of the Hamiltonian ${\mathcal{H}}_{\mathrm{at}}$,
eq. (\ref{H=H0+HB-atom-append}), are described by the magnetic quantum
numbers $f$, but not by the total atomic spin $F$.

In order to find eigenenergies of the Hamiltonian (\ref{H=H0+HB-atom-append}),
we find the matric elements of ${\mathcal{H}}_{B}$,
\begin{eqnarray}
  V_{F,F'}^{(f)}
  &=&
  \big\langle
      F,f
  \big|
      {\mathcal{H}}_{B}
  \big|
      F',f
  \big\rangle =
  \nonumber \\ &=&
  -g
  \mu_B
  B~
  {\mathcal{C}}_{F,F'}^{(f)},
  \label{matrix-elements-HB-append}
\end{eqnarray}
where
\begin{eqnarray*}
  {\mathcal{C}}_{F,F'}^{(f)}
  &=&
  \sum_{i,j}
  j
  C^{F,f}_{I,i;J,j}
  C^{F',f}_{I,i;J,j},
\end{eqnarray*}
where $f=-\frac{9}{2},-\frac{7}{2},\ldots,\frac{9}{2}$
and ${f}\leq{F,F'}\leq\frac{9}{2}$.
Then the eigenenergies of ${\mathcal{H}}_{\mathrm{at}}$
are found from diagonalization of the matrices $\hat{h}^{(f)}$
with matrix elements $h_{F,F'}^{(f)}$ given by,
\begin{eqnarray}
  h_{F,F'}^{(f)} &=&
  {\mathcal{E}}_{F}^{(0)}~
  \delta_{F,F'}+
  V_{F,F'}^{(f)}.
  \label{matrix-elements-H-atom-append}
\end{eqnarray}

The eigenvalues of the Hamiltonian (\ref{H=H0+HB-atom-append})
as functions of the magnetic field are shown in Fig.
\ref{Fig-Zeeman-gen}. It is seen that for weak magnetic field
[when $g \mu_B B$ is small with respect to the hyperfine
splitting], every energy level ${\mathcal{E}}_{F}^{(0)}$ splits
into $2F+1$ spectral lines with energies ${\mathcal{E}}_{F,f}$
given by
\begin{eqnarray}
  {\mathcal{E}}_{F,f} &=&
  {\mathcal{E}}_{F}^{(0)}-
  g
  \mu_B
  B~
  {\mathcal{C}}_{F,F}^{(f)}.
  \label{En-F-fz-weak}
\end{eqnarray}

For strong magnetic field  [when $g \mu_B B$ is large with
respect to the hyperfine splitting], the $^{3}$P$_{2}$ energy
level splits into 5 levels with $j_z=0,\pm1,\pm2$, and every
level splits into six levels by the hyperfine interaction.

Splitting of the $F=\frac{3}{2}$ energy level (that we are
interested in) is,
\begin{eqnarray}
  {\mathcal{E}}_{\frac{3}{2},f} &=&
  {\mathcal{E}}_{F}^{(0)}-
  \frac{13}{30}~
  g
  \mu_B
  B-
  f~
  g
  \mu_B
  B.
  \label{En-3/2-fz-weak}
\end{eqnarray}
Energies ${\mathcal{E}}_{\frac{3}{2},f}$ calculated numerically
by diagonalization of the matrices (\ref{matrix-elements-H-atom-append})
are shown in Fig. \ref{Fig-Zeeman-small} as functions of the magnetic field.
The energies are almost linear with the magnetic field which agrees
with equation (\ref{En-3/2-fz-weak}).

\section{Averaged Dipole, Quadrupole and Octupole Moments}
  \label{append-average-D-Q-O}

The density matrix of the impurity atom placed in the magnetic field
${\mathbf{B}}=B{\mathbf{e}}_{z}$ is,
\begin{eqnarray}
  \hat\varrho_{\mathrm{i}} &=&
  \frac{1}{Z_{\mathrm{i}}}
  \sum_{f}
  e^{\beta g \mu_B B f}~
  X^{f,f},
  \label{DensityMatrixImpurity-append}
\end{eqnarray}
where $\beta=\frac{1}{T}$,
\begin{eqnarray*}
  Z_{\mathrm{i}} &=&
  \sum_{f}
  e^{\beta g \mu_B B f}.
\end{eqnarray*}

Expectation value of an operator $\hat{\mathcal{O}}$ acting in
the Hilbert state of quantum states of the isolated impurity is,
\begin{eqnarray*}
  \big\langle
      \hat{\mathcal{O}}
  \big\rangle
  &=&
  \frac{1}{Z_{\mathrm{i}}}
  \sum_{f}
  {\mathcal{O}}_{f,f},
\end{eqnarray*}
where ${\mathcal{O}}_{f,f}=\langle{f}|\hat{\mathcal{O}}|{f}\rangle$.

{\textbf{1}}. Expectation value of the magnetic dipole angular momentum
operator is,
\begin{eqnarray}
  \big\langle
      \hat{F}^{\alpha}
  \big\rangle
  &=&
  -{\mathcal{F}}_{\mathrm{d}}~
  \delta^{\alpha,z},
  \label{dipole-average-append}
\end{eqnarray}
where
\begin{eqnarray*}
  {\mathcal{F}}_{\mathrm{d}}
  &=&
  \frac{1}{2}~
  \tanh
  \bigg(
       \frac{g \mu_B B}{2T}
  \bigg)+
  \tanh
  \bigg(
       \frac{g \mu_B B}{T}
  \bigg).
\end{eqnarray*}

When $g\mu_B B \ll T$, ${\mathcal{F}}_{d}$ can be written in
the linear with $B$ approximation as,
\begin{eqnarray*}
  {\mathcal{F}}_{\mathrm{d}}
  &=&
  \frac{5}{4}~
  \frac{g \mu_B B}{T}+
  O\bigg(\frac{\mu_{B}^{3}B^3}{T^3}\bigg).
\end{eqnarray*}

{\textbf{2}}. Expectation value of the magnetic quadrupole angular momentum
operator is,
\begin{eqnarray}
  \big\langle
      \hat{F}^{\alpha,\alpha'}
  \big\rangle
  &=&
  -{\mathcal{F}}_{\mathrm{q}}~
  \delta^{\alpha,\alpha'}~
  \Big\{
      \delta^{\alpha,x}+
      \delta^{\alpha,y}-
      2
      \delta^{\alpha,z}
  \Big\},
  \label{quadrupole-average-append}
\end{eqnarray}
where
\begin{eqnarray*}
  {\mathcal{F}}_{\mathrm{q}}
  &=&
  \frac{2~
        \sinh^{2}
        \big(
            \frac{g \mu_B B}{2T}
        \big)}
       {\cosh
        \big(
            \frac{g \mu_B B}{2T}
        \big)}.
\end{eqnarray*}

When $g\mu_B B \ll T$, ${\mathcal{F}}_{q}$ can be expanded with
$B$ as,
\begin{eqnarray*}
  {\mathcal{F}}_{\mathrm{q}}
  &=&
  \frac{1}{2}~
  \frac{\big(g \mu_B B\big)^{2}}{T^2}+
  O\bigg(\frac{\mu_{B}^{4}B^4}{T^4}\bigg).
\end{eqnarray*}

{\textbf{3}}. Expectation value of the magnetic octupole angular momentum
operator is,
\begin{eqnarray}
  \big\langle
      \hat{F}^{\alpha,\alpha',\alpha''}
  \big\rangle
  &=&
  {\mathcal{F}}_{\mathrm{o}}~
  \Big\{
      \delta^{\alpha,\alpha'}
      \delta^{\alpha'',z}+
      \delta^{\alpha,\alpha''}
      \delta^{\alpha',z}+
  \nonumber \\ &+&
      \delta^{\alpha',\alpha''}
      \delta^{\alpha,z}-
      5~
      \delta^{\alpha,z}
      \delta^{\alpha',z}
      \delta^{\alpha'',z}
  \Big\},
  \label{octupole-average-append}
\end{eqnarray}
where
\begin{eqnarray*}
  {\mathcal{F}}_{\mathrm{o}} &=&
  \frac{36}{5}~
  \frac{2~
        \sinh^{4}
        \big(
            \frac{g \mu_B B}{2T}
        \big)}
       {\sinh
        \big(
            \frac{2 g \mu_B B}{2T}
        \big)}.
\end{eqnarray*}

When $g\mu_B B \ll T$, ${\mathcal{F}}_{o}$ can be expanded with
$B$ as,
\begin{eqnarray*}
  {\mathcal{F}}_{\mathrm{o}}
  &=&
  \frac{9}{40}~
  \frac{\big(g \mu_B B\big)^{3}}{T^3}+
  O\bigg(\frac{\mu_{B}^{5}B^5}{T^4}\bigg).
\end{eqnarray*}

{\textbf{4}}. Expectation value of $F^{\alpha_1}F^{\alpha_2}$ is,
\begin{eqnarray}
  \frac{1}{2}~
  \big\langle
      \hat{F}^{\alpha_1}
      \hat{F}^{\alpha_2}+
      \hat{F}^{\alpha_2}
      \hat{F}^{\alpha_1}
  \big\rangle
  =
  \bigg(
       \frac{5}{4}+
       {\mathcal{F}}_{\mathrm{d,d}}^{\alpha_1}
  \bigg)~
  \delta^{\alpha_1,\alpha_2},
  \label{dipole-dipole-average-append}
\end{eqnarray}
where
\begin{eqnarray*}
  {\mathcal{F}}_{\mathrm{d,d}}^{\alpha} &=&
  \Big(
      3~
      \delta^{\alpha,z}-
      1
  \Big)~
  \frac{2~
        \sinh^{2}
        \big(
            \frac{g \mu_B B}{2T}
        \big)}
       {\cosh
        \big(
            \frac{g \mu_B B}{2T}
        \big)}.
\end{eqnarray*}

When $g\mu_B B \ll T$, ${\mathcal{F}}_{d,d}$ can be expanded with
$B$ as,
\begin{eqnarray*}
  {\mathcal{F}}_{\mathrm{d,d}}^{\alpha} &=&
  \Big(
      3~
      \delta^{\alpha,z}-
      1
  \Big)~
  \frac{1}{2}~
  \frac{\big(g \mu_B B\big)^{2}}{T^2}+
  O\bigg(\frac{\mu_{B}^{4}B^4}{T^4}\bigg).
\end{eqnarray*}

{\textbf{5}}. Expectation value of $F^{\alpha_1}F^{\alpha_2,\alpha'_2}$ is,
\begin{eqnarray}
  &&
  \frac{1}{2}~
  \big\langle
      \hat{F}^{\alpha_1}
      \hat{F}^{\alpha_2,\alpha'_2}+
      \hat{F}^{\alpha_2,\alpha'_2}
      \hat{F}^{\alpha_1}
  \big\rangle
  =
  \nonumber \\ && ~~~~~ =
  \Big(
      \delta^{\alpha_1,\alpha_2}
      \delta^{\alpha'_2,z}+
      \delta^{\alpha_1,\alpha'_2}
      \delta^{\alpha_2,z}-
  \nonumber \\ && ~~~~~ ~~~ -
      2~
      \delta^{\alpha_1,z}
      \delta^{\alpha_2,z}
      \delta^{\alpha'_2,z}
  \Big)~
  {\mathcal{F}}_{\mathrm{d,q}}^{(1)}+
  \nonumber
  \\ && ~~~~~ +
  \delta^{\alpha_2,\alpha'_2}~
  \Big(
      \delta^{\alpha_1,z}-
      2~
      \delta^{\alpha_1,z}
      \delta^{\alpha_2,z}
  \Big)~
  {\mathcal{F}}_{\mathrm{d,q}}^{(2)},
  \label{dipole-quadrupole-average-append}
\end{eqnarray}
where
\begin{eqnarray*}
  {\mathcal{F}}_{\mathrm{d,q}}^{(1)}
  &=&
  -\frac{3}{2}~
  \tanh
  \bigg(
       \frac{g \mu_B B}{T}
  \bigg),
  \\
  {\mathcal{F}}_{\mathrm{d,q}}^{(2)}
  &=&
  \tanh
  \bigg(
       \frac{g \mu_B B}{2T}
  \bigg)-
  \frac{1}{2}~
  \tanh
  \bigg(
       \frac{g \mu_B B}{T}
  \bigg).
\end{eqnarray*}

When $g\mu_B B \ll T$, ${\mathcal{F}}_{d,q}^{(1,2)}$ can be expanded with
$B$ as,
\begin{eqnarray*}
  {\mathcal{F}}_{\mathrm{d,q}}^{(1)} &=&
  -\frac{3}{2}~
  \frac{g \mu_B B}{T}+
  O\bigg(\frac{\mu_{B}^{3}B^3}{T^3}\bigg),
  \\
  {\mathcal{F}}_{\mathrm{d,q}}^{(2)} &=&
  \frac{g \mu_B B}{T}+
  O\bigg(\frac{\mu_{B}^{3}B^3}{T^3}\bigg).
\end{eqnarray*}

{\textbf{6}}. Expectation value of $F^{\alpha_1}F^{\alpha_2,\alpha'_2,\alpha''_2}$ is,
\begin{eqnarray}
  &&
  \frac{1}{2}~
  \big\langle
      \hat{F}^{\alpha_1}
      \hat{F}^{\alpha_2,\alpha'_2,\alpha''_2}+
      \hat{F}^{\alpha_2,\alpha'_2,\alpha''_2}
      \hat{F}^{\alpha_1}
  \big\rangle
  =
  \nonumber \\ && ~~~~~ =
  {\mathcal{A}}^{\alpha_1;\alpha_2,\alpha'_2,\alpha''_2}
  {\mathcal{F}}_{\mathrm{d,o}},
  \label{dipole-octupole-average-append}
\end{eqnarray}
where ${\mathcal{A}}^{\alpha_1;\alpha_2,\alpha'_2,\alpha''_2}$
is symmetric with $\alpha_2,\alpha'_2,\alpha''_2$ tensor which does
not depend on temperature or magnetic field,
\begin{eqnarray*}
  {\mathcal{F}}_{\mathrm{d,o}} &=&
  \frac{2~
        \sinh^{2}
        \big(
            \frac{g \mu_B B}{2T}
        \big)}
       {\cosh
        \big(
            \frac{g \mu_B B}{2T}
        \big)}.
\end{eqnarray*}

When $g\mu_B B \ll T$, ${\mathcal{F}}_{d,o}$ can be expanded with
$B$ as,
\begin{eqnarray*}
  {\mathcal{F}}_{\mathrm{d,o}} &=&
  \frac{1}{2}~
  \frac{\big(g \mu_B B\big)^{2}}{T^2}+
  O\bigg(\frac{\mu_{B}^{4}B^4}{T^4}\bigg).
\end{eqnarray*}


\end{document}